\documentclass[10pt]{article}
\usepackage{mathrsfs}
\usepackage{subfigure}
\usepackage{graphicx}
\usepackage{amsmath}
\usepackage{amssymb}
\usepackage{caption2}
\usepackage{tabularx}
\usepackage{dcolumn}
\usepackage{array}
\usepackage{setspace}
\usepackage{booktabs}
\usepackage[figuresright]{rotating}
\begin{document}
\bibliographystyle{prsty}
\begin{center}
{\large {\bf \sc{  Study of isospin eigenstates of the pentaquark molecular states with strangeness }}} \\[2mm]
Xiu-Wu Wang$^{*\dagger}$,
Zhi-Gang  Wang$^*$\footnote{E-mail: zgwang@aliyun.com.  } \\
 Department of Physics, North China Electric Power University, Baoding 071003, P. R. China$^*$\\
 School of Nuclear Science and Engineering, North China Electric Power University, Beijing 102206, P. R. China$^\dagger$
\end{center}

\begin{abstract}
In the paper, we construct eight color singlet-singlet type five-quark currents with distinguished isospins  to study the $\bar{D}\Xi^{\prime}$, $\bar{D}\Xi_c^{*}$, $\bar{D}^{*}\Xi_c^{\prime}$ and $\bar{D}^{*}\Xi_c^{*}$ molecular states with strangeness via the QCD sum rules.  Numerical results show that the central values of the pentaquark masses  with higher (lower) isospin are slightly above (below) the thresholds of the corresponding   meson-baryon pairs, and support assigning the $P_{cs}(4459)$ as the $\bar{D}\Xi_c^{*}$ molecular state with the quantum numbers $IJ^P=0{\frac{3}{2}}^-$. The other predictions can be confronted to the experimental data in the future.
\end{abstract}

 PACS number: 12.39.Mk, 14.20.Lq, 12.38.Lg

Key words: Pentaquark states with strangeness, QCD sum rules

\section{Introduction}
In the past few decades, many exotic $X$, $Y$, $Z$ particles have been observed at the Belle, BaBar, BESIII and LHCb collaborations \cite{PDG}, the intriguing fact is that many of their masses are near the meson-meson  thresholds, which shed light on the possible hadronic molecule interpretations \cite{Guo1}. In 2015, the LHCb collaboration observed two hidden-charm pentaquark candidates in the $\Lambda_b^0\rightarrow J/\psi pK^-$ decay process \cite{RAaij1}, namely, the $P_c(4380)$ and $P_c(4450)$. In 2019, the observations were updated and the $P_c(4312)$, $P_c(4440)$ and $P_c(4457)$ were reported by the LHCb collaboration \cite{RAaij2}, it reported that the  $P_c(4450)$ is actually the overlapping peak of the $P_c(4440)$ and $P_c(4457)$. In the present work, we focus on the observation reported by the LHCb collaboration in 2020 for the hidden-charm strange pentaquark $P_{cs}(4459)$ in the $J/\psi\Lambda$ mass spectrum from amplitude analysis of the $\Xi^-_b\rightarrow J/\psi \Lambda K^-$ decay \cite{RAaij3}, the state's mass and width are $4458.8\pm2.9^{+4.7}_{-1.1}\,\rm{MeV}$ and $17.3\pm6.5^{+8.0}_{-5.7}\,\rm{MeV}$, respectively.

Due to the exotic hadronic structures, the $P_c$ states have been attracting lots of interests in the strong interaction area \cite{Guo1,Esposito,Lebed,YRliu}.  Now, for those $P_c$ states, a typical interpretation is that they are the S-wave hidden-charm meson--baryon molecules with definite isospin $I$, spin $J$ and parity $P$ \cite{Manuel,MingZhu,FuLai,MengLin,JunHe}, inspired by the interpretation of the exotic $P_c$ states, many theoretic groups interpret the newly discovered $P_{cs}(4459)$ in a similar way. For example, in Ref.\cite{FYang}, Yang, Huang and Zhu  assume the $P_{cs}(4459)$ as the $\bar{D}^*\Xi_c$ molecular state and study its strong decays  via  considering   its $J^P$ as ${\frac{3}{2}}^-$ and ${\frac{1}{2}}^-$. In the framework of the QCD sum rules \cite{HXChenN}, the investigation  supports  assigning the $P_{cs}(4459)$ as the $\bar{D}^*\Xi_c$ hadronic molecular state with either $J^P={\frac{1}{2}}^-$ or ${\frac{3}{2}}^-$. Applying the quasi-potential Bethe-Salpeter equation approach \cite{JTZhuN}, Zhu, Song and He interpret the $P_{cs}(4459)$ as the $\bar{D}^*\Xi_c$ molecule with the $J^P={\frac{3}{2}}^-$. Under the one-boson-exchange model, Chen concludes  that this exotic state is not the pure molecular state \cite{RChenN}. As for the other arguments about the properties of the $P_{cs}(4459)$, one can consult the Refs.\cite{CWXiao,WZG-Pcs,KAzizi,Uozdem,YHuang,PPShi} and so on.

Since the spin-parity $J^P$ of the $P_{cs}(4459)$ has not been determined experimentally yet,  the nature of this exotic state is still under hot debate. In Ref.\cite{WangZG-Xin-CPC}, our group apply the color singlet-singlet type pentaquark currents to study the $P_c$ and $P_{cs}$ states in a systemic way via the QCD sum rules, and assign the $P_{cs}(4459)$ with the $J^P$ either to be ${\frac{1}{2}}^-$ or to be ${\frac{3}{2}}^-$, in that paper, the color singlet-singlet type pentaquark currents being the isospin eigenstates are proposed.
In Ref.\cite{wangxiuwuN}, the isospins are unambiguously distinguished to study the hadronic molecules in the framework of the QCD sum rules  in details for the first time and the $P_c(4312)$, $P_c(4380)$, $P_c(4440)$ and $P_c(4457)$ are assigned as the $\bar{D}^{(*)}\Sigma_c^{(*)}$ molecules with the isospin $I=\frac{1}{2}$. Intrigued by our previous works, we are very interested to investigate the present topic: what about the situation if we distinguish  the isospins of  the pentaquark states with strangeness? Can we clearly determine the nature of $P_{cs}(4459)$? What about the  properties of the other possible pentaquark states with strangeness?

Among the popular theoretic methods, the QCD sum rules approach is a powerful theoretical tool to study the strong interactions  \cite{SVZ1,SVZ2,Reinders}. It has achieved many successful descriptions, such as the tetraquark states \cite{WZGXZ1-1,WZGXZ1-2,WZGXZ1-3,ZSLXZ1,CHXXZ1,QCFXZ1,WZGXZ11}, tetraquark molecular states \cite{ZhangJR-Zc3900,WZGXZ3,WZGXZ4,WZGXZ13-mole}, pentaquark states \cite{WZG-Pcs,KAzizi,Uozdem,WZGNNN1,WZGXZ5,Pimikov-Penta}, pentaquark molecular states \cite{HXChenN,Uozdem,WangZG-Xin-CPC,wangxiuwuN,CHXXZ2,CHXXZ3,WZGXZ7,WZGXZ9}, dibaryon and baryonium \cite{KodamaXZ1,ChenXZ4,WZGXZ10,WanXZ1,wangxiuwu3} and so on. However, the isospins of the states are seldom distinguished  except our previous calculation \cite{wangxiuwuN}, it shows that the mass of the higher isospin state is a few dozens of $\rm{MeV}$ above that of the lower  one. As is known, deviation of a few dozens of $\rm{MeV}$ is enough to confuse the assignment of the state, thus, we argue that distinguishing  the isospin may be one of the key preconditions for the accurate assignment.

The article is organized as follows: in Sect.2, the QCD sum rules for the pentaquark molecular states are derived; the numerical results and discussions are given in Sect.3; Sect.4 is reserved for our conclusions.

\section{QCD sum rules for the pentaquark molecular  states}
In the isospin space, the $u$ and $d$ quarks have the isospin eigenvalues $\frac{1}{2}$ and $-\frac{1}{2}$, respectively, thus the $\bar{D}^0$, $\bar{D}^{*0}$, $\bar{D}^-$, $\bar{D}^{*-}$, $\Xi_c^{\prime 0}$, $\Xi_c^{*0}$, $\Xi_c^{\prime +}$ and $\Xi_c^{*+}$ correspond to the isospin eigenstates $|\frac{1}{2},\frac{1}{2}\rangle$, $|\frac{1}{2},\frac{1}{2}\rangle$, $|\frac{1}{2},-\frac{1}{2}\rangle$, $|\frac{1}{2},-\frac{1}{2}\rangle$, $|\frac{1}{2},-\frac{1}{2}\rangle$, $|\frac{1}{2},-\frac{1}{2}\rangle$, $|\frac{1}{2},\frac{1}{2}\rangle$ and $|\frac{1}{2},\frac{1}{2}\rangle$, respectively. We can apply the following color-singlet currents to interpolate the above mesons and baryons,
\begin{eqnarray}
J^{\bar{D}^0}(x)&=&\bar{c}(x)i\gamma_5u(x)\, ,\nonumber \\
J^{\bar{D}^-}(x)&=&\bar{c}(x)i\gamma_5d(x)\, ,\nonumber \\
J^{\bar{D}^{*0}}_{\mu}(x)&=&\bar{c}(x)\gamma_{\mu} u(x)\, , \nonumber\\
J^{\bar{D}^{*-}}_{\mu}(x)&=&\bar{c}(x)\gamma_{\mu} d(x)\, , \nonumber\\
J^{\Xi_c^{\prime 0}}(x)&=&\varepsilon^{ijk}d^{iT}(x)C\gamma_{\mu}s^j(x)\gamma^{\mu}\gamma_5c^k(x)\, ,\nonumber\\
J^{\Xi_c^{*0}}_{\mu}(x)&=&\varepsilon^{ijk}d^{iT}(x)C\gamma_{\mu}s^j(x)c^k(x)\, , \nonumber\\
J^{\Xi_c^{\prime +}}(x)&=&\varepsilon^{ijk}u^{iT}(x)C\gamma_{\mu}s^j(x)\gamma^{\mu}\gamma_5c^k(x)\, , \nonumber\\
J^{\Xi_c^{*+}}_{\mu}(x)&=&\varepsilon^{ijk}u^{iT}(x)C\gamma_{\mu}s^j(x)c^k(x)\, ,
\end{eqnarray}
where the superscripts $i, j, k$ are the color indices and the $C$ represents the charge conjugation matrix. Based on the above currents for the mesons and baryons, we construct the color singlet-singlet type five-quark currents to study the $\bar{D}\Xi_c^{\prime}$, $\bar{D}\Xi_c^{*}$, $\bar{D}^*\Xi_c^{\prime}$ and $\bar{D}^*\Xi_c^{*}$ pentaquark (molecular) sates,
\begin{eqnarray}
J_{0}^{\bar{D}\Xi_c^{\prime}}(x)&=&\frac{1}{\sqrt{2}}J^{\bar{D}^0}(x)J^{\Xi_c^{\prime 0}}(x)-\frac{1}{\sqrt{2}}J^{\bar{D}^-}(x)J^{\Xi_c^{\prime +}}(x) \, , \nonumber\\
J_{1}^{\bar{D}\Xi_c^{\prime}}(x)&=&\frac{1}{\sqrt{2}}J^{\bar{D}^0}(x)J^{\Xi_c^{\prime0}}(x)+\frac{1}{\sqrt{2}}J^{\bar{D}^-}(x)J^{\Xi_c^{\prime+}}(x)\, ,\nonumber\\
J_{0;\mu}^{\bar{D}\Xi_c^*}(x)&=&\frac{1}{\sqrt{2}}J^{\bar{D}^0}(x)J^{\Xi_c^{*0}}_{\mu}(x)-\frac{1}{\sqrt{2}}J^{\bar{D}^-}(x)J^{\Xi_c^{*+}}_{\mu}(x)\, ,\nonumber\\
J_{1;\mu}^{\bar{D}\Xi_c^*}(x)&=&\frac{1}{\sqrt{2}}J^{\bar{D}^0}(x)J^{\Xi_c^{*0}}_{\mu}(x)+\frac{1}{\sqrt{2}}J^{\bar{D}^-}(x)J^{\Xi_c^{*+}}_{\mu}(x)\, ,\nonumber\\
J_{0;\mu}^{\bar{D}^{*}\Xi_c^{\prime}}(x)&=&\frac{1}{\sqrt{2}}J^{\bar{D}^{*0}}_{\mu}(x)J^{\Xi_c^{\prime0}}(x)-\frac{1}{\sqrt{2}}J^{\bar{D}^{*-}}_{\mu}(x)J^{\Xi_c^{\prime+}}(x)\, ,\nonumber\\
J_{1;\mu}^{\bar{D}^{*}\Xi_c^{\prime}}(x)&=&\frac{1}{\sqrt{2}}J^{\bar{D}^{*0}}_{\mu}(x)J^{\Xi_c^{\prime0}}(x)+\frac{1}{\sqrt{2}}J^{\bar{D}^{*-}}_{\mu}(x)J^{\Xi_c^{\prime+}}(x)\, ,\nonumber\\
J_{0;\mu\nu}^{\bar{D}^{*}\Xi_c^*}(x)&=&\frac{1}{\sqrt{2}}J^{\bar{D}^{*0}}_{\mu}(x)J^{\Xi_c^{*0}}_{\nu}(x)-\frac{1}{\sqrt{2}}J^{\bar{D}^{*-}}_{\mu}(x)J^{\Xi_c^{*+}}_{\nu}(x)+(\mu\leftrightarrow\nu)\, ,\nonumber\\
J_{1;\mu\nu}^{\bar{D}^{*}\Xi_c^*}(x)&=&\frac{1}{\sqrt{2}}J^{\bar{D}^{*0}}_{\mu}(x)J^{\Xi_c^{*0}}_{\nu}(x)+\frac{1}{\sqrt{2}}J^{\bar{D}^{*-}}_{\mu}(x)J^{\Xi_c^{*+}}_{\nu}(x)+(\mu\leftrightarrow\nu)\, ,
\end{eqnarray}
where the subscripts $0$ and $1$ stand for the isospins $I=0$ and $1$, respectively \cite{WangZG-Xin-CPC}, and those currents are isospin eigenstates, either $|0,0\rangle$ or $|1,0\rangle$. Considering the parity operator $\widehat{P}$, we can show that the above eight currents have the negative parity, since $\widehat{P}\psi\widehat{P}^{-1}=\gamma^0\psi$ and $\widehat{P}\bar{\psi}\widehat{P}^{-1}=\bar{\psi}\gamma^0$, where the $\psi$ are the quark fields.

  The two-point correlation functions are then written as,
\begin{eqnarray}
 \Pi(p)&=&i\int d^4x e^{ip\cdot x}\langle 0 |T\left\{ J (x) \bar{J}(0) \right\}| 0\rangle \, ,\nonumber\\
\Pi_{\mu\nu}(p)&=&i\int d^4x e^{ip\cdot x}\langle 0 |T\left\{ J_{\mu} (x) \bar{J}_{\nu}(0) \right\}| 0\rangle \, ,\nonumber\\
\Pi_{\mu\nu\alpha\beta}(p)&=&i\int d^4x e^{ip\cdot x}\langle 0 |T\left\{ J_{\mu\nu} (x) \bar{J}_{\alpha\beta}(0) \right\}| 0\rangle \, ,
\end{eqnarray}
where the currents
\begin{eqnarray}
J(x)&=&J_{0}^{\bar{D}\Xi_c^{\prime}}(x)\, , \,\,\,J_{1}^{\bar{D}\Xi_c^{\prime}}(x)\, ,\nonumber\\
J_{\mu} (x)&=&J_{0;\mu}^{\bar{D}\Xi_c^*}(x)\, , \,\,\, J_{1;\mu}^{\bar{D}\Xi_c^*}(x)\, ,\,\,\, J_{0;\mu}^{\bar{D}^{*}\Xi_c^{\prime}}(x)\,, \,\,\,J_{1;\mu}^{\bar{D}^{*}\Xi_c^{\prime}}(x)\, , \nonumber\\
 J_{\mu\nu} (x)&=&J_{0;\mu\nu}^{\bar{D}^{*}\Xi_c^*}(x)\, , \, \,\, J_{1;\mu\nu}^{\bar{D}^{*}\Xi_c^*}(x)\, .
  \end{eqnarray}

Note that, the currents $J(x)$, $J_\mu(x)$ and $J_{\mu\nu}(x)$ can couple potentially to  the pentaquark molecular  states with not only the negative parity
but also the positive parity. At the hadron side, we isolate the contributions of the ground states and write the correlation functions as,
\begin{eqnarray}
  \Pi(p) & = & {\lambda^{-}_{\frac{1}{2}}}^2  {\!\not\!{p}+ M_{-} \over M_{-}^{2}-p^{2}  } +  {\lambda^{+}_{\frac{1}{2}}}^2  {\!\not\!{p}- M_{+} \over M_{+}^{2}-p^{2}  } +\cdots  \, ,
\end{eqnarray}
\begin{eqnarray}
   \Pi_{\mu\nu}(p) & = & {\lambda^{-}_{\frac{3}{2}}}^2  {\!\not\!{p}+ M_{-} \over M_{-}^{2}-p^{2}  } \left(- g_{\mu\nu}+\frac{\gamma_\mu\gamma_\nu}{3}+\frac{2p_\mu p_\nu}{3p^2}-\frac{p_\mu\gamma_\nu-p_\nu \gamma_\mu}{3\sqrt{p^2}}
\right)\nonumber\\
&&+  {\lambda^{+}_{\frac{3}{2}}}^2  {\!\not\!{p}- M_{+} \over M_{+}^{2}-p^{2}  } \left(- g_{\mu\nu}+\frac{\gamma_\mu\gamma_\nu}{3}+\frac{2p_\mu p_\nu}{3p^2}-\frac{p_\mu\gamma_\nu-p_\nu \gamma_\mu}{3\sqrt{p^2}}
\right)   \nonumber \\
& &+ {f^{+}_{\frac{1}{2}}}^2  {\!\not\!{p}+ M_{+} \over M_{+}^{2}-p^{2}  } p_\mu p_\nu+  {f^{-}_{\frac{1}{2}}}^2  {\!\not\!{p}- M_{-} \over M_{-}^{2}-p^{2}  } p_\mu p_\nu  +\cdots  \, ,
\end{eqnarray}
\begin{eqnarray}
\Pi_{\mu\nu\alpha\beta}(p) & = & {\lambda^{-}_{\frac{5}{2}}}^2  {\!\not\!{p}+ M_{-} \over M_{-}^{2}-p^{2}  } \left[\frac{ \widetilde{g}_{\mu\alpha}\widetilde{g}_{\nu\beta}+\widetilde{g}_{\mu\beta}\widetilde{g}_{\nu\alpha}}{2}-\frac{\widetilde{g}_{\mu\nu}\widetilde{g}_{\alpha\beta}}{5}-\frac{1}{10}\left( \gamma_{\mu}\gamma_{\alpha}+\frac{\gamma_{\mu}p_{\alpha}-\gamma_{\alpha}p_{\mu}}{\sqrt{p^2}}-\frac{p_{\mu}p_{\alpha}}{p^2}\right)\widetilde{g}_{\nu\beta}\right.\nonumber\\
&&\left.-\frac{1}{10}\left( \gamma_{\nu}\gamma_{\alpha}+\frac{\gamma_{\nu}p_{\alpha}-\gamma_{\alpha}p_{\nu}}{\sqrt{p^2}}-\frac{p_{\nu}p_{\alpha}}{p^2}\right)\widetilde{g}_{\mu\beta}
+\cdots\right]\nonumber\\
&&+ {\lambda^{+}_{\frac{5}{2}}}^2  {\!\not\!{p}- M_{+} \over M_{+}^{2}-p^{2}  } \left[\frac{ \widetilde{g}_{\mu\alpha}\widetilde{g}_{\nu\beta}+\widetilde{g}_{\mu\beta}\widetilde{g}_{\nu\alpha}}{2}
-\frac{\widetilde{g}_{\mu\nu}\widetilde{g}_{\alpha\beta}}{5}-\frac{1}{10}\left( \gamma_{\mu}\gamma_{\alpha}+\frac{\gamma_{\mu}p_{\alpha}-\gamma_{\alpha}p_{\mu}}{\sqrt{p^2}}-\frac{p_{\mu}p_{\alpha}}{p^2}\right)\widetilde{g}_{\nu\beta}\right.\nonumber\\
&&\left.
-\frac{1}{10}\left( \gamma_{\nu}\gamma_{\alpha}+\frac{\gamma_{\nu}p_{\alpha}-\gamma_{\alpha}p_{\nu}}{\sqrt{p^2}}-\frac{p_{\nu}p_{\alpha}}{p^2}\right)\widetilde{g}_{\mu\beta}
 +\cdots\right]   \nonumber\\
 && +{f^{+}_{\frac{3}{2}}}^2  {\!\not\!{p}+ M_{+} \over M_{+}^{2}-p^{2}  } \left[ p_\mu p_\alpha \left(- g_{\nu\beta}+\frac{\gamma_\nu\gamma_\beta}{3}+\frac{2p_\nu p_\beta}{3p^2}-\frac{p_\nu\gamma_\beta-p_\beta \gamma_\nu}{3\sqrt{p^2}}
\right)+\cdots \right]\nonumber\\
&&+  {f^{-}_{\frac{3}{2}}}^2  {\!\not\!{p}- M_{-} \over M_{-}^{2}-p^{2}  } \left[ p_\mu p_\alpha \left(- g_{\nu\beta}+\frac{\gamma_\nu\gamma_\beta}{3}+\frac{2p_\nu p_\beta}{3p^2}-\frac{p_\nu\gamma_\beta-p_\beta \gamma_\nu}{3\sqrt{p^2}}
\right)+\cdots \right]   \nonumber \\
& &+ {g^{-}_{\frac{1}{2}}}^2  {\!\not\!{p}+ M_{-} \over M_{-}^{2}-p^{2}  } p_\mu p_\nu p_\alpha p_\beta+  {g^{+}_{\frac{1}{2}}}^2  {\!\not\!{p}- M_{+} \over M_{+}^{2}-p^{2}  } p_\mu p_\nu p_\alpha p_\beta  +\cdots \, ,
\end{eqnarray}
where $\widetilde{g}_{\mu\nu}=g_{\mu\nu}-\frac{p_{\mu}p_{\nu}}{p^2}$.
In calculations, we have taken account of the current-pentaquark coupling constants $\lambda$, $f$ and $g$,
\begin{eqnarray}
\langle 0| J (0)|P_{\frac{1}{2}}^{-}(p)\rangle &=&\lambda^{-}_{\frac{1}{2}} U^{-}(p,s) \, , \nonumber \\
\langle 0| J (0)|P_{\frac{1}{2}}^{+}(p)\rangle &=&\lambda^{+}_{\frac{1}{2}}i\gamma_5 U^{+}(p,s) \, ,
\end{eqnarray}
\begin{eqnarray}
\langle 0| J_{\mu} (0)|P_{\frac{1}{2}}^{+}(p)\rangle &=&f^{+}_{\frac{1}{2}}p_\mu U^{+}(p,s) \, , \nonumber \\
\langle 0| J_{\mu} (0)|P_{\frac{1}{2}}^{-}(p)\rangle &=&f^{-}_{\frac{1}{2}}p_\mu i\gamma_5 U^{-}(p,s) \, , \nonumber\\
\langle 0| J_{\mu} (0)|P_{\frac{3}{2}}^{-}(p)\rangle &=&\lambda^{-}_{\frac{3}{2}} U^{-}_\mu(p,s) \, , \nonumber \\
\langle 0| J_{\mu} (0)|P_{\frac{3}{2}}^{+}(p)\rangle &=&\lambda^{+}_{\frac{3}{2}}i\gamma_5 U^{+}_{\mu}(p,s) \, , \end{eqnarray}
\begin{eqnarray}
\langle 0| J_{\mu\nu} (0)|P_{\frac{1}{2}}^{-}(p)\rangle &=&g^{-}_{\frac{1}{2}}p_\mu p_\nu U^{-}(p,s) \, , \nonumber\\
\langle 0| J_{\mu\nu} (0)|P_{\frac{1}{2}}^{+}(p)\rangle &=&g^{+}_{\frac{1}{2}}p_\mu p_\nu i\gamma_5 U^{+}(p,s) \, , \nonumber\\
\langle 0| J_{\mu\nu} (0)|P_{\frac{3}{2}}^{+}(p)\rangle &=&f^{+}_{\frac{3}{2}} \left[p_\mu U^{+}_{\nu}(p,s)+p_\nu U^{+}_{\mu}(p,s)\right] \, , \nonumber\\
\langle 0| J_{\mu\nu} (0)|P_{\frac{3}{2}}^{-}(p)\rangle &=&f^{-}_{\frac{3}{2}} i\gamma_5\left[p_\mu U^{-}_{\nu}(p,s)+p_\nu U^{-}_{\mu}(p,s)\right] \, , \nonumber\\
\langle 0| J_{\mu\nu} (0)|P_{\frac{5}{2}}^{-}(p)\rangle &=&\lambda^{-}_{\frac{5}{2}} U^{-}_{\mu\nu}(p,s) \, ,\nonumber\\
\langle 0| J_{\mu\nu} (0)|P_{\frac{5}{2}}^{+}(p)\rangle &=&\lambda^{+}_{\frac{5}{2}}i\gamma_5 U^{+}_{\mu\nu}(p,s) \, ,
\end{eqnarray}
and the summations of the Dirac/Rarita-Schwinger spinors \cite{WZGXZ5},
\begin{eqnarray}
\sum_s U \overline{U}&=&\left(\!\not\!{p}+M_{\pm}\right) \,  ,  \nonumber \\
\sum_s U_\mu \overline{U}_\nu&=&\left(\!\not\!{p}+M_{\pm}\right)\left( -g_{\mu\nu}+\frac{\gamma_\mu\gamma_\nu}{3}+\frac{2p_\mu p_\nu}{3p^2}-\frac{p_\mu
\gamma_\nu-p_\nu \gamma_\mu}{3\sqrt{p^2}} \right) \,  ,
\end{eqnarray}
\begin{eqnarray}
\sum_s U_{\mu\nu}\overline {U}_{\alpha\beta}&=&\left(\!\not\!{p}+M_{\pm}\right)\left\{\frac{\widetilde{g}_{\mu\alpha}\widetilde{g}_{\nu\beta}+\widetilde{g}_{\mu\beta}\widetilde{g}_{\nu\alpha}}{2} -\frac{\widetilde{g}_{\mu\nu}\widetilde{g}_{\alpha\beta}}{5}-\frac{1}{10}\left( \gamma_{\mu}\gamma_{\alpha}+\frac{\gamma_{\mu}p_{\alpha}-\gamma_{\alpha}p_{\mu}}{\sqrt{p^2}}-\frac{p_{\mu}p_{\alpha}}{p^2}\right)\widetilde{g}_{\nu\beta}\right. \nonumber\\
&&-\frac{1}{10}\left( \gamma_{\nu}\gamma_{\alpha}+\frac{\gamma_{\nu}p_{\alpha}-\gamma_{\alpha}p_{\nu}}{\sqrt{p^2}}-\frac{p_{\nu}p_{\alpha}}{p^2}\right)\widetilde{g}_{\mu\beta}
-\frac{1}{10}\left( \gamma_{\mu}\gamma_{\beta}+\frac{\gamma_{\mu}p_{\beta}-\gamma_{\beta}p_{\mu}}{\sqrt{p^2}}-\frac{p_{\mu}p_{\beta}}{p^2}\right)\widetilde{g}_{\nu\alpha}\nonumber\\
&&\left.-\frac{1}{10}\left( \gamma_{\nu}\gamma_{\beta}+\frac{\gamma_{\nu}p_{\beta}-\gamma_{\beta}p_{\nu}}{\sqrt{p^2}}-\frac{p_{\nu}p_{\beta}}{p^2}\right)\widetilde{g}_{\mu\alpha} \right\} \, ,
\end{eqnarray}
and $p^2=M^2_{\pm}$ on the mass-shell, the subscripts $\frac{1}{2}$, $\frac{3}{2}$ and $\frac{5}{2}$ are the spins of the pentaquark molecular  states, the subscripts/superscripts $\pm$ of the $\lambda$ and $M$ denote the positive-parity and negative-parity, respectively.

We can rewrite the correlation functions $\Pi_{\mu\nu}(p)$ and $\Pi_{\mu\nu\alpha\beta}(p)$ into another form according to Lorentz covariance \cite{WZGXZ5},
\begin{eqnarray}
\Pi_{\mu\nu}(p)&=&\Pi_{\frac{3}{2}}(p)\,\left(- g_{\mu\nu}\right)+\Pi_{\frac{3}{2}}^1(p)\,\gamma_\mu \gamma_\nu+\Pi_{\frac{3}{2}}^2(p)\,\left(p_\mu\gamma_\nu-p_\nu \gamma_\mu\right) +\Pi_{\frac{1}{2},\frac{3}{2}}(p)\, p_\mu p_\nu\, , \\
\Pi_{\mu\nu\alpha\beta}(p)&=&\Pi_{\frac{5}{2}}(p)\,\left(g_{\mu\alpha}g_{\nu\beta}+g_{\mu\beta}g_{\nu\alpha} \right)+\Pi_{\frac{5}{2}}^1(p)\, g_{\mu\nu}g_{\alpha\beta}+\Pi_{\frac{5}{2}}^2(p)\, \left(g_{\mu\nu}p_{\alpha}p_{\beta}+g_{\alpha\beta}p_{\mu}p_{\nu}\right) \nonumber\\
&&+\Pi_{\frac{5}{2}}^3(p)\,\left(  g_{\mu \alpha} \gamma_\nu \gamma_\beta+ g_{\mu \beta} \gamma_\nu \gamma_\alpha+ g_{\nu \alpha} \gamma_\mu \gamma_\beta+ g_{\nu \beta} \gamma_\mu \gamma_\alpha \right) \nonumber\\
&&+\Pi_{\frac{5}{2}}^4(p)\,\left[  g_{\nu \beta}\left(\gamma_{\mu} p_{\alpha}- \gamma_{\alpha}p_{\mu}\right) +
g_{\nu \alpha}\left(\gamma_{\mu}p_{ \beta}-\gamma_{ \beta}p_{\mu}\right) + g_{\mu \beta}\left(\gamma_{\nu} p_{\alpha}- \gamma_{\alpha}p_{\nu}\right)\right.\nonumber\\
&&\left.+ g_{\mu \alpha} \left(\gamma_{\nu}p_{ \beta}-\gamma_{ \beta}p_{\nu} \right)\right]\nonumber\\
&&+\Pi_{\frac{3}{2},\frac{5}{2}}^1(p)\,\left(  g_{\mu \alpha} p_\nu p_\beta+ g_{\mu \beta} p_\nu p_\alpha+ g_{\nu \alpha} p_\mu p_\beta+ g_{\nu \beta} p_\mu p_\alpha \right) \nonumber\\
&&+\Pi_{\frac{3}{2},\frac{5}{2}}^2(p)\,\left(  \gamma_{\mu} \gamma_{\alpha} p_\nu p_\beta+ \gamma_{\mu}\gamma_{ \beta} p_\nu p_\alpha+ \gamma_{\nu} \gamma_{\alpha} p_\mu p_\beta+ \gamma_{\nu}\gamma_{ \beta} p_\mu p_\alpha \right) \nonumber\\
&&+\Pi_{\frac{3}{2},\frac{5}{2}}^3(p)\,\left[  \left(\gamma_{\mu} p_{\alpha}- \gamma_{\alpha}p_{\mu}\right) p_\nu p_\beta+
\left(\gamma_{\mu}p_{ \beta}-\gamma_{ \beta}p_{\mu}\right) p_\nu p_\alpha+ \left(\gamma_{\nu} p_{\alpha}- \gamma_{\alpha}p_{\nu}\right) p_\mu p_\beta\right.\nonumber\\
&&\left.+ \left(\gamma_{\nu}p_{ \beta}-\gamma_{ \beta}p_{\nu} \right)p_\mu p_\alpha \right] +\Pi_{\frac{1}{2},\frac{3}{2},\frac{5}{2}}(p)\,p_\mu p_\nu p_\alpha p_\beta \, ,
\end{eqnarray}
 where the subscripts $\frac{1}{2}$, $\frac{3}{2}$ and $\frac{5}{2}$ in the components $\Pi_{\frac{3}{2}}(p)$, $\Pi_{\frac{3}{2}}^1(p)$, $\Pi_{\frac{3}{2}}^2(p)$,  $\Pi_{\frac{1}{2},\frac{3}{2}}(p)$, $\Pi_{\frac{5}{2}}(p)$, $\Pi_{\frac{5}{2}}^1(p)$, $\Pi_{\frac{5}{2}}^2(p)$, $\Pi_{\frac{5}{2}}^3(p)$, $\Pi_{\frac{5}{2}}^4(p)$, $\Pi_{\frac{3}{2},\frac{5}{2}}^1(p)$, $\Pi_{\frac{3}{2},\frac{5}{2}}^2(p)$, $\Pi_{\frac{3}{2},\frac{5}{2}}^3(p)$ and
$\Pi_{\frac{1}{2},\frac{3}{2},\frac{5}{2}}(p)$ stand for  the spins of the molecular states. The components $\Pi_{\frac{1}{2},\frac{3}{2}}(p)$, $\Pi_{\frac{3}{2},\frac{5}{2}}^1(p)$, $\Pi_{\frac{3}{2},\frac{5}{2}}^2(p)$, $\Pi_{\frac{3}{2},\frac{5}{2}}^3(p)$ and $\Pi_{\frac{1}{2},\frac{3}{2},\frac{5}{2}}(p)$ receive contributions from more than one molecular state, for this reason they can be neglected in analysis. We can rewrite $\gamma_\mu \gamma_\nu=g_{\mu\nu}-i\sigma_{\mu\nu}$, then the components $\Pi_{\frac{3}{2}}^1(p)$, $\Pi_{\frac{3}{2}}^2(p)$, $\Pi_{\frac{5}{2}}^3(p)$ and $\Pi_{\frac{5}{2}}^4(p)$ are companied with tensor structures which are antisymmetric in the Lorentz indexes  $\mu$, $\nu$, $\alpha$ or $\beta$. In calculations, we observe that such antisymmetric properties lead to smaller (negligible) intervals of dimensions of the vacuum condensates, therefore worse QCD sum rules, for this reason the components $\Pi_{\frac{3}{2}}^1(p)$, $\Pi_{\frac{3}{2}}^2(p)$, $\Pi_{\frac{5}{2}}^3(p)$ and $\Pi_{\frac{5}{2}}^4(p)$ can also be  neglected. On the other hand, if we make  the replacement $J_{\mu\nu}(x)\to \widehat{J}_{\mu\nu}(x)=J_{\mu\nu}(x)-\frac{1}{4}g_{\mu\nu}J_\alpha{}^\alpha(x)$ to subtract the contributions of the $J=\frac{1}{2}$ molecular states, a lot of terms   $\propto g_{\mu\nu}$, $g_{\alpha\beta}$ would  disappear at the QCD side, and result in smaller intervals of dimensions of the vacuum condensates, for this reason the components  $\Pi_{\frac{5}{2}}^1(p)$ and $\Pi_{\frac{5}{2}}^2(p)$ are not the optimal choices  to study the $J=\frac{5}{2}$ molecular  states. Now only the components  $\Pi_{\frac{3}{2}}(p)$ and $\Pi_{\frac{5}{2}}(p)$ are left.

In this paper, we pick out the structures $\!\not\!{p}$,  $1$,   $\!\not\!{p}g_{\mu\nu}$, $g_{\mu\nu}$ and $\!\not\!{p}\left(g_{\mu\alpha}g_{\nu\beta}+g_{\mu\beta}g_{\nu\alpha}\right)$, $g_{\mu\alpha}g_{\nu\beta}+g_{\mu\beta}g_{\nu\alpha}$ for the correlation functions $\Pi(p)$, $\Pi_{\mu\nu}(p)$ and $\Pi_{\mu\nu\alpha\beta}(p)$ respectively to study the $J^P={\frac{1}{2}}^\mp$, ${\frac{3}{2}}^\mp$ and ${\frac{5}{2}}^\mp$ pentaquark molecular states,
\begin{eqnarray}\label{Tensor-Use}
\Pi(p)&=&\Pi_{\frac{1}{2}}^1(p^2)\!\not\!{p}+\Pi_{\frac{1}{2}}^0(p^2)\, , \nonumber\\
\Pi_{\mu\nu}(p)&=&-\Pi_{\frac{3}{2}}^1(p^2)\!\not\!{p}\,g_{\mu\nu}-\Pi_{\frac{3}{2}}^0(p^2)\,g_{\mu\nu}+\cdots\, , \nonumber\\
\Pi_{\mu\nu\alpha\beta}(p)&=&\Pi_{\frac{5}{2}}^1(p^2)\!\not\!{p}\left( g_{\mu\alpha}g_{\nu\beta}+g_{\mu\beta}g_{\nu\alpha}\right)+\Pi_{\frac{5}{2}}^0(p^2)\,\left( g_{\mu\alpha}g_{\nu\beta}+g_{\mu\beta}g_{\nu\alpha}\right)+ \cdots \, .
\end{eqnarray}
In calculations, we select the terms proportional to those tensor structures one by one, it is difficult to construct projection operators to project out the relevant components straightforwardly.

 It is straightforward and easy to obtain the hadron spectral densities  through dispersion relation,
\begin{eqnarray}
\frac{{\rm Im}\Pi_{j}^{1}(s)}{\pi}&=&{\lambda^{-}_{j}}^2 \delta\left(s-M_{-}^2\right)+{\lambda^{+}_{j}}^2 \delta\left(s-M_{+}^2\right)= \rho^1_{j,H}(s) \, , \nonumber \\
\frac{{\rm Im}\Pi_{j}^0(s)}{\pi}&=&M_{-}{\lambda^{-}_{j}}^2 \delta\left(s-M_{-}^2\right)-M_{+}{\lambda^{+}_{j}}^2 \delta\left(s-M_{+}^2\right)=\rho^0_{j,H}(s) \, ,
\end{eqnarray}
where the subscripts $j=\frac{1}{2}$, $\frac{3}{2}$, $\frac{5}{2}$ according to Eq.\eqref{Tensor-Use}, we add  the subscript $H$ to stand for  the hadron side.
The components $\Pi_j^{1}(p^2)$ and $\Pi_j^{0}(p^2)$  receive contributions from both the negative-parity and positive-parity molecular states. We separate the  negative-parity and positive-parity molecular states explicitly by resorting to the special combinations $\sqrt{s}\rho^1_{j,H}(s)\pm\rho^0_{j,H}(s)$.
 Then we introduce the weight function $\exp\left(-\frac{s}{T^2}\right)$ to acquire  the QCD sum rules at  hadron side,
\begin{eqnarray}
\Pi_j(T^2,s_0)&=&\int_{4m_c^2}^{s_0}ds \left[\sqrt{s}\rho^1_{j,H}(s)+\rho^0_{j,H}(s)\right]\exp\left( -\frac{s}{T^2}\right)\nonumber\\
&=&2M_{-}{\lambda^{-}_{j}}^2\exp\left( -\frac{M_{-}^2}{T^2}\right) \, ,
\end{eqnarray}
where the $s_0$ are the continuum threshold parameters and the $T^2$ are the Borel parameters.

At the QCD side, we contract the $u$, $d$, $s$ and $c$ quark fields in the correlation functions $\Pi(p)$, $\Pi_{\mu\nu}(p)$ and $\Pi_{\mu\nu\alpha\beta}(p)$  at the quark-gluon level via the Wick's theorem, then, express them in terms of the full quark propagators,
 \begin{eqnarray}
U/D_{ij}(x)&=& \frac{i\delta_{ij}\!\not\!{x}}{ 2\pi^2x^4}-\frac{\delta_{ij}\langle
\bar{q}q\rangle}{12} -\frac{\delta_{ij}x^2\langle \bar{q}g_s\sigma Gq\rangle}{192} -\frac{ig_sG^{a}_{\alpha\beta}t^a_{ij}(\!\not\!{x}
\sigma^{\alpha\beta}+\sigma^{\alpha\beta} \!\not\!{x})}{32\pi^2x^2} -\frac{\delta_{ij}x^4\langle \bar{q}q \rangle\langle g_s^2 GG\rangle}{27648} \nonumber\\
&&  -\frac{1}{8}\langle\bar{q}_j\sigma^{\mu\nu}q_i \rangle \sigma_{\mu\nu}-\frac{1}{4}\langle\bar{q}_j\gamma^{\mu}q_i\rangle \gamma_{\mu }+\cdots \, ,
\end{eqnarray}
\begin{eqnarray}
S_{ij}(x)&=& \frac{i\delta_{ij}\!\not\!{x}}{ 2\pi^2x^4}
-\frac{\delta_{ij}m_s}{4\pi^2x^2}-\frac{\delta_{ij}\langle
\bar{s}s\rangle}{12} +\frac{i\delta_{ij}\!\not\!{x}m_s
\langle\bar{s}s\rangle}{48}-\frac{\delta_{ij}x^2\langle \bar{s}g_s\sigma Gs\rangle}{192}+\frac{i\delta_{ij}x^2\!\not\!{x} m_s\langle \bar{s}g_s\sigma
 Gs\rangle }{1152}\nonumber\\
&& -\frac{ig_s G^{a}_{\alpha\beta}t^a_{ij}(\!\not\!{x}
\sigma^{\alpha\beta}+\sigma^{\alpha\beta} \!\not\!{x})}{32\pi^2x^2} -\frac{\delta_{ij}x^4\langle \bar{s}s \rangle\langle g_s^2 GG\rangle}{27648}-\frac{1}{8}\langle\bar{s}_j\sigma^{\mu\nu}s_i \rangle \sigma_{\mu\nu}  -\frac{1}{4}\langle\bar{s}_j\gamma^{\mu}s_i\rangle \gamma_{\mu }+\cdots \, , \nonumber \\
\end{eqnarray}
\begin{eqnarray}
C_{ij}(x)&=&\frac{i}{(2\pi)^4}\int d^4k e^{-ik \cdot x} \left\{
\frac{\delta_{ij}}{\!\not\!{k}-m_c}
-\frac{g_sG^n_{\alpha\beta}t^n_{ij}}{4}\frac{\sigma^{\alpha\beta}(\!\not\!{k}+m_c)+(\!\not\!{k}+m_c)
\sigma^{\alpha\beta}}{(k^2-m_c^2)^2}\right.\nonumber\\
&&\left. -\frac{g_s^2 (t^at^b)_{ij} G^a_{\alpha\beta}G^b_{\mu\nu}(f^{\alpha\beta\mu\nu}+f^{\alpha\mu\beta\nu}+f^{\alpha\mu\nu\beta}) }{4(k^2-m_c^2)^5}+\cdots\right\} \, ,\nonumber\\
f^{\alpha\beta\mu\nu}&=&(\!\not\!{k}+m_c)\gamma^\alpha(\!\not\!{k}+m_c)\gamma^\beta(\!\not\!{k}+m_c)\gamma^\mu(\!\not\!{k}+m_c)\gamma^\nu(\!\not\!{k}+m_c)\, ,
\end{eqnarray}
and  $t^n=\frac{\lambda^n}{2}$, the $\lambda^n$ is the Gell-Mann matrix
\cite{Reinders,Shifman-OPE,Grozin-OPE,BaxiG-OPE,Pimikov-OPE,Pascual-1984,WangHuang3900}.
We retain the possible operators $\langle\bar{q}_j\sigma_{\mu\nu}q_i \rangle$,  $\langle\bar{s}_j\sigma_{\mu\nu}s_i \rangle$, $\langle\bar{q}_j\gamma_{\mu}q_i \rangle$,  $\langle\bar{s}_j\gamma_{\mu}s_i \rangle$   from the Fierz transformations of the quark operators
$\langle q_i \bar{q}_j\rangle$ and $\langle s_i \bar{s}_j\rangle$ (before the Wick's contractions) to  absorb the gluons  emitted from other quark lines to  obtain  the additional mixed condensates  $\langle\bar{q}g_s\sigma G q\rangle$ and $\langle\bar{s}g_s\sigma G s\rangle$ and four-quark condensates $g_s^2\langle\bar{q} q\rangle^2$, $g_s^2\langle\bar{q} q\rangle\langle\bar{s} s\rangle$ and $g_s^2\langle\bar{s} s\rangle^2$, respectively \cite{WangHuang3900}. For detailed derivations of other terms in the full-quark propagators, one can consult Refs.\cite{Reinders,Shifman-OPE,Grozin-OPE,BaxiG-OPE,Pimikov-OPE,Pascual-1984}. Then we compute  all the integrals in the coordinate space and momentum space sequentially to obtain the representations at the quark-gluon  level, and  pick out the structures $\!\not\!{p}$,  $1$,   $\!\not\!{p}g_{\mu\nu}$, $g_{\mu\nu}$ and $\!\not\!{p}\left(g_{\mu\alpha}g_{\nu\beta}+g_{\mu\beta}g_{\nu\alpha}\right)$, $g_{\mu\alpha}g_{\nu\beta}+g_{\mu\beta}g_{\nu\alpha}$ for the correlation functions $\Pi(p)$, $\Pi_{\mu\nu}(p)$ and $\Pi_{\mu\nu\alpha\beta}(p)$ respectively  one by one to match with the hadron side. Again, we acquire the QCD spectral densities through dispersion relation,
 \begin{eqnarray}
\frac{{\rm Im}\Pi_{j}^{1}(s)}{\pi}&=& \rho^1_{j,QCD}(s) \, , \nonumber \\
\frac{{\rm Im}\Pi_{j}^0(s)}{\pi}&=&\rho^0_{j,QCD}(s) \, ,
\end{eqnarray}
where we add the subscript QCD to stand for the QCD side.

If each charm-quark line emits a gluon and each light-quark line contributes  a quark-antiquark   pair, we acquire  a quark-gluon  operator $g_s^2G_{\alpha\beta}G^{\alpha\beta}\bar{q}q \bar{q}q \bar{q}q$ (with $q=u$, $d$ or $s$)  of dimension 13, as a consequence,  we have to  take account of  the vacuum  condensates at least
up to dimension 13 to testify  the possible  behaviors  of the operator product expansion, as the vacuum condensates are vacuum expectations of the quark-gluon operators in the QCD vacuum. In calculations, we assume vacuum saturation for the higher dimensional vacuum condensates, just like in our previous works \cite{WZG-Pcs,WangZG-Xin-CPC,wangxiuwuN,WZGXZ1-1,WZGXZ1-2,WZGXZ1-3,WZGXZ3,WZGXZ4,WZGXZ13-mole,WZGNNN1,WZGXZ5},
 the vacuum saturation works well, large deviations from the vacuum saturation cannot lead to good QCD sum rules.  For detailed discussions of this subject, one can consult Ref.\cite{WZG-Vacuum}. In fact, the concept of "full quark propagators" is acquired by assuming vacuum saturation (or factorization) tacitly \cite{Reinders,Shifman-OPE,Grozin-OPE,BaxiG-OPE,Pimikov-OPE,Pascual-1984}.

We carefully analyze the contributions of all the related terms of the vacuum condensates after accomplishing  the operator product expansion. The highest dimensional vacuum condensates determined by the leading  order Feynman diagrams are $\langle\frac{\alpha_s}{\pi}GG\rangle\langle\overline{q}q\rangle^3$ and $\langle\overline{q}g_s\sigma Gq\rangle^2\langle\overline{q}q\rangle$ with dimension $13$. The vacuum condensates proportional to the strong fine-structure constant $\alpha_s^k $ with $k\leq 1$ are selected for calculations \cite{wangxiuwu}. Thus, in this work, there are solid reasons for us to choose the terms $\langle\bar{q}q\rangle$, $\langle\frac{\alpha_s}{\pi}GG\rangle$, $\langle\overline{q}g_s\sigma Gq\rangle$, $\langle\overline{q}q\rangle^2$, $\langle\frac{\alpha_s}{\pi}GG\rangle\langle\bar{q}q\rangle$, $\langle\overline{q}g_s\sigma Gq\rangle\langle\overline{q}q\rangle$, $\langle\bar{q}q\rangle^3$, $\langle\overline{q}g_s\sigma Gq\rangle^2$, $\langle\frac{\alpha_s}{\pi}GG\rangle\langle\overline{q}q\rangle^2$, $\langle\overline{q}g_s\sigma Gq\rangle\langle\bar{q}q\rangle^2$, $\langle\overline{q}q\rangle^4$, $\langle\overline{q}g_s\sigma Gq\rangle^2\langle\overline{q}q\rangle$ and $\langle\frac{\alpha_s}{\pi}GG\rangle\langle\overline{q}q\rangle^3$, where $q=u$, $d$ or $s$. Considering the masses of the light quarks $u$ and $d$ are too small to make significant difference, we set their masses to be zero and keep the terms of the vacuum condensates proportional to the $s$-quark mass $m_s$,  and we throw away the terms related to $m_s^k$ for $k\geq 2$ due to their tiny contributions.

 In the case of the color singlet-singlet type four-quark currents, Lucha, Melikhov and Sazdjian assert that the disconnected (connected) Feynman diagrams in the color space only make  contributions to two-meson states (tetraquark  states),
 the contributions  at the order $\mathcal{O}(\alpha_s^k)$ with $k\leq1$ in the operator product expansion, which are factorizable/disconnected in the color space, are exactly  canceled out   by the two-meson states (in other words, meson-meson scattering states) at the hadron side, the tetraquark (molecular) states begin to receive contributions at the order $\mathcal{O}(\alpha_s^2)$ \cite{ChuSh-PRD,ChuSh-2208}.

 Direct calculations indicate that the meson-meson scattering states alone  cannot saturate the
 disconnected Feynman diagrams, we have to introduce the tetraquark molecular states, the intermediate meson-loops amount to give a finite width to modify dispersion relation \cite{WZG-Landau,WZG-XicXicc,WZG-DvDvDv}, the tetraquark (molecular) states begin to receive contributions at the leading order $\mathcal{O}(\alpha_s^0)$. If the widths are not larger than $400\,\rm{MeV}$, the net effects of the intermediate meson-loops can be safely absorbed into the pole residues, and cannot affect the predicted tetraquark (molecule) masses \cite{WZG-Z4200}.

The conclusions acquired in Refs.\cite{WZG-Landau,WZG-XicXicc,WZG-DvDvDv} are applicable to all the multiquark states, and we should take account of all  the connected and disconnected Feynman diagrams, and the multiquark states alone can saturate the QCD sum rules.
 We maybe worry  that the five-quark currents also couple  potentially to the  baryon-meson scattering states if they have the same quantum numbers, as the quantum field theory cannot exclude such a possibility.
We should bear in mind that we choose the local currents, while the traditional mesons and baryons are spatial extended objects and have average  spatial sizes $\sqrt{\langle r^2\rangle} \neq 0$, for example,
 $\sqrt{\langle r^2\rangle}=0.5\sim 0.8\,\rm{fm}$   for the charmed baryons and $\sqrt{\langle r^2\rangle}\sim 0.5\,\rm{fm}$ for the charmed mesons \cite{WangZG-Xin-CPC}. The charmed baryon-meson pairs should have the average size $\sqrt{\langle r^2\rangle}>1\,\rm{fm}$. In this work, we choose the local currents $J(x)$, $J_\mu(x)$ and $J_{\mu\nu}(x)$, which couple potentially to the compact objects having the average spatial sizes as that of the typical charmed baryons, rather than to the baryon-meson scattering states, due to the small overlapping of the wave-functions. The net effects of the intermediate baryon-meson loops can be absorbed into the pole residues safely and cannot affect the predicted molecule  masses.
  Though we refer to  the color singlet-singlet type pentaquark states as the pentaquark molecular states, they are compact objects in the QCD sum rules.

 Now we take the quark-hadron duality below the continuum thresholds $s_0$ and get the QCD sum rules for the pentaquark molecular states,
\begin{eqnarray}\label{QCDN}
2M_{-}\lambda^{-2}\exp\left( -M_{-}^2\tau\right)
&=& \int_{4m_c^2}^{s_0}ds \left[\sqrt{s}\rho^1_{QCD}(s)+\rho^0_{QCD}(s)\right]\exp\left( -\tau s\right)\, ,
\end{eqnarray}
\begin{eqnarray}\label{QCDSR-M}
 M^2_{-} &=& \frac{-\frac{d}{d \tau}\int_{4m_c^2}^{s_0}ds \,\left[\sqrt{s}\,\rho^1_{QCD}(s)+\,\rho^0_{QCD}(s)\right]\exp\left(- \tau s\right)}{\int_{4m_c^2}^{s_0}ds \left[\sqrt{s}\,\rho_{QCD}^1(s)+\,\rho^0_{QCD}(s)\right]\exp\left( -\tau s\right)}\, ,
 \end{eqnarray}
where $\tau=\frac{1}{T^2}$. For simplicity, the detailed expressions of the complicated spectral densities $\rho^1_{QCD}(s)$ and $\rho^0_{QCD}(s)$ are not shown here, one can contact us via Email. However, we give the explicit QCD spectral densities for the current $J_{0}^{\bar{D}\Xi_c^{\prime}}(x)$ in the Appendix as an example to illustrate the form.

\section{Numerical results and discussions}
We apply the standard values of the vacuum condensates $\langle\overline{q}q\rangle=-(0.24\pm0.01\;{\rm GeV})^3$, $\langle\bar{s}s\rangle=(0.8\pm0.1)\langle\bar{q}q\rangle$, $\langle\overline{q}g_s\sigma Gq\rangle=m_0^2\langle\overline{q}q\rangle$, $\langle\overline{s}g_s\sigma Gs\rangle=m_0^2\langle\overline{s}s\rangle$, $m_0^2=(0.8\pm0.1)\;{\rm GeV}^2$, $\langle\frac{\alpha_s}{\pi}GG\rangle=(0.012\pm0.004)\;{\rm GeV}^4$ at the energy scale $\mu=1\;{\rm GeV}$ \cite{SVZ1,SVZ2,Reinders,ColangeloReview}, and choose the $
\overline{MS}$ masses  $m_c(m_c)=(1.275\pm0.025)\;{\rm GeV}$ and $m_s(\mu=2\;{\rm GeV})=(0.095\pm0.005)\;{\rm GeV}$ from the Particle Data Group \cite{PDG}. We consider the energy-scale dependence of those parameters,
\begin{eqnarray}
\notag \langle\overline{q}q\rangle(\mu)&&=\;\;\langle\overline{q}q\rangle(1{\rm GeV})\left[\frac{\alpha_s(1{\rm GeV})}{\alpha_s(\mu)}\right]^{\frac{12}{33-2n_f}}\, ,\\
\notag \langle\overline{s}s\rangle(\mu)&&=\;\;\langle\overline{s}s\rangle(1{\rm GeV})\left[\frac{\alpha_s(1{\rm GeV})}{\alpha_s(\mu)}\right]^{\frac{12}{33-2n_f}}\, ,\\
\notag \langle\overline{q}g_s\sigma Gq\rangle(\mu)&& =\;\;\langle\overline{q}g_s\sigma Gq\rangle(1{\rm GeV})\left[\frac{\alpha_s(1{\rm GeV})}{\alpha_s(\mu)}\right]^{\frac{2}{33-2n_f}}\, ,\\
\notag \langle\overline{s}g_s\sigma Gs\rangle(\mu)&& =\;\;\langle\overline{s}g_s\sigma Gs\rangle(1{\rm GeV})\left[\frac{\alpha_s(1{\rm GeV})}{\alpha_s(\mu)}\right]^{\frac{2}{33-2n_f}}\, ,\\
\notag  m_c(\mu)&&=\;\;m_c(m_c)\left[\frac{\alpha_s(\mu)}{\alpha_s(m_c)}\right]^{\frac{12}{33-2n_f}}\, ,\\
\notag  m_s(\mu)&&=\;\;m_s(2{\rm GeV})\left[\frac{\alpha_s(\mu)}{\alpha_s(2{\rm GeV})}\right]^{\frac{12}{33-2n_f}}\, ,\\
\notag \alpha_s(\mu)&&=\;\;\frac{1}{b_0t}\left[1-\frac{b_1}{b_0^2}\frac{\rm{log}\emph{t}}{t}+\frac{b_1^2(\rm{log}^2\emph{t}-\rm{log}\emph {t}-1)+\emph{b}_0\emph{b}_2}{b_0^4t^2}\right]\, ,
\end{eqnarray}
where, $t=\rm{log}\frac{\mu^2}{\Lambda_{\emph{QCD}}^2}$, $\emph b_0=\frac{33-2\emph{n}_\emph{f}}{12\pi}$, $b_1=\frac{153-19n_f}{24\pi^2}$, $b_2=\frac{2857-\frac{5033}{9}n_f+\frac{325}{27}n_f^2}{128\pi^3}$
and $\Lambda_{QCD}=213$ MeV, $296$ MeV, $339$ MeV for the flavors $n_f=5,\,4,\,3$, respectively \cite{PDG,Narison}. Since there are $u,\;d,\;s$ and $c$ quarks for the pentaquark (molecular) states with strangeness in the present study, we set the flavor number $n_f=4$. As for the energy scales, we take account of the light flavor $SU(3)$ mass breaking effect and apply the modified energy scale formula \cite{WangZG-Xin-CPC,WZG-mole-IJMPA},
\begin{eqnarray}
\mu=\sqrt{M_{X/Y/Z/P}^2-4\mathbb{M}_c^2}-k\mathbb{M}_s\, ,
\end{eqnarray}
to evolve the QCD spectral densities to the optimal energy scales $\mu$ to extract the hadron masses, where the $M_{X/Y/Z/P}$ represent the hidden-charm tetraquark (molecular) states $X$, $Y$, $Z$ and pentaquark (molecular) states $P$, the $\mathbb{M}_c$ represents the effective charm quark mass, we choose the updated value $\mathbb{M}_c=1.85\pm0.01$ \rm GeV \cite{WangZG-Xin-CPC}, the $s$ quark number $k=1$, and the effective $s$-quark mass $\mathbb{M}_s=0.2\;\rm{GeV}$ \cite{WangZG-Xin-CPC,WZG-mole-IJMPA}.

We choose the Borel parameters $T^2$ and continuum threshold parameters $s_0$ to satisfy the four criteria:

$\bf 1.$ Pole dominance at the hadron side;

$\bf 2.$ Convergence of the operator product expansion;

$\bf 3.$ Appearance of the Borel platforms;

$\bf 4.$ Satisfying the modified energy scale formula,\\
via trial  and error.
The pole dominance and convergence of the operator product expansion are the basic criteria of the QCD sum rules. In order to testify whether or not the QCD sum rules satisfy those two basic criteria, we define the pole contributions and the contributions of the vacuum condensates of dimension $n$, which are listed as,
\begin{eqnarray}
{\rm Pole\,\, Contributions\,\, (PC)}&=& \frac{\Pi_{\frac{1}{2}/\frac{3}{2}/\frac{5}{2}}(T^2,s_0)}{\Pi_{\frac{1}{2}/\frac{3}{2}/\frac{5}{2}}(T^2,\infty)} \, ,\nonumber\\ &=&\frac{\int_{4m_c^2}^{s_0}ds\left[\sqrt{s}\rho_{QCD}^1(s)+\rho_{QCD}^0(s)\right]\exp\left(-\frac{s}{T^2}\right)}
{\int_{4m_c^2}^{\infty}ds\left[\sqrt{s}\rho_{QCD}^1(s)+\rho_{QCD}^0(s)\right]\exp\left(-\frac{s}{T^2}\right)}\, ,
\end{eqnarray}
\begin{eqnarray}
D(n)&=&\frac{\int_{4m_c^2}^{s_0}ds\left[\sqrt{s}\rho_{QCD;n}^1(s)+\rho_{QCD;n}^0(s)\right]\exp\left(-\frac{s}{T^2}\right)}
{\int_{4m_c^2}^{s_0}ds\left[\sqrt{s}\rho_{QCD}^1(s)+\rho_{QCD}^0(s)\right]\exp\left(-\frac{s}{T^2}\right)}\, ,
\end{eqnarray}
where the $\rho_{QCD;n}^1(s)$ and $\rho_{QCD;n}^0(s)$ are the spectral densities involving  the vacuum condensates of dimension $n$ picked out from the $\rho_{QCD}^1(s)$ and $\rho_{QCD}^0(s)$, respectively. In this paper, we also give a detailed technical discussion about the contributions of the vacuum condensates proportional to the $s$ quark mass $m_s$. Now we define the $D(m_s,n)$ as,
\begin{eqnarray}
D(m_s,n)&=&\frac{\int_{4m_c^2}^{s_0}ds\left[\sqrt{s}\rho_{QCD;n;m_s}^1(s)+\rho_{QCD;n;m_s}^0(s)\right]\exp\left(-\frac{s}{T^2}\right)}
{\int_{4m_c^2}^{s_0}ds\left[\sqrt{s}\rho_{QCD}^1(s)+\rho_{QCD}^0(s)\right]\exp\left(-\frac{s}{T^2}\right)}\, ,
\end{eqnarray}
where the $\rho_{QCD;n;m_s}^1(s)$ and $\rho_{QCD;n;m_s}^0(s)$ refer to the spectral densities proportional to the $m_s$ picked out from the $\rho_{QCD;n}^1(s)$ and $\rho_{QCD;n}^0(s)$,  respectively.

After trial and error, we acquire the Borel windows, continuum threshold parameters, optimal energy scales of the QCD spectral densities, and pole contributions,  which are displayed in the Table \ref{table}. From the table, we can see clearly that the pole contributions for all the considered states are about $(40-60)\%$ with the central values larger than $50\%$, the criterion {\bf 1} is satisfied very well. Just like in our previous works, we take the uniform pole contributions $(40-60)\%$ \cite{WZG-Pcs,WangZG-Xin-CPC,wangxiuwuN,WZGXZ1-1,WZGXZ1-2,WZGXZ1-3,WZGXZ3,WZGXZ4,WZGXZ13-mole,WZGNNN1,WZGXZ5}.

\begin{figure}
 \centering
 \includegraphics[totalheight=5cm,width=7cm]{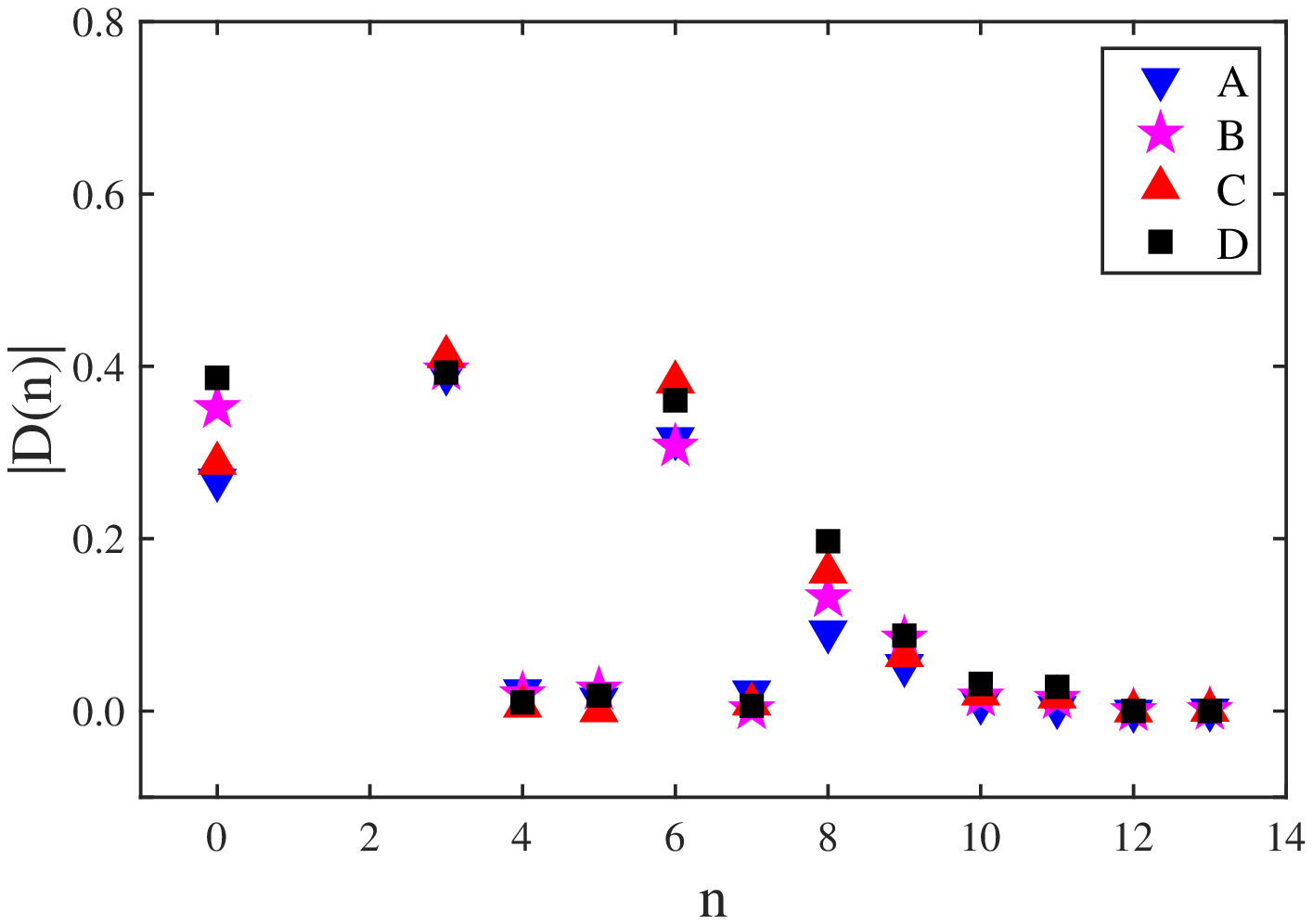}
 \includegraphics[totalheight=5cm,width=7cm]{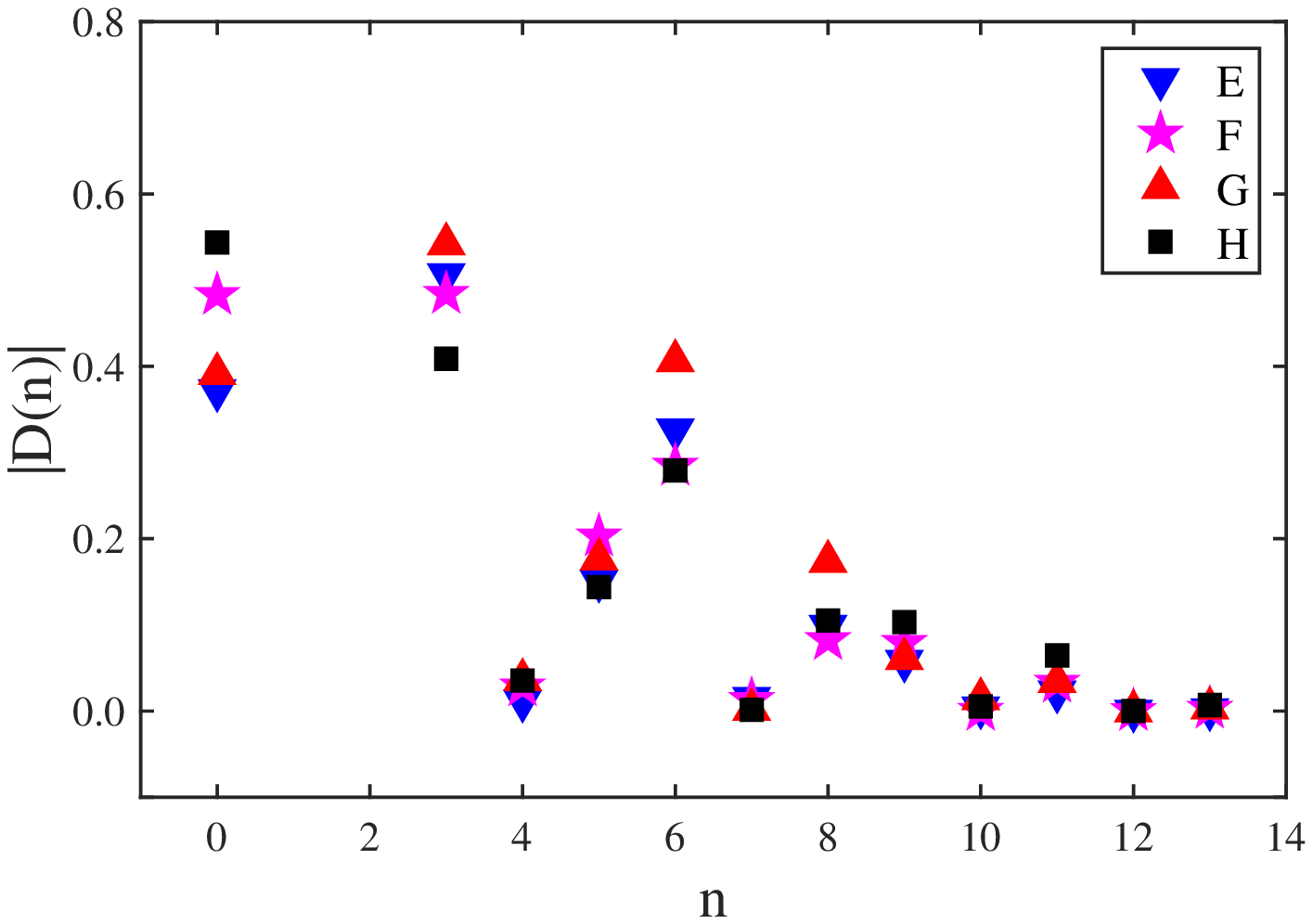}
 \caption{ The contributions of the vacuum condensates with the central values of the input parameters, where the $A$, $B$, $C$, $D$, $E$, $F$, $G$ and $H$ denote the molecular states $\bar{D}\Xi_c^{\prime}$ with $I=0$, $\bar{D}\Xi_c^{\prime}$ with
$I=1$, $\bar{D}\Xi_c^*$ with $I=0$, $\bar{D}\Xi_c^*$ with $I=1$, $\bar{D}^*\Xi_c^{\prime}$ with
$I=0$, $\bar{D}^*\Xi_c^{\prime}$ with $I=1$, $\bar{D}^*\Xi_c^*$ with $I=0$ and
$\bar{D}^*\Xi_c^*$ with $I=1$, sequentially.}\label{fig1}
 \end{figure}

 In Fig.\ref{fig1}, we plot the absolute values of the $D(n)$, and find that the  contributions of the higher dimensional vacuum condensates  play a tiny role, the contributions  $|D(12)|$ and $|D(13)|$ are less than $0.7\,\%$ for all the eight states, thus the convergence of the operator product expansion  holds well, the criterion {\bf 2} is satisfied very well.
  The most important contributions  come from the vacuum condensates $\langle \bar{q}q\rangle$ and $\langle \bar{q}q\rangle^2$, where $q=u$, $d$, or $s$. Furthermore, contributions of the gluon condensate $\langle\frac{\alpha_s}{\pi}GG\rangle$ are small for all the molecular states, $|D(4)|<4\%$.

\begin{figure}
 \centering
 \includegraphics[totalheight=5cm,width=7cm]{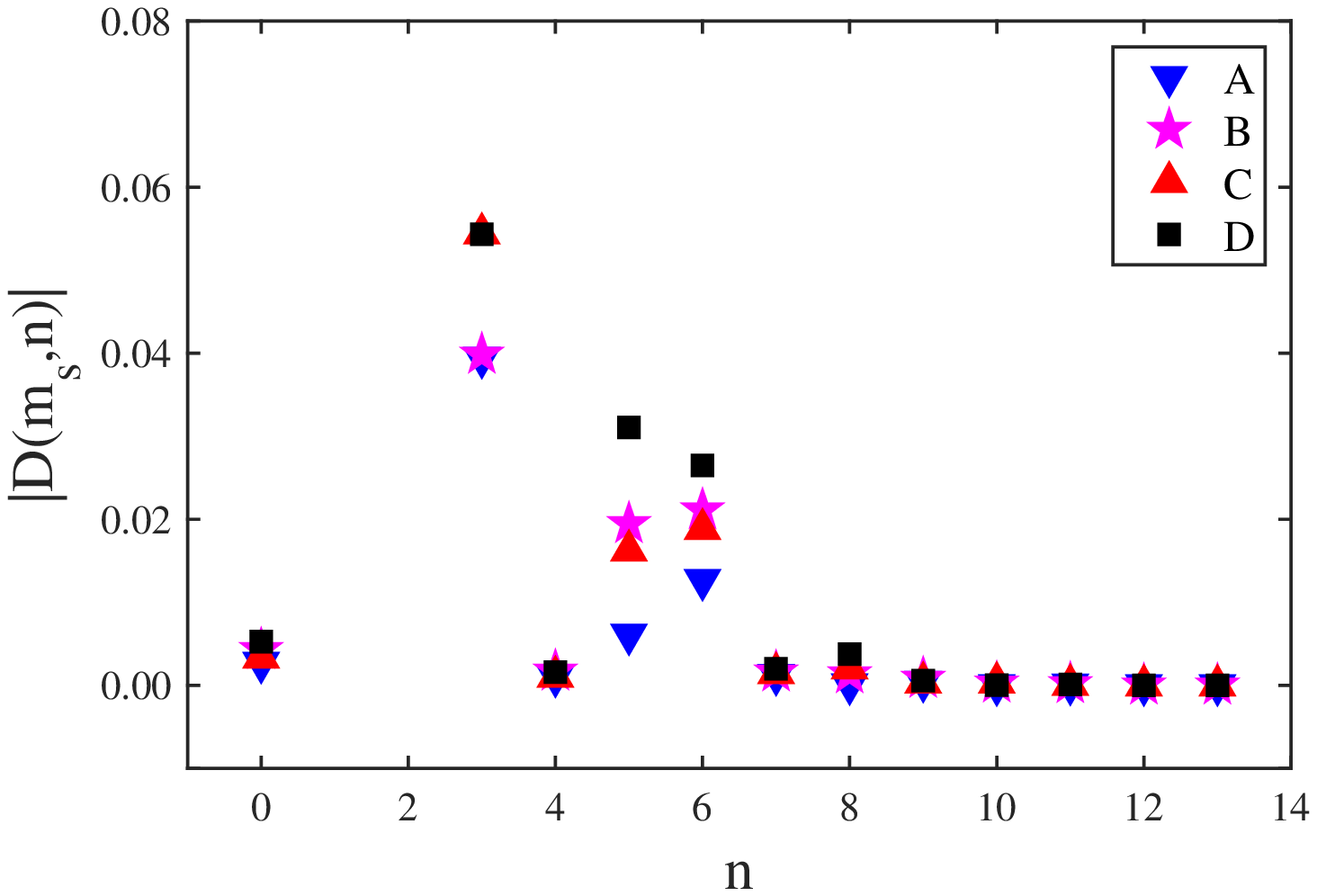}
 \includegraphics[totalheight=5cm,width=7cm]{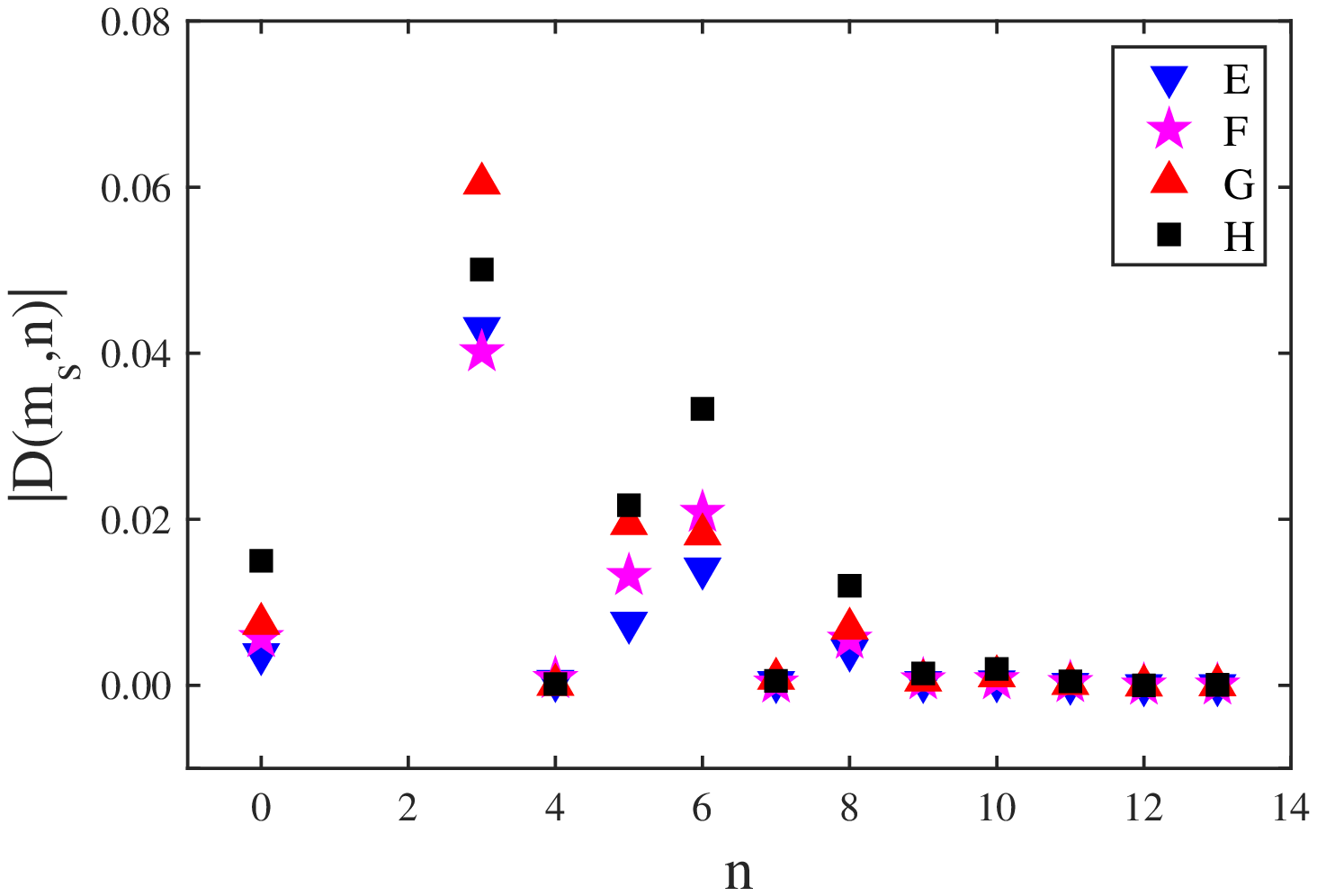}
 \caption{ The contributions of the vacuum condensates proportional to the $m_s$ with the central values of the input parameters,
where the $A$, $B$, $C$, $D$, $E$, $F$, $G$ and $H$ denote the molecular  states $\bar{D}\Xi_c^{\prime}$ with $I=0$, $\bar{D}\Xi_c^{\prime}$ with
$I=1$, $\bar{D}\Xi_c^*$ with $I=0$, $\bar{D}\Xi_c^*$ with $I=1$, $\bar{D}^*\Xi_c^{\prime}$ with
$I=0$, $\bar{D}^*\Xi_c^{\prime}$ with $I=1$, $\bar{D}^*\Xi_c^*$ with $I=0$ and
$\bar{D}^*\Xi_c^*$ with $I=1$, sequentially.}\label{fig2}
  \end{figure}

  We carry out a detailed calculation about the contributions of the terms which are proportional to the $m_s$ picked out from the spectral densities, and we find the total contributions $\sum_n D(m_s,n)$ for all the eight states are about $5\%$, thus it is accurate enough for us to consider the vacuum condensates  proportional to the $m_s^k$ up to $k=1$. Interestingly, for those terms of the vacuum condensates proportional to the $m_s$, $|D(m_s,\,4)|$, $|D(m_s,\,7)|$, $|D(m_s,\,9)|$, $|D(m_s,\,10)|$, $|D(m_s,\,11)|$, $|D(m_s,\,12)|$ and $|D(m_s,\,13)|$ are less than $0.5\%$, especially for the higher  dimensional terms, their contributions are even smaller, thus  the worthy calculations are of  the leading order, such as the $m_s\langle \bar{q}q\rangle$, $m_s\langle\overline{q}g_s\sigma Gq\rangle$, $m_s\langle \bar{q}q\rangle^2$ and $m_s\langle \bar{q}q\rangle$$\langle\overline{q}g_s\sigma Gq\rangle$, where $q=u,\,d$ or $s$. In Fig.\ref{fig2}, we draw the dimensional contributions of the vacuum condensates proportional to the $m_s$ explicitly.

\begin{sidewaystable}[thp]
\begin{center}
\begin{tabular}{|c|c|c|c|c|c|c|c|c|c|c|c|c|}\hline\hline
                         &$IJ^P$                        &$T^2({\rm GeV}^2)$   &$\sqrt{s_0}({\rm GeV})$    &$\mu ({\rm GeV})$   &$\rm PC$  &$M({\rm GeV})$  &$\lambda(10^{-3}{\rm GeV}^6)$     & Assignments  &Thresholds (MeV)\\ \hline

$\bar{D}\Xi_c^{\prime}$  &$0{\frac{1}{2}}^-$    &$3.4-4.0$  &$5.12\pm0.10$  &$2.2$   &$(41-58)\% $ &$4.43^{+0.07}_{-0.07}$  &$3.02^{+0.39}_{-0.37}$   &molecular state  &$4446$ \\ \hline
$\bar{D}\Xi_c^{\prime}$  &$1{\frac{1}{2}}^-$    &$3.2-3.8$  &$5.14\pm0.10$  &$2.3$   &$(43-61)\% $ &$4.45^{+0.07}_{-0.08}$  &$2.50^{+0.33}_{-0.31}$   &resonance state  &$4446$ \\ \hline

$\bar{D}\Xi_c^*$         &$0{\frac{3}{2}}^-$    &$3.4-4.0$  &$5.15\pm0.10$  &$2.3$   &$(43-60)\% $ &$4.46^{+0.07}_{-0.07}$  &$1.71^{+0.22}_{-0.21}$   &$P_{cs}(4459)$   &$4513$ \\ \hline
$\bar{D}\Xi_c^*$         &$1{\frac{3}{2}}^-$    &$3.3-3.9$  &$5.22\pm0.10$  &$2.4$   &$(44-62)\% $ &$4.53^{+0.07}_{-0.07}$  &$1.56^{+0.20}_{-0.19}$   &resonance state  &$4513$ \\ \hline

$\bar{D}^*\Xi_c^{\prime}$&$0{\frac{3}{2}}^-$    &$3.5-4.1$  &$5.26\pm0.10$  &$2.5$   &$(42-59)\% $ &$4.57^{+0.07}_{-0.07}$  &$3.41^{+0.43}_{-0.41}$   &molecular state &$4588$\\ \hline
$\bar{D}^*\Xi_c^{\prime}$&$1{\frac{3}{2}}^-$    &$3.4-4.0$  &$5.31\pm0.10$  &$2.6$   &$(43-60)\% $ &$4.62^{+0.08}_{-0.08}$  &$3.05^{+0.39}_{-0.37}$   &resonance state &$4588$\\ \hline

$\bar{D}^*\Xi_c^*$       &$0{\frac{5}{2}}^-$    &$3.6-4.2$  &$5.31\pm0.10$  &$2.6$   &$(42-58)\% $ &$4.64^{+0.07}_{-0.07}$  &$4.36^{+0.54}_{-0.51}$   &molecular state &$4655$\\ \hline
$\bar{D}^*\Xi_c^*$       &$1{\frac{5}{2}}^-$    &$3.4-4.0$  &$5.35\pm0.10$  &$2.6$   &$(44-61)\% $ &$4.67^{+0.08}_{-0.08}$  &$3.25^{+0.41}_{-0.39}$   &resonance state &$4655$\\
\hline\hline
$\bar{D}\Sigma_c$        &$\frac{1}{2}{\frac{1}{2}}^-$    &$3.2-3.8$  &$5.00\pm0.10$  &$2.2$   &$(42-60)\% $ &$4.31^{+0.07}_{-0.07}$  &$3.25^{+0.43}_{-0.41}$   &$P_c(4312)$  &$4321$ \\ \hline
$\bar{D}\Sigma_c$        &$\frac{3}{2}{\frac{1}{2}}^-$    &$2.8-3.4$  &$4.98\pm0.10$  &$2.2$   &$(44-65)\% $ &$4.33^{+0.09}_{-0.08}$  &$1.97^{+0.28}_{-0.26}$   &resonance state  &$4321$ \\ \hline

$\bar{D}\Sigma_c^*$      &$\frac{1}{2}{\frac{3}{2}}^-$    &$3.3-3.9$  &$5.06\pm0.10$  &$2.3$   &$(42-60)\% $ &$4.38^{+0.07}_{-0.07}$  &$1.97^{+0.26}_{-0.24}$   &$P_c(4380)$  &$4385$ \\ \hline
$\bar{D}\Sigma_c^*$      &$\frac{3}{2}{\frac{3}{2}}^-$    &$2.9-3.5$  &$5.03\pm0.10$  &$2.4$   &$(44-64)\% $ &$4.41^{+0.08}_{-0.08}$  &$1.24^{+0.17}_{-0.16}$   &resonance state &$4385$\\ \hline

$\bar{D}^*\Sigma_c$      &$\frac{1}{2}{\frac{3}{2}}^-$    &$3.3-3.9$  &$5.12\pm0.10$  &$2.5$   &$(42-60)\% $ &$4.44^{+0.07}_{-0.08}$  &$3.60^{+0.47}_{-0.44}$   &$P_c(4440)$ &$4462$\\ \hline
$\bar{D}^*\Sigma_c$      &$\frac{3}{2}{\frac{3}{2}}^-$    &$3.0-3.6$  &$5.10\pm0.10$  &$2.5$   &$(41-61)\% $ &$4.47^{+0.09}_{-0.09}$  &$2.31^{+0.33}_{-0.31}$   &resonance state &$4462$\\ \hline

$\bar{D}^*\Sigma_c^*$    &$\frac{1}{2}{\frac{5}{2}}^-$    &$3.2-3.8$  &$5.08\pm0.10$  &$2.5$   &$(43-60)\% $ &$4.46^{+0.08}_{-0.08}$  &$4.05^{+0.54}_{-0.50}$   &$P_c(4457)$  &$4527$\\ \hline
$\bar{D}^*\Sigma_c^*$    &$\frac{3}{2}{\frac{5}{2}}^-$    &$3.0-3.6$  &$5.24\pm0.10$  &$2.8$   &$(42-61)\% $ &$4.62^{+0.09}_{-0.09}$  &$2.40^{+0.37}_{-0.35}$   &resonance state &$4527$\\
\hline\hline
\end{tabular}
\end{center}
\caption{The Borel parameters, continuum threshold parameters, energy scales of the spectral densities, pole contributions, masses and pole residues of the pentaquark molecular states with possible assignments.  The predictions of the molecular  states without strangeness are also presented \cite{wangxiuwuN}. And thresholds of the corresponding meson-baryon pairs are listed.   }\label{table}
\end{sidewaystable}

\begin{figure}
 \centering
 \includegraphics[totalheight=5cm,width=7cm]{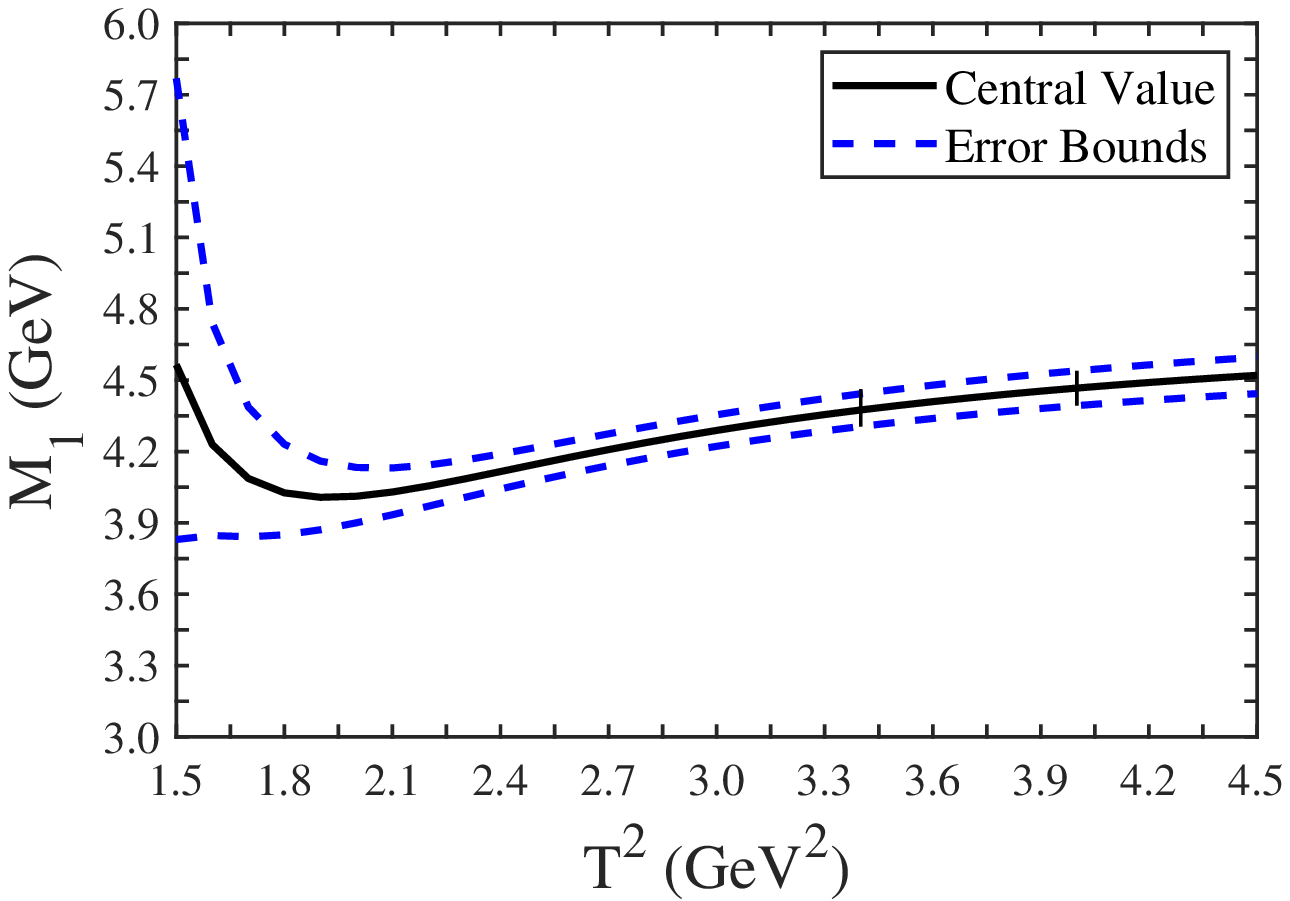}
 \includegraphics[totalheight=5cm,width=7cm]{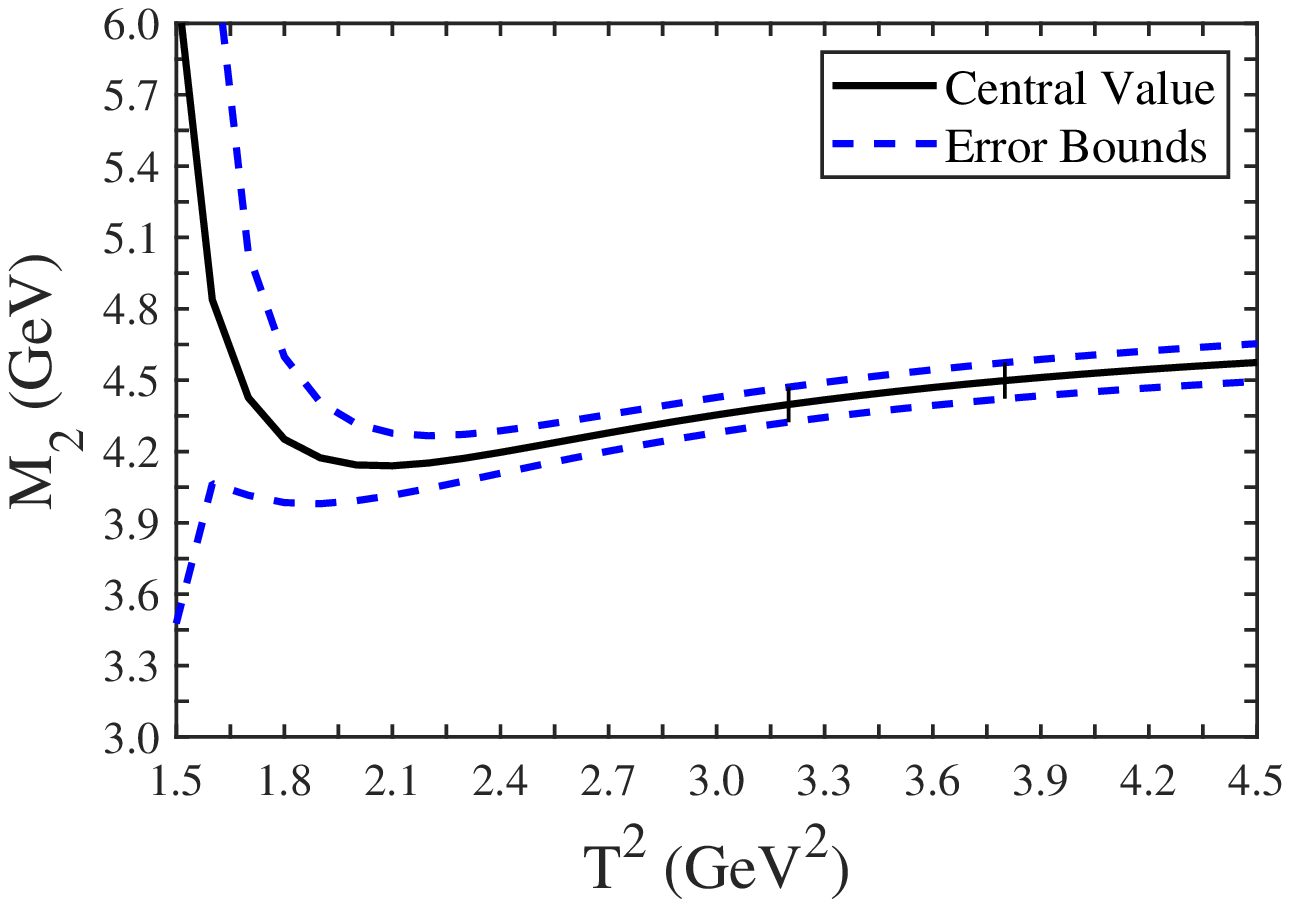}
 \includegraphics[totalheight=5cm,width=7cm]{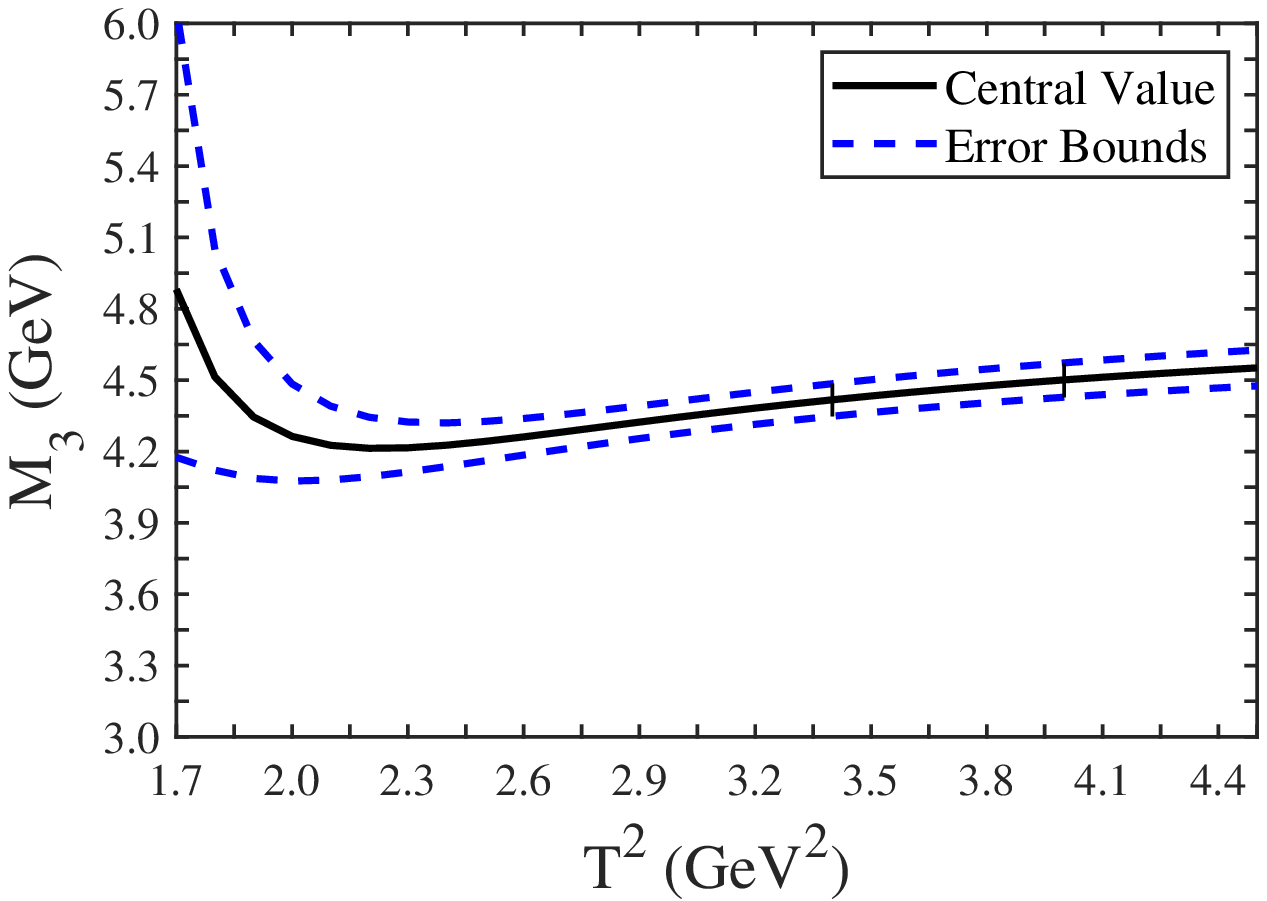}
 \includegraphics[totalheight=5cm,width=7cm]{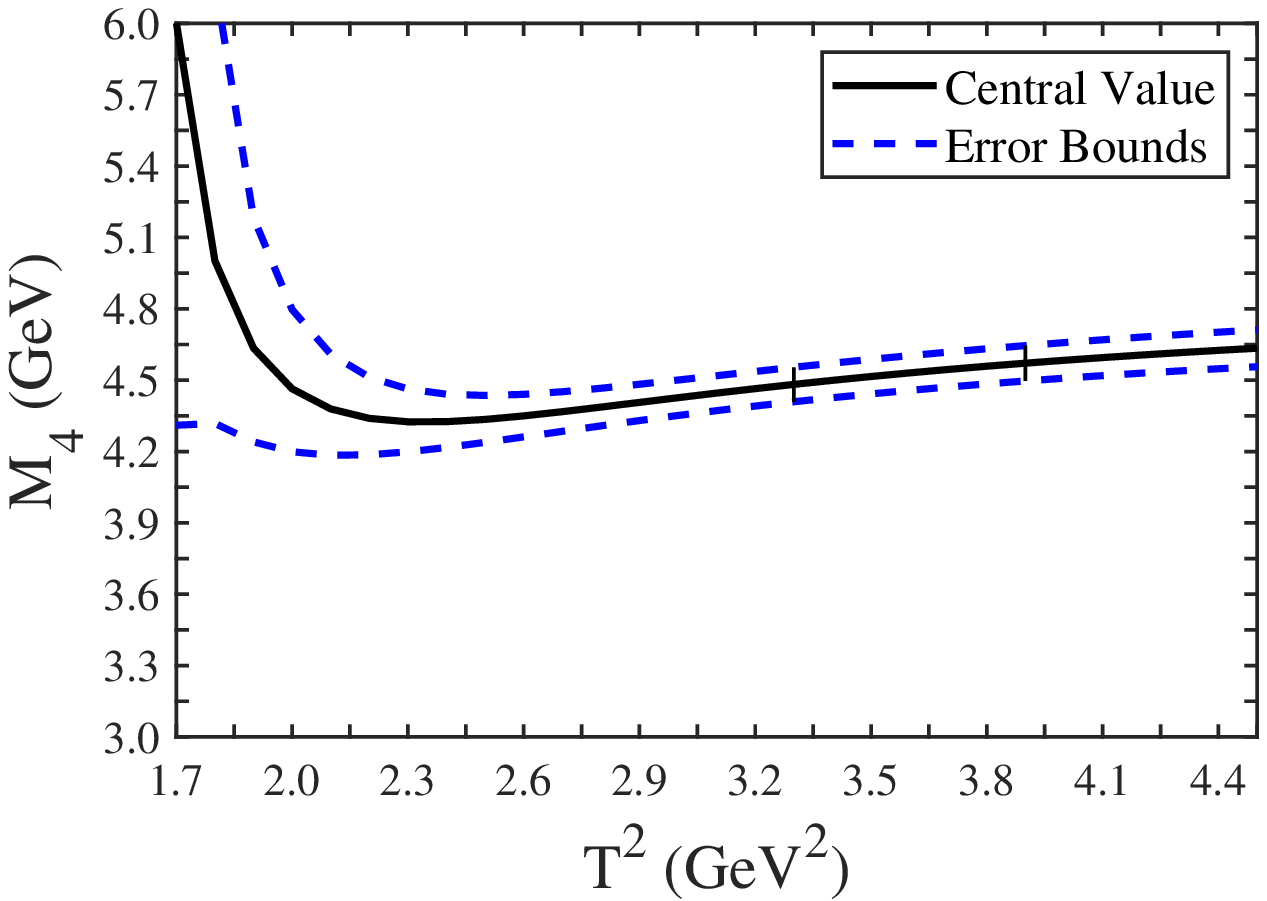}
 \includegraphics[totalheight=5cm,width=7cm]{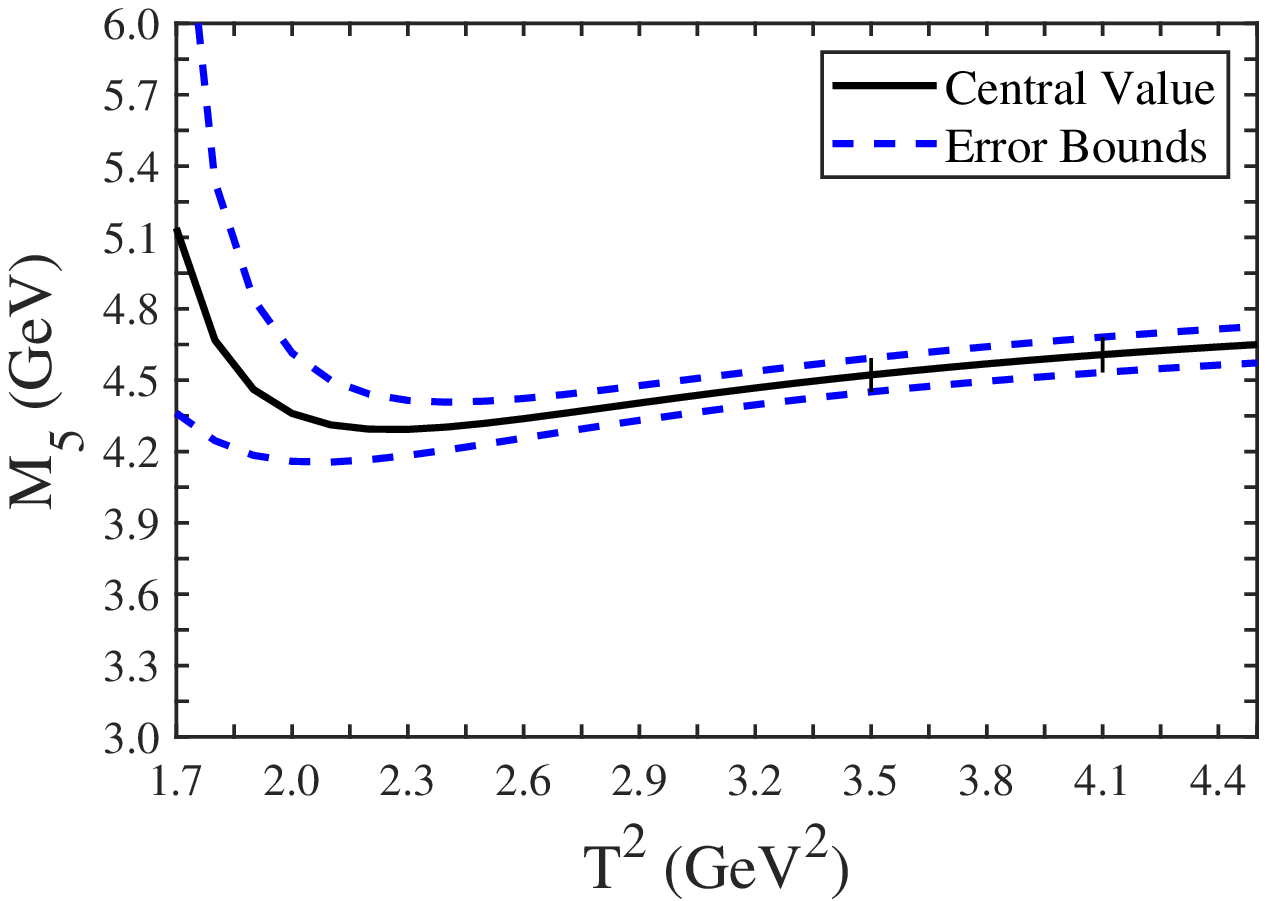}
 \includegraphics[totalheight=5cm,width=7cm]{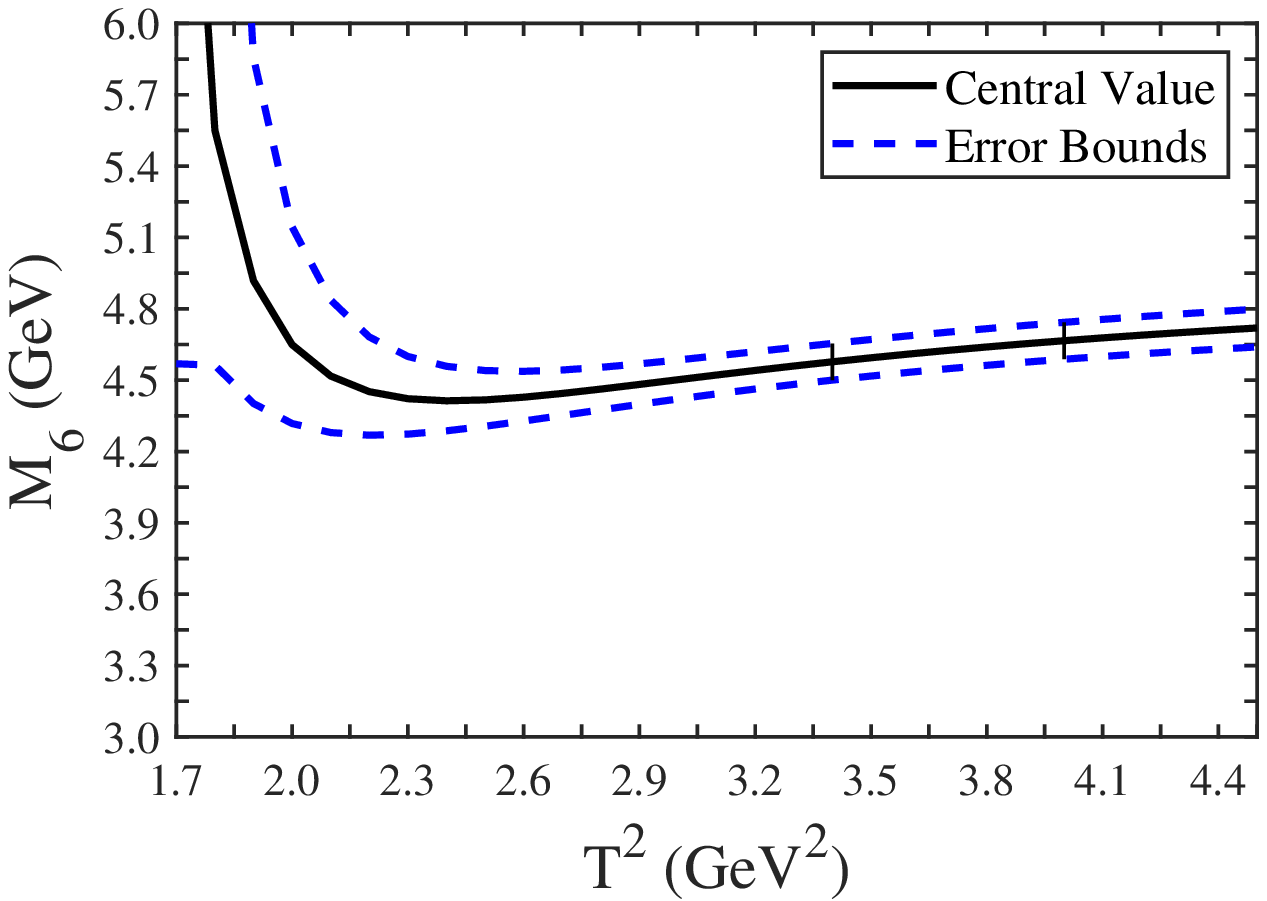}
 \includegraphics[totalheight=5cm,width=7cm]{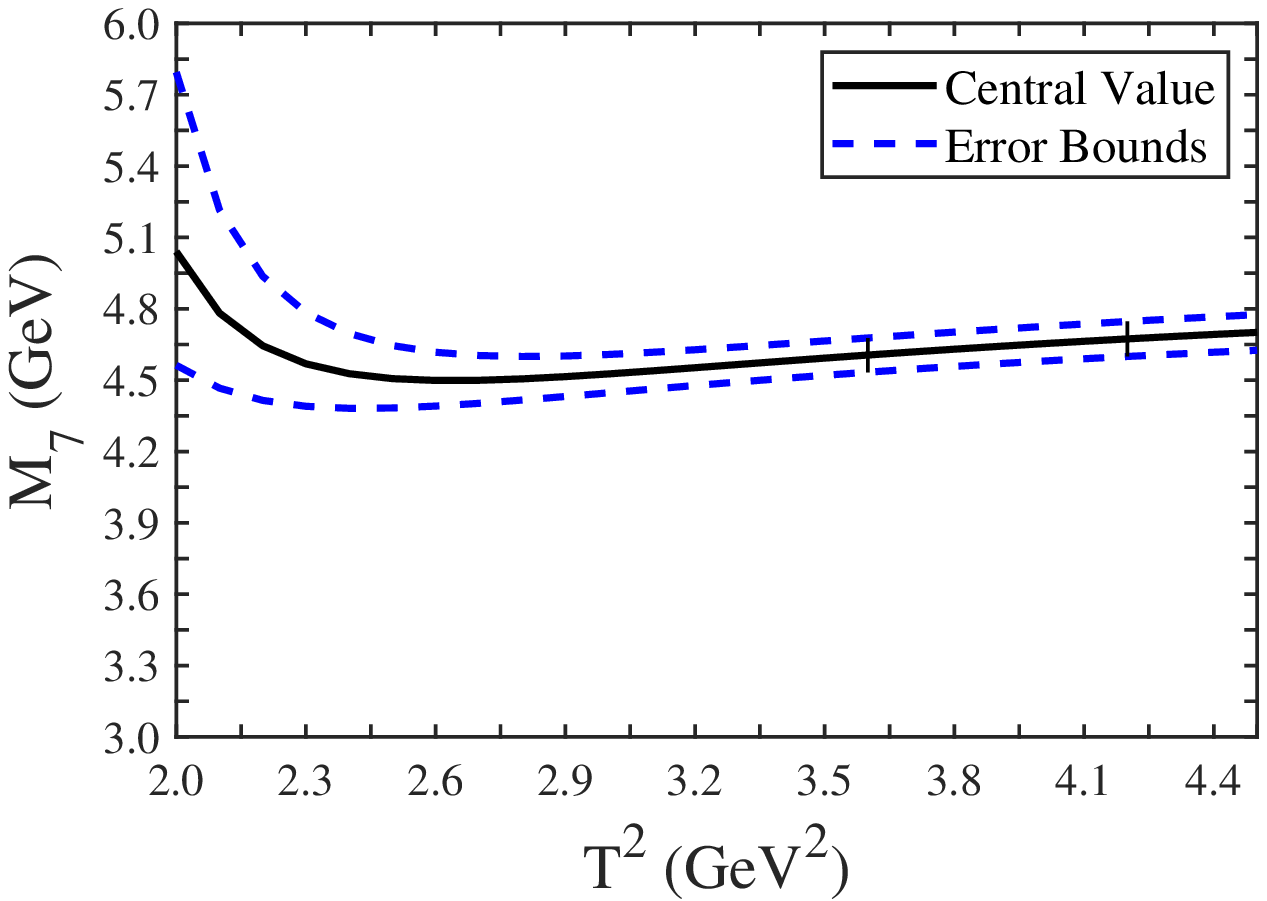}
 \includegraphics[totalheight=5cm,width=7cm]{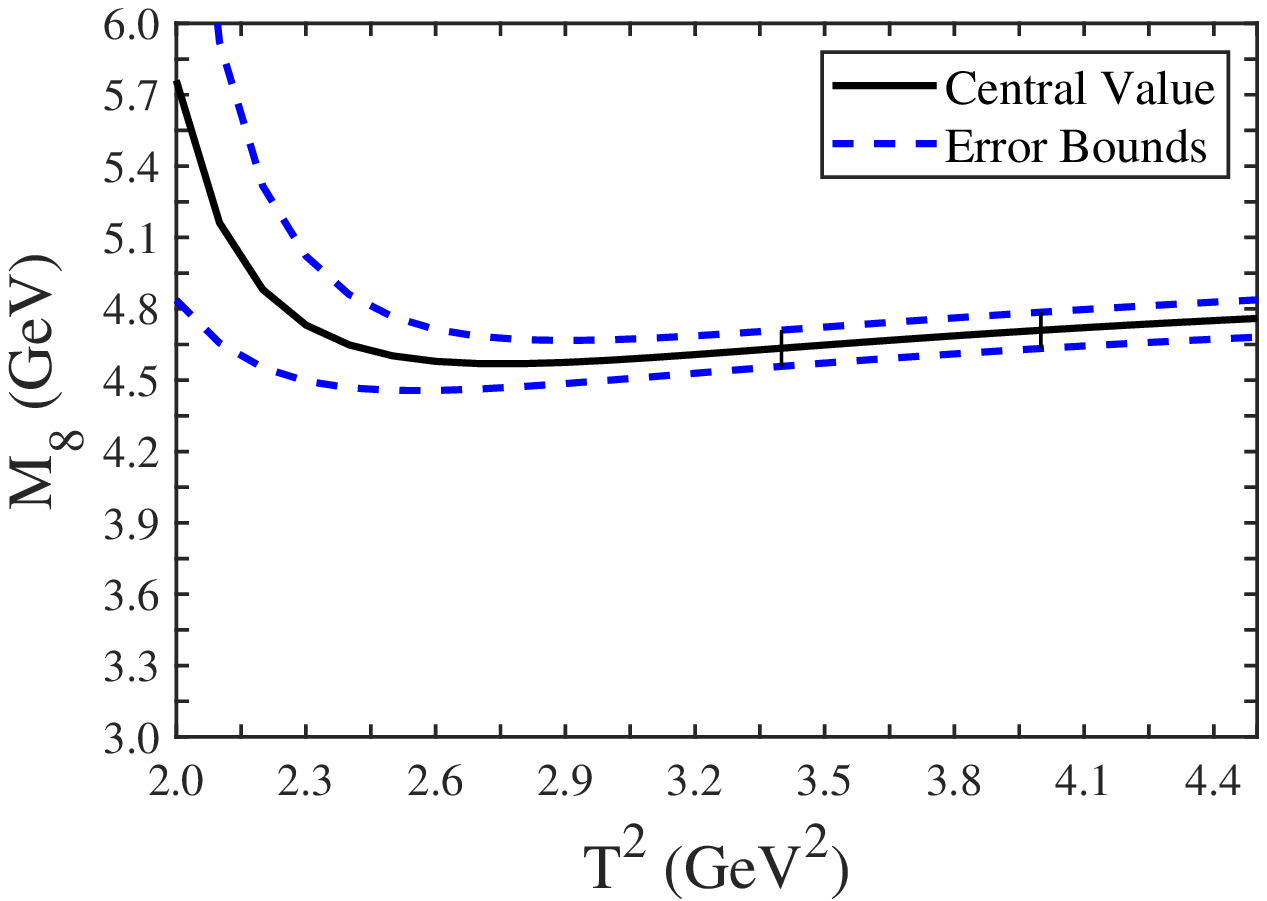}
 \caption{ The masses $M_i$ of the pentaquark molecular states  with variations of the Borel parameters $T^2$, where the $M_i\,(i=1,2,\cdot\cdot\cdot,\,8)$ denote the masses of the $\bar{D}\Xi^{\prime}_c$ with
 $I=0$, $\bar{D}\Xi^{\prime}_c$ with $I=1$, $\bar{D}\Xi_c^*$ with $I=0$, $\bar{D}\Xi_c^*$ with
 $I=1$, $\bar{D}^*\Xi^{\prime}_c$ with $I=0$, $\bar{D}^*\Xi^{\prime}_c$ with $I=1$, $\bar{D}^*\Xi_c^*$
 with $I=0$ and $\bar{D}^*\Xi_c^*$ with $I=1$, sequentially.}\label{mass-Borel}
\end{figure}

\begin{figure}
 \centering
 \includegraphics[totalheight=5cm,width=7cm]{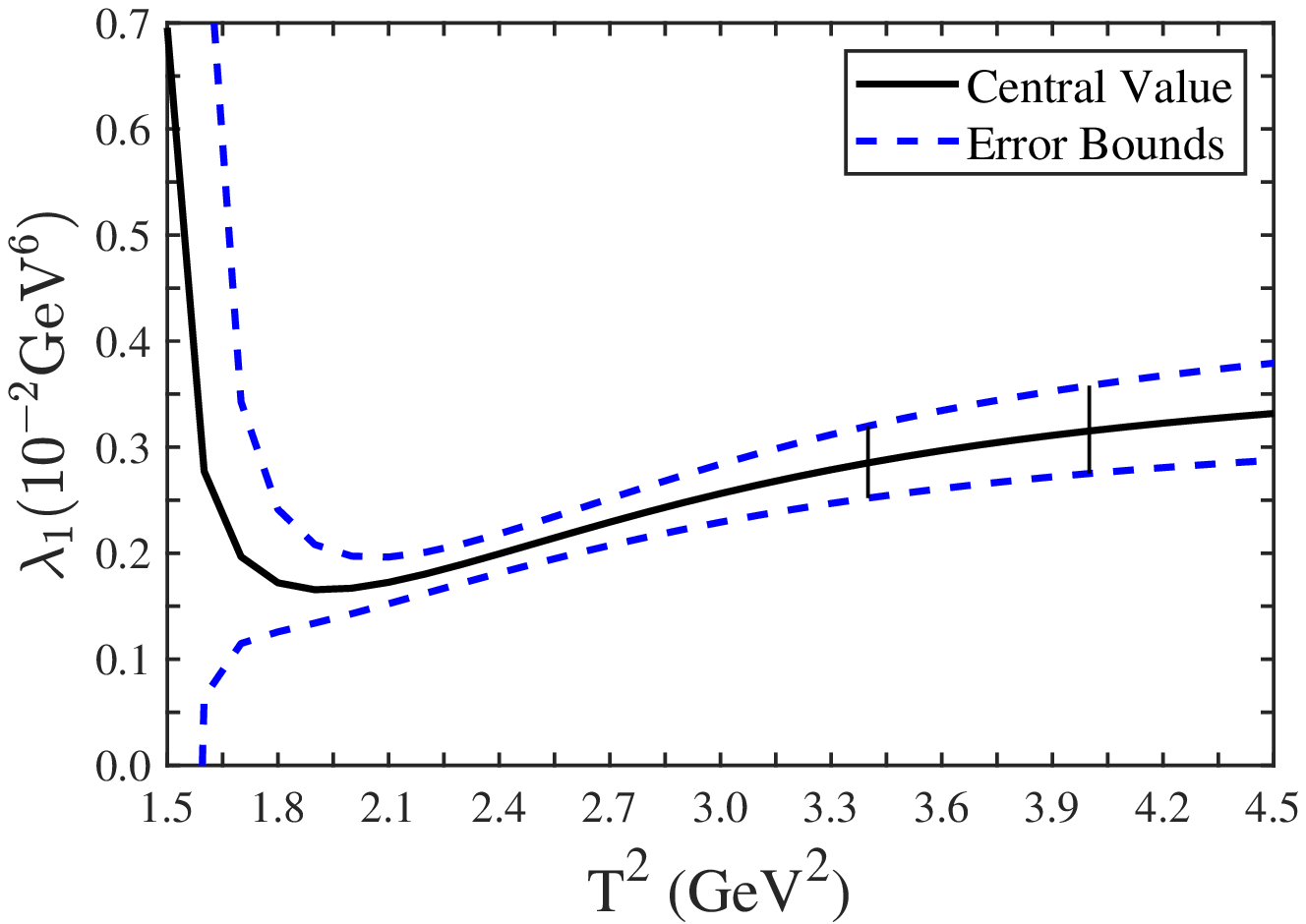}
 \includegraphics[totalheight=5cm,width=7cm]{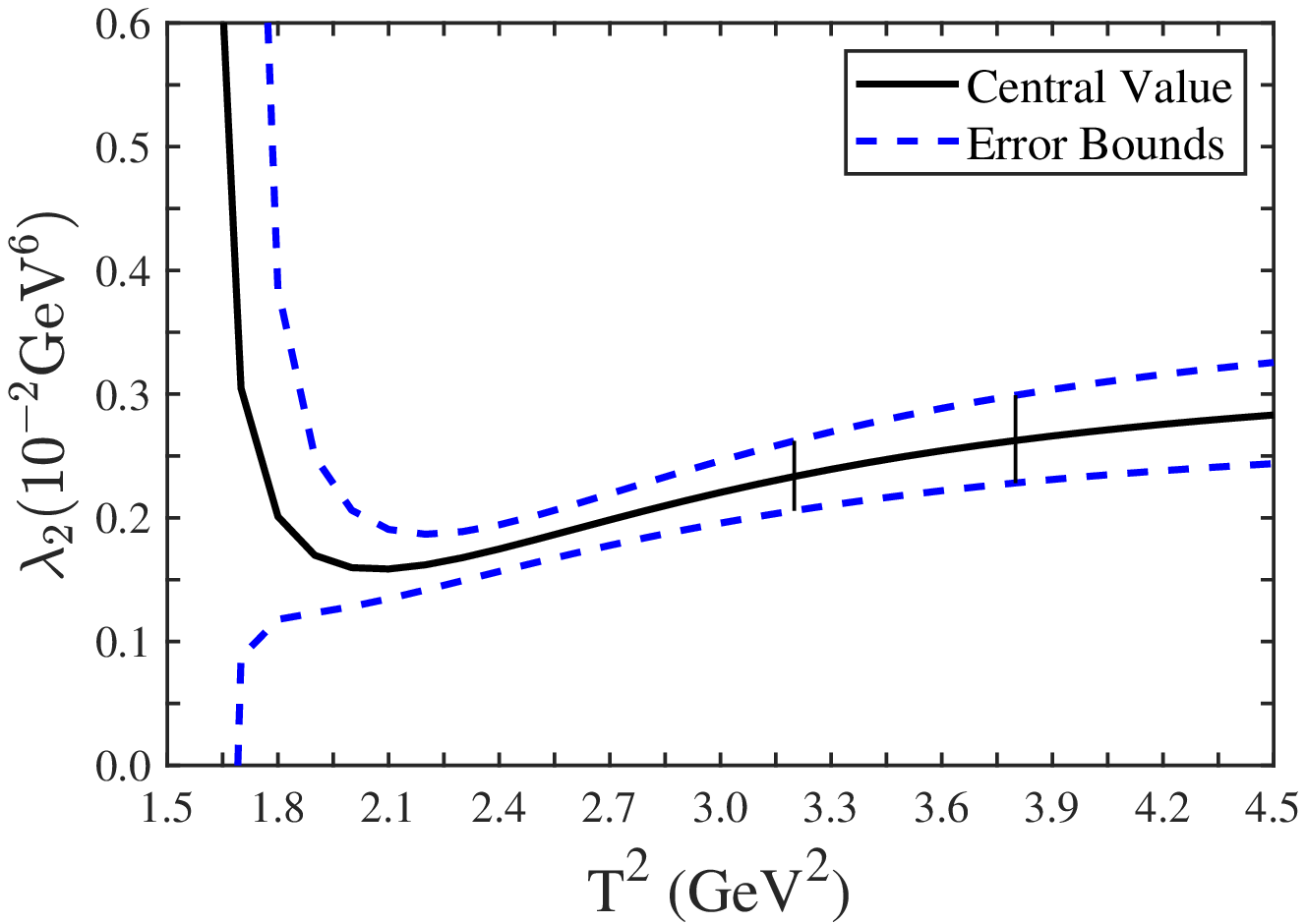}
 \includegraphics[totalheight=5cm,width=7cm]{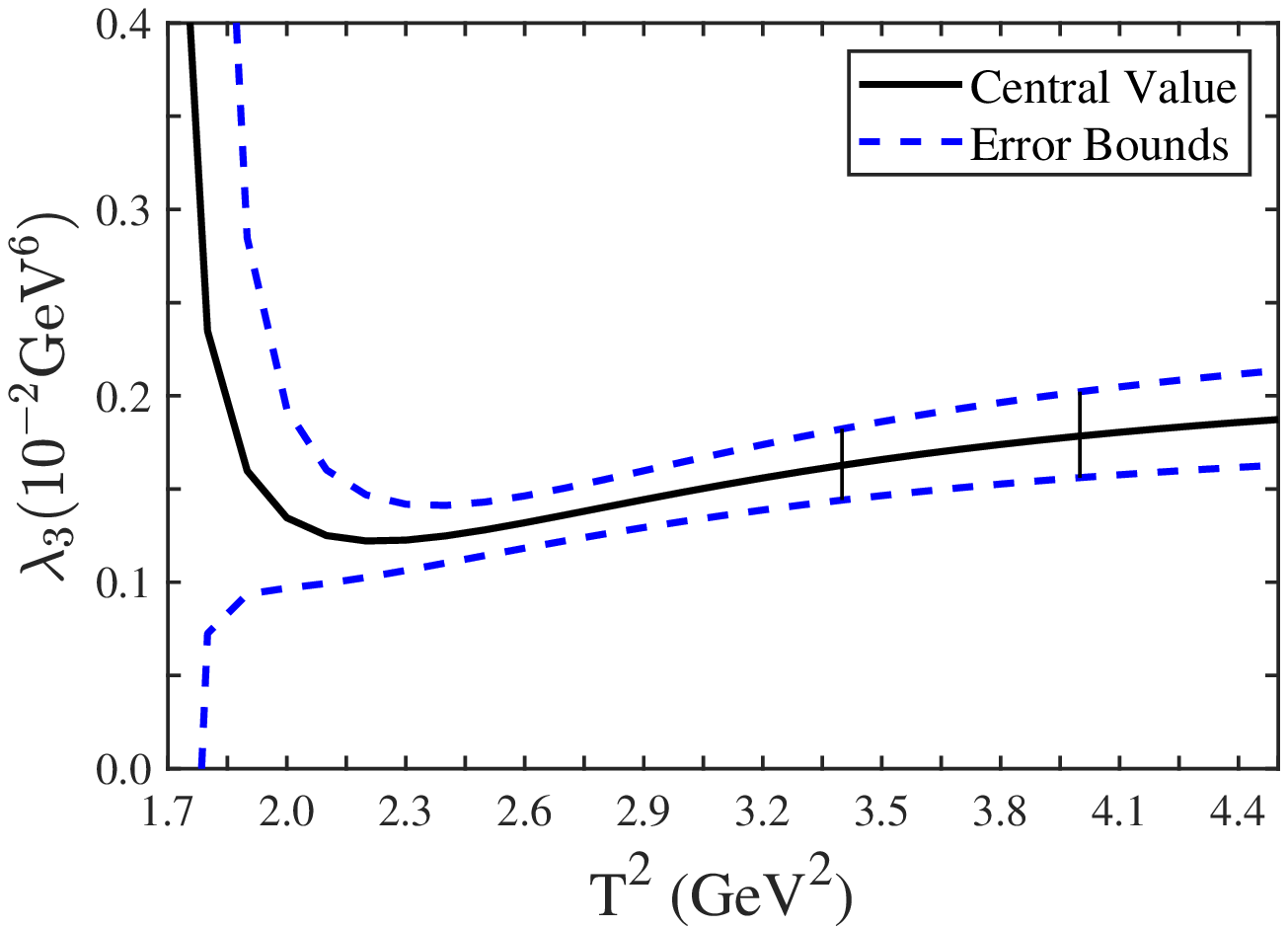}
 \includegraphics[totalheight=5cm,width=7cm]{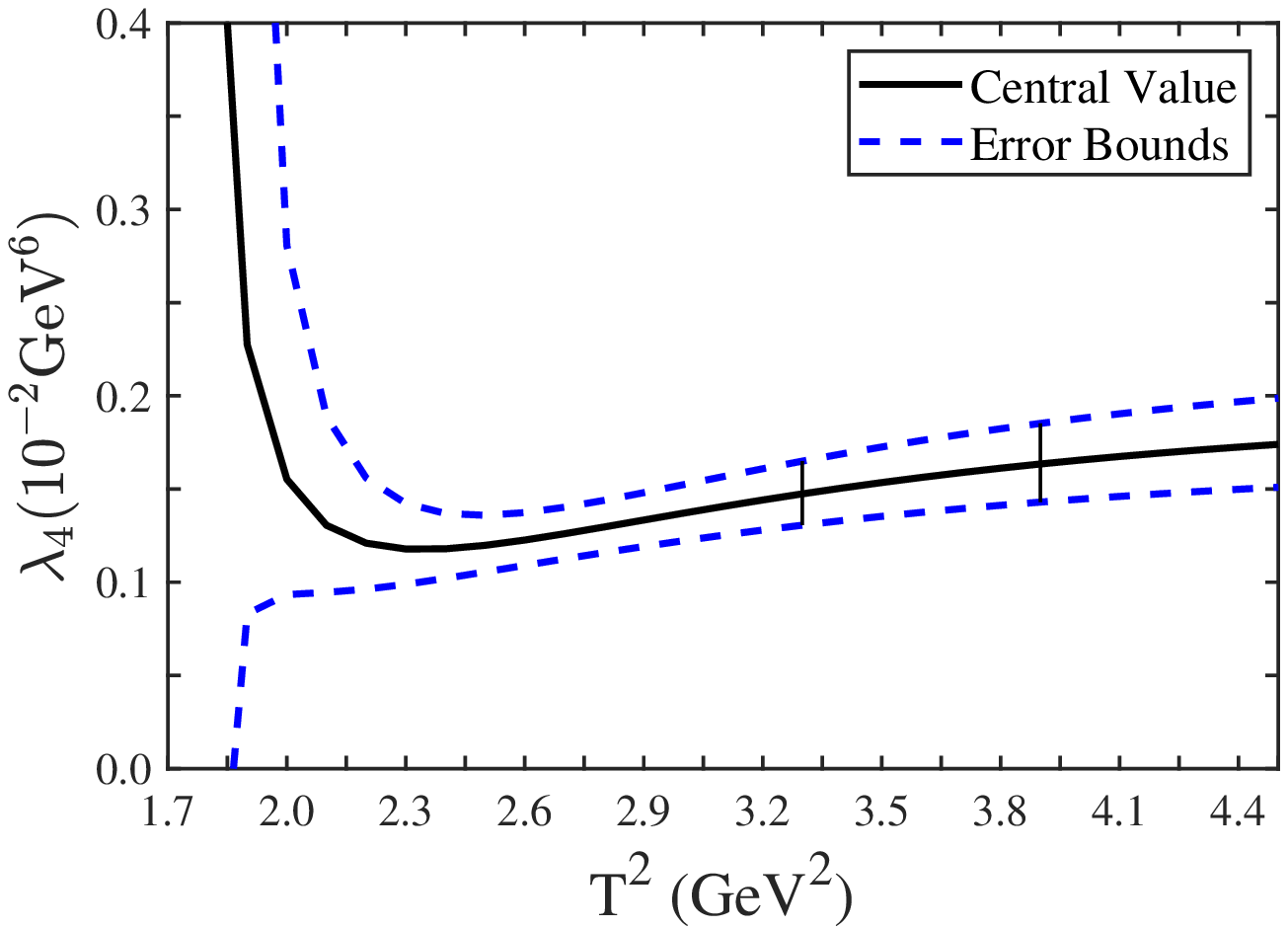}
 \includegraphics[totalheight=5cm,width=7cm]{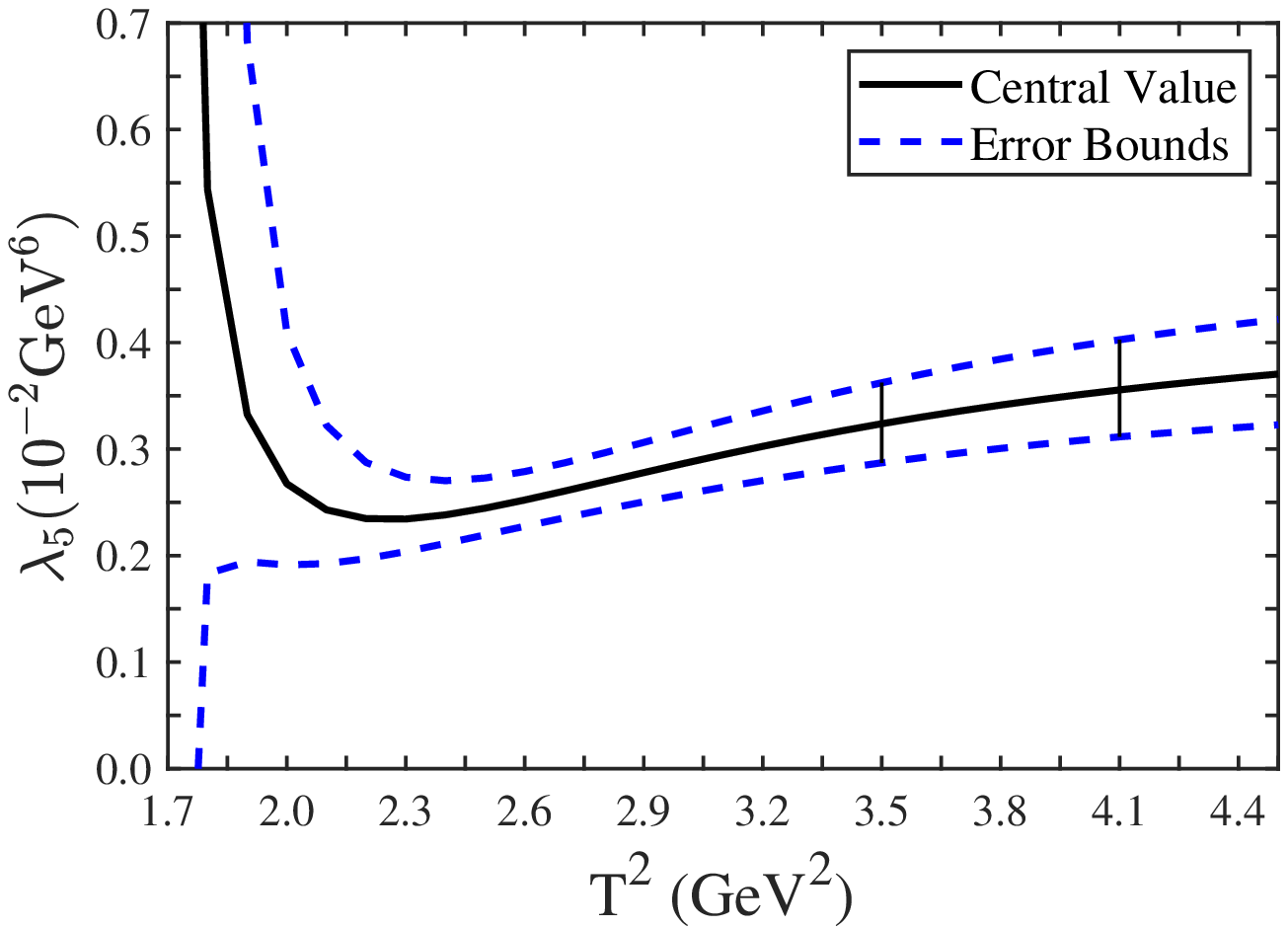}
 \includegraphics[totalheight=5cm,width=7cm]{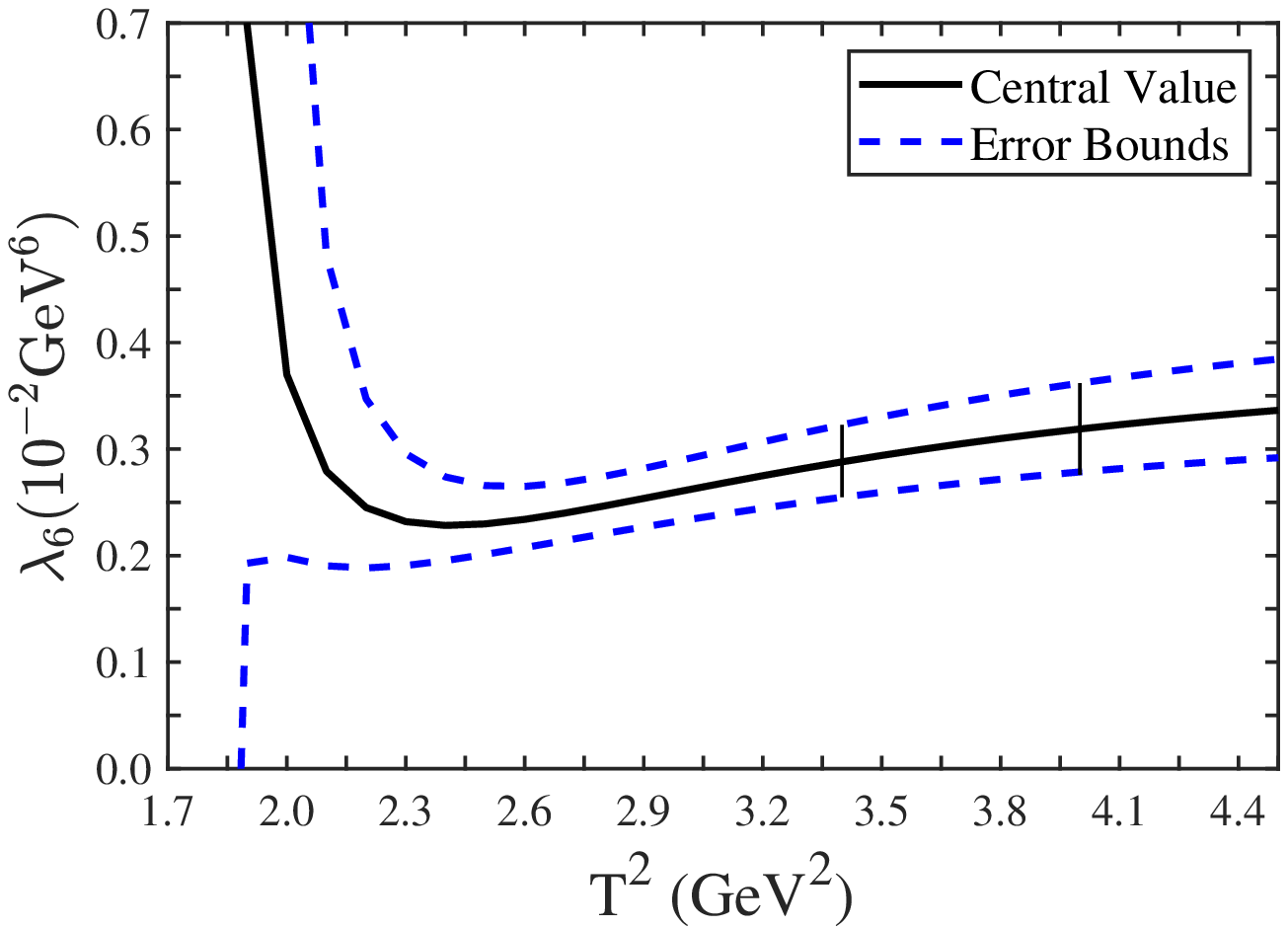}
 \includegraphics[totalheight=5cm,width=7cm]{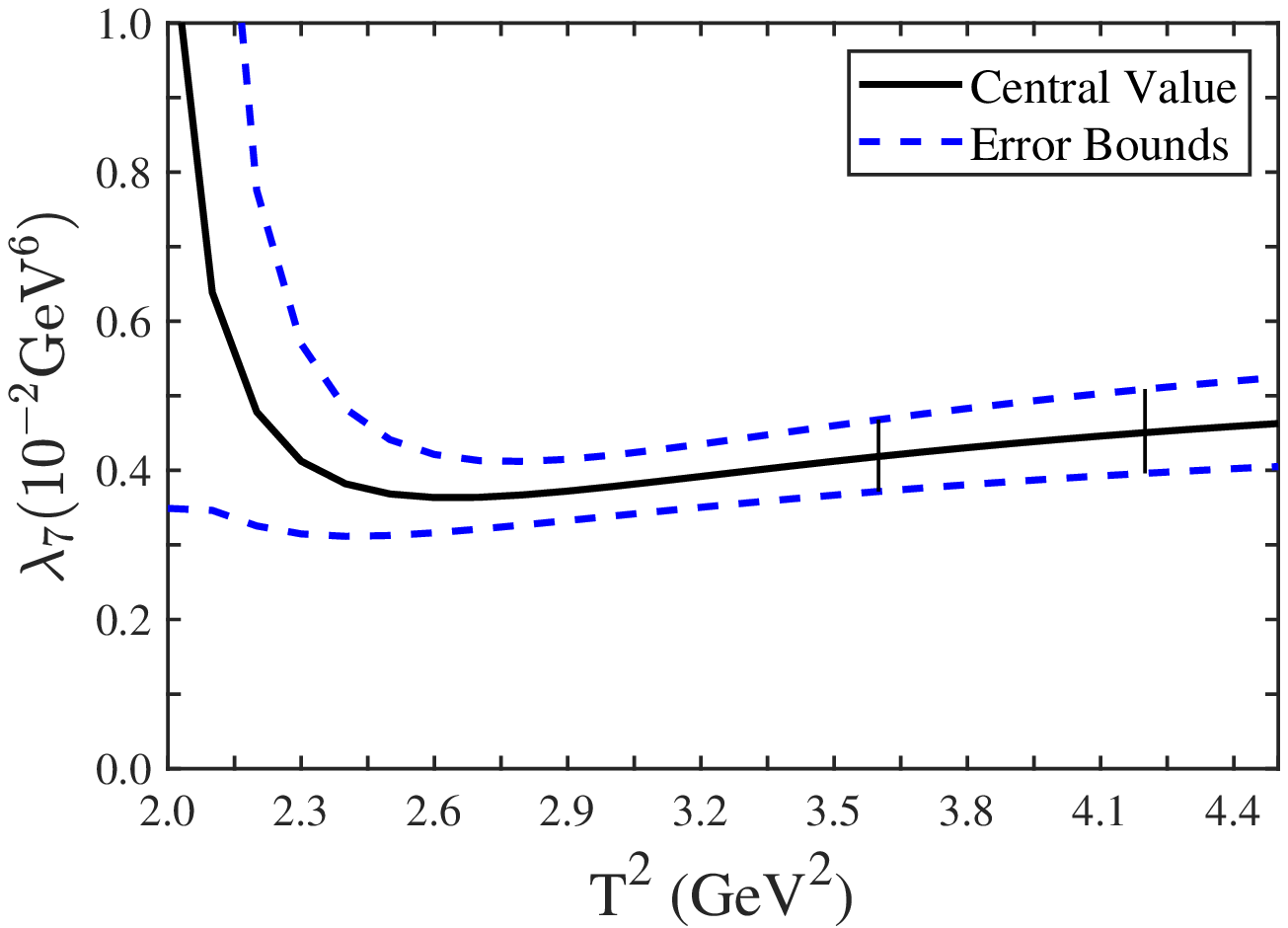}
 \includegraphics[totalheight=5cm,width=7cm]{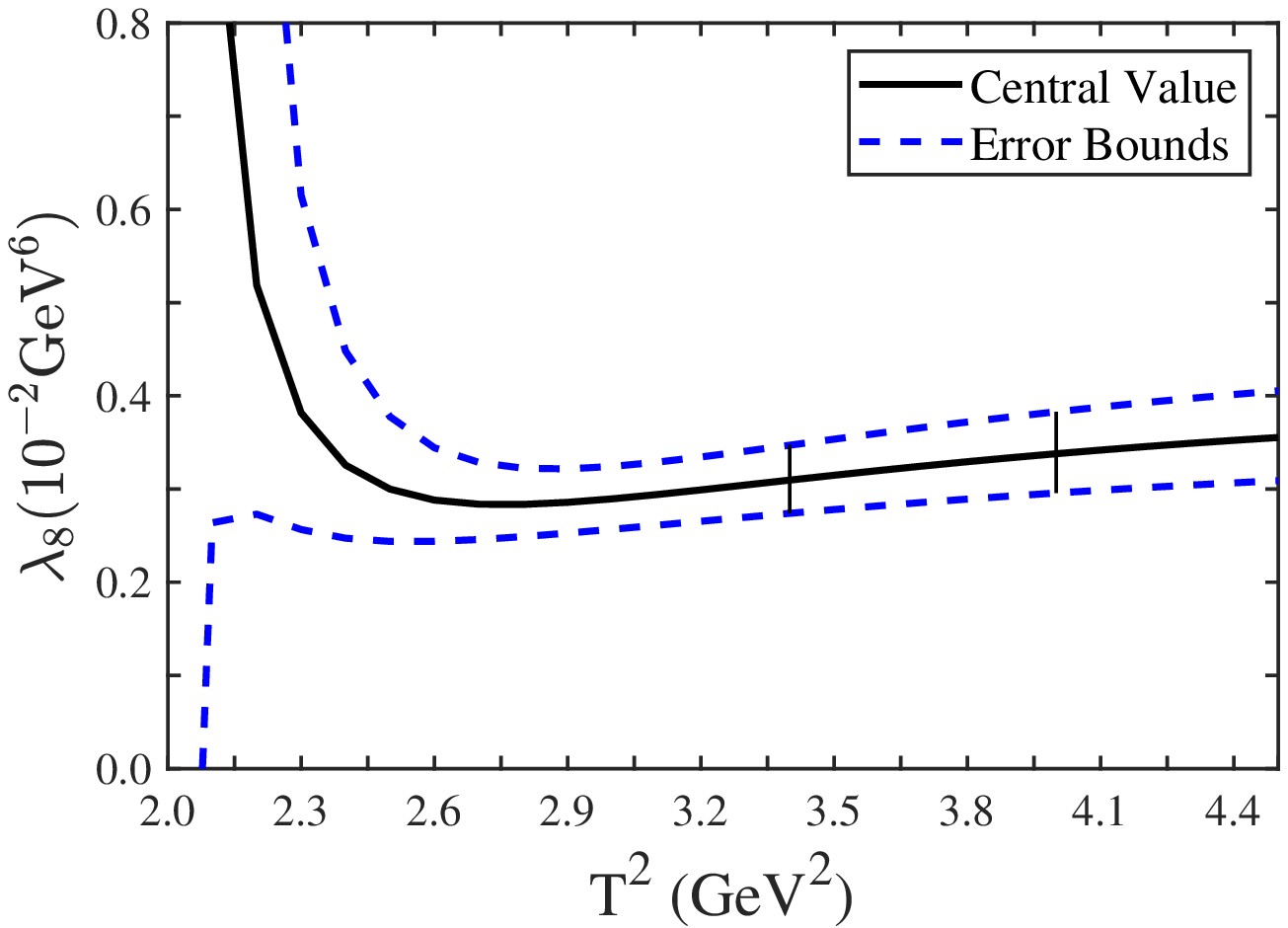}
 \caption{ The  pole residues $\lambda_i$ of the pentaquark molecular states with variations of the Borel parameters $T^2$, where the $\lambda_i\,(i=1,2,\cdot\cdot\cdot,\,8)$ denote the pole residues of the $\bar{D}\Xi^{\prime}_c$ with
 $I=0$, $\bar{D}\Xi^{\prime}_c$ with $I=1$, $\bar{D}\Xi_c^*$ with $I=0$, $\bar{D}\Xi_c^*$ with
 $I=1$, $\bar{D}^*\Xi^{\prime}_c$ with $I=0$, $\bar{D}^*\Xi^{\prime}_c$ with $I=1$, $\bar{D}^*\Xi_c^*$
 with $I=0$ and $\bar{D}^*\Xi_c^*$ with $I=1$, sequentially.}\label{residue-Borel}
\end{figure}

We take account of all the uncertainties (including the Borel parameters) of the input parameters, and acquire  the numerical values of the masses and pole residues, which are shown in the Table \ref{table} (also in
Figs.\ref{mass-Borel}-\ref{residue-Borel}). The present study belongs to the systematic research of the color singlet-singlet type pentaquark molecular states being the eigenstates of isospins, the predictions  of the $P_c$ states from our previous study \cite{wangxiuwuN} are also presented  in the Table \ref{table} for the sake of completeness.

In Figs.\ref{mass-Borel}-\ref{residue-Borel}, we plot the masses and pole residues with variations of the Borel parameters, respectively. From the figures, we can see clearly that there appear platforms in the Borel windows, which are the regions between the two short vertical lines. The criterion {\bf 3} is satisfied very well. On the other hand, from Table \ref{table}, we can verify that the modified energy scale formula is obeyed, the criterion {\bf 4} is also satisfied very well. Now the four criteria are all satisfied, and we expect to make reliable predictions.

The masses of the molecular states with the isospin $I=0$ are slightly below the  thresholds of the corresponding meson-baryon pairs, for the ones with the isospin $I=1$, they are a few dozens of $\rm{MeV}$ above the  thresholds of the corresponding meson-baryon pairs. If we take the central values as the milestone,  we tentatively denote  the states with the isospin $I=0$ as the molecular states, and the states with the isospin $I=1$ as the resonance states. Especially,  the masses of the $\bar{D}^*\Xi_c^{\prime}$ and $\bar{D}^*\Xi_c^{*}$ molecular states lie  about $0.1\,\rm{GeV}$  and  $0.2\,\rm{GeV}$  above the $P_{cs}(4459)$ respectively, the $P_{cs}(4459)$ is unlikely to be the $\bar{D}^*\Xi_c^{\prime}$ or $\bar{D}^*\Xi_c^{*}$ molecular state. The mass of the $\bar{D}\Xi^{\prime}$ molecular state with the isospin $I=1$ is $4.45_{-0.08}^{+0.07}\,\rm{GeV}$, which is near the value $4459\,\rm{MeV}$, but this state lies slightly above the corresponding  meson-baryon threshold, it is reasonable for us to assign it as the resonance state, more significantly, from the observed decay mode $P_{cs}(4459)\to J/\psi \Lambda$ \cite{RAaij3}, the isospin of the exotic state $P_{cs}(4459)$ is zero, which also excludes assigning the $P_{cs}(4459)$ as the $\bar{D}\Xi_c^{*}$ molecular state with the isospin $I=1$. The mass of the $\bar{D}\Xi_c^{*}$ molecular state with the isospin $I=0$ is $4.46^{+0.07}_{-0.07}\,\rm{GeV}$, which is in very good agreement with the experimental mass of the $P_{cs}(4459)$, thus, it is nice for us to assign the $P_{cs}(4459)$ as the $\bar{D}\Xi_c^{*}$ molecular state with the isospin $I=0$, then we tentatively determine the $J^P$ of the $P_{cs}(4459)$ to be ${\frac{3}{2}}^-$.

We usually choose the diquark-diquark-antiquark type \cite{WZG-Pcs,KAzizi,Uozdem,WZGNNN1,WZGXZ5}, color singlet-singlet type \cite{HXChenN,Uozdem,WangZG-Xin-CPC,wangxiuwuN,CHXXZ2,CHXXZ3,WZGXZ7,WZGXZ9} and color octet-octet type \cite{Pimikov-Penta}
five-quark currents with the definite quantum numbers, which correspond to several  pentaquark states or a pentaquark state with several  Fock components. It is not odd to acquire different predicted pentaquark masses.
On the other hand, the pentaquark masses are extracted in the Borel windows, there are several schemes in determining the Borel windows, which depend on how to truncate the operator product expansion, how to judge the convergence, how to choose the pole contributions, etc. Even for the same current,  different Borel windows lead to different predictions. The central values of the masses of the lowest pentaquark states with the valence quarks $qqsc\bar{c}$
vary in the range $4.3\sim 4.6\,\rm{GeV}$ \cite{HXChenN,WZG-Pcs,KAzizi,Uozdem,Pimikov-Penta}. All the predictions should be confronted to the experimental data to select the optimal QCD sum rules. And the pole residues can be used to calculate the strong decays of those pentaquark (molecular) states via the QCD sum rules using the three-point correlation functions.
At the present time, the experimental data are far from enough, we can only obtain the conclusion that the mass of the $P_{cs}(4459)$ can be reproduced  both in the picture of the pentaquark state and molecular state, it maybe have both the diquark-diquark-antiquark type and color singlet-singlet type Fock components \cite{HXChenN,WZG-Pcs,KAzizi,Uozdem}.

\section{Conclusions}
In this paper, we construct the color singlet-singlet type five-quark currents with strangeness to study the pentaquark (molecular) states being the isospin eigenstates $I=0$ and $I=1$ via the QCD sum rules. After detailed analysis, we acquire the masses of the $\bar{D}\Xi^{\prime}$, $\bar{D}\Xi_c^{*}$, $\bar{D}^{*}\Xi_c^{\prime}$ and $\bar{D}^{*}\Xi_c^{*}$ molecular states with the isospins $I=0$ and $I=1$, and observe that the molecular states with the lower isospin lie a few dozens of $\rm{MeV}$ below the corresponding higher isospin ones, what's more, they are slightly below the  thresholds of the corresponding meson-baryon pairs. While the molecular states with the higher isospin lie above the  thresholds of the corresponding meson-baryon pairs. The mass of the $\bar{D}\Xi_c^{*}$ molecular state with the isospin $I=0$ coincides well with that of the exotic state $P_{cs}(4459)$,  it is natural and reasonable to assign the $P_{cs}(4459)$ as the $\bar{D}\Xi_c^{*}$ molecular state with the quantum numbers $IJ^P=0{\frac{3}{2}}^-$. The numerical results  present a reference for the experimental search of the other pentaquark molecular states with strangeness besides the $P_{cs}(4459)$, and shed light on the low-energy QCD dynamics.

\section*{Appendix}
The detailed QCD spectral densities for the current $J_{0}^{\bar{D}\Xi_c^{\prime}}(x)$,

\begin{eqnarray}
\notag
\rho_{QCD}^1(s)&=&\sum\limits_{n}\left[\rho^1_{a}(n)+\rho^1_{b}(n)+\rho^1_{c}(n)\delta(s-\widetilde{m}_c^2)+\rho^1_{d}(n)
\delta(s-\overline{m}_c^2)\right]\, ,\\
\notag \rho_{QCD}^0(s)&=&\sum\limits_{n}\left[\rho^0_{a}(n)+\rho^0_{b}(n)+\rho^0_{c}(n)\delta(s-\widetilde{m}_c^2)+
\rho^0_{d}(n)\delta(s-\overline{m}_c^2)\right]\, ,
\end{eqnarray}
 where the $a$, $b$, $c$ and $d$ refer to four types of integrals, the $n$ are the dimension of the vacuum condensates.
In the integrals, we introduce the notations  $\widetilde{m}_c^2=\frac{m_c^2}{y(1-y)}$, $\overline{m}_c^2=\frac{(y+z)m_c^2}{y z}$, $y+z-1=\xi$, $1-y=\zeta$ and $s-\overline{m}_c^2=\omega$.
 For the types $a$ and $b$, $y_i=\frac{1}{2}\left(1-\sqrt{1-4m_c^2/s}\right)$, $y_f=\frac{1}{2}\left(1+\sqrt{1-4m_c^2/s}\right)$ and $z_i=\frac{y m_c^2}{y s-m_c^2}$. For the types $c$ and $d$, $y_i=0$, $y_f=1$ and $z_i=0$.

The $a$ type integrals for $\rho_{QCD}^1(s)$,

\begin{eqnarray}
\notag \rho^1_a(10)&=&\frac{\langle\bar{q}g_s\sigma Gq\rangle\left[2 \langle\bar{q}g_s\sigma Gq\rangle + 13 \langle\bar{s}g_s\sigma Gs\rangle\right]}{4096 \pi^4} \int_{y_i}^{y_f}dy \,  y \zeta    +\frac{ \langle g_s^2GG\rangle \langle\bar{q}q\rangle \left[\langle\bar{q}q\rangle  + 7  \langle\bar{s}s\rangle \right]}{9216 \pi^4} \int_{y_i}^{y_f}dy \,   y \zeta     \, ,
\end{eqnarray}

\begin{eqnarray}
\notag \rho^1_a(9)&=&\frac{m_s \langle\bar{q}q\rangle^2 \langle\bar{s}s\rangle}{192 \pi^2} \int_{y_i}^{y_f}dy \,  y \zeta   -\frac{13 m_c \langle\bar{q}q\rangle^2 \langle\bar{s}s\rangle}{576 \pi^2} \int_{y_i}^{y_f}dy \,  \zeta   \\
\notag &&-\frac{m_s g_s^2 \langle\bar{q}q\rangle^2 \left[7 \langle\bar{q}q\rangle - 13  \langle\bar{s}s\rangle \right]}{10368 \pi^4} \int_{y_i}^{y_f}dy \,  y \zeta   \\
\notag &&-\frac{m_c g_s^2 \langle\bar{q}q\rangle \left[14 \langle\bar{q}q\rangle^2 + \langle\bar{q}q\rangle \langle\bar{s}s\rangle + 14 \langle\bar{s}s\rangle^2\right]}{31104 \pi^4}  \int_{y_i}^{y_f}dy \,  \zeta     \, ,
\end{eqnarray}

\begin{eqnarray}
\notag \rho^1_a(8)&=&\frac{m_c m_s \left[-39 \langle\bar{q}q\rangle \langle\bar{q}g_s\sigma Gq\rangle + 18 \langle\bar{q}g_s\sigma Gq\rangle \langle\bar{s}s\rangle + 14 \langle\bar{q}q\rangle \langle\bar{s}g_s\sigma Gs\rangle\right]}{4608 \pi^4} \int_{y_i}^{y_f}dy \,  \zeta     \, .
\end{eqnarray}

The $b$ type integrals for $\rho_{QCD}^1(s)$,

\begin{eqnarray}
\notag \rho^1_b(10)&=&-\frac{ \langle g_s^2GG\rangle \langle\bar{q}q\rangle \left[\langle\bar{q}q\rangle + 2 \langle\bar{s}s\rangle\right]}{12288 \pi^4}  \int_{y_i}^{y_f}dy\int_{z_i}^{\zeta}dz  \,  \xi   +\frac{ \langle g_s^2GG\rangle  \langle\bar{q}q\rangle \left[\langle\bar{q}q\rangle + 30 \langle\bar{s}s\rangle\right]}{12288 \pi^4} \int_{y_i}^{y_f}dy\int_{z_i}^{\zeta}dz  \,  z   \\
\notag &&-\frac{\langle\bar{q}g_s\sigma Gq\rangle \left[ 124 \langle\bar{q}g_s\sigma Gq\rangle + 377 \langle\bar{s}g_s\sigma Gs\rangle \right]}{98304 \pi^4}   \int_{y_i}^{y_f}dy\int_{z_i}^{\zeta}dz  \,  z   \\
\notag &&-\frac{\langle\bar{q}g_s\sigma Gq\rangle \left[ 6 \langle\bar{q}g_s\sigma Gq\rangle + 5 \langle\bar{s}g_s\sigma Gs\rangle \right]}{24576 \pi^4}  \int_{y_i}^{y_f}dy\int_{z_i}^{\zeta}dz  \,  \xi     \, ,
\end{eqnarray}

\begin{eqnarray}
\notag \rho^1_b(8)&=&-\frac{4 \langle\bar{q}q\rangle \langle\bar{q}g_s\sigma Gq\rangle + 12 \langle\bar{q}g_s\sigma Gq\rangle \langle\bar{s}s\rangle + 15 \langle\bar{q}q\rangle \langle\bar{s}g_s\sigma Gs\rangle}{3072 \pi^4}  \int_{y_i}^{y_f}dy\int_{z_i}^{\zeta}dz  \,  y z (s + 3 \omega)   \\
\notag &&-\frac{5 \langle\bar{q}q\rangle \langle\bar{q}g_s\sigma Gq\rangle + 9 \langle\bar{q}g_s\sigma Gq\rangle \langle\bar{s}s\rangle + 6 \langle\bar{q}q\rangle \langle\bar{s}g_s\sigma Gs\rangle }{3072 \pi^4}  \int_{y_i}^{y_f}dy\int_{z_i}^{\zeta}dz  \,  z \xi (s + 3 \omega)   \\
\notag &&+\frac{7 m_c m_s \langle\bar{q}g_s\sigma Gq\rangle \left[ \langle\bar{q}q\rangle - \langle\bar{s}s\rangle \right]}{768 \pi^4}  \int_{y_i}^{y_f}dy\int_{z_i}^{\zeta}dz  \,  \frac{z}{y}   \\
\notag &&+\frac{m_c m_s \langle\bar{q}g_s\sigma Gq\rangle \left[ 14 \langle\bar{q}q\rangle - \langle\bar{s}s\rangle \right] }{6144 \pi^4}  \int_{y_i}^{y_f}dy\int_{z_i}^{\zeta}dz  \,      \, ,
\end{eqnarray}

\begin{eqnarray}
\notag \rho^1_b(7)&=&\frac{m_c^3  \langle g_s^2GG\rangle  \left[ 28 \langle\bar{q}q\rangle + \langle\bar{s}s\rangle \right]}{147456 \pi^6} \int_{y_i}^{y_f}dy\int_{z_i}^{\zeta}dz  \, \frac{\xi^2}{y^2}   \\
\notag &&+\frac{m_c^2 m_s  \langle g_s^2GG\rangle \left[ 14 \langle\bar{q}q\rangle-13 \langle\bar{s}s\rangle\right]}{24576 \pi^6}  \int_{y_i}^{y_f}dy\int_{z_i}^{\zeta}dz  \,   \frac{z \xi^2}{y^2}   \\
\notag &&+\frac{m_c  \langle g_s^2GG\rangle  \left[28 \langle\bar{q}q\rangle+\langle\bar{s}s\rangle\right] }{147456 \pi^6}  \int_{y_i}^{y_f}dy\int_{z_i}^{\zeta}dz  \,  \frac{z \xi^2}{y^2} \left[s y+(-2+3 y) \omega\right]   \\
\notag &&+\frac{m_c  \langle g_s^2GG\rangle  \left[2 \langle\bar{q}q\rangle+\langle\bar{s}s\rangle\right]}{98304 \pi^6}  \int_{y_i}^{y_f}dy\int_{z_i}^{\zeta}dz  \,  \frac{(8 z-\xi) \xi \omega}{y} \\
\notag &&+\frac{m_c  \langle g_s^2GG\rangle  \left[14 \langle\bar{q}q\rangle + \langle\bar{s}s\rangle\right] }{24576 \pi^6}  \int_{y_i}^{y_f}dy\int_{z_i}^{\zeta}dz  \,  \xi \omega   \\
\notag &&-\frac{7 m_c  \langle g_s^2GG\rangle  \left[4 \langle\bar{q}q\rangle + \langle\bar{s}s\rangle\right] }{147456 \pi^6}  \int_{y_i}^{y_f}dy\int_{z_i}^{\zeta}dz  \,  z \omega   \\
\notag &&-\frac{m_s  \langle g_s^2GG\rangle  \left[ 7 \langle\bar{q}q\rangle - 3 \langle\bar{s}s\rangle \right] }{73728 \pi^6}  \int_{y_i}^{y_f}dy\int_{z_i}^{\zeta}dz  \,   y z (s + 3 \omega)   \\
\notag &&-\frac{m_s   \langle g_s^2GG\rangle  \left[ 4 \langle\bar{q}q\rangle - \langle\bar{s}s\rangle \right] }{196608 \pi^6}  \int_{y_i}^{y_f}dy\int_{z_i}^{\zeta}dz  \,  \xi^2 (s + 3 \omega)   \\
\notag &&+\frac{5 m_s  \langle g_s^2GG\rangle  \left[ 6 \langle\bar{q}q\rangle - 7 \langle\bar{s}s\rangle \right]}{49152 \pi^6}  \int_{y_i}^{y_f}dy\int_{z_i}^{\zeta}dz  \,  z \xi (s + 3 \omega)  \, ,
\end{eqnarray}

\begin{eqnarray}
\notag \rho^1_b(6)&=&\frac{m_c m_s \langle\bar{q}q\rangle \left[ 13 \langle\bar{q}q\rangle - 14  \langle\bar{s}s\rangle \right]}{768 \pi^4}  \int_{y_i}^{y_f}dy\int_{z_i}^{\zeta}dz  \,  z \omega   \\
\notag &&-\frac{\langle\bar{q}q\rangle \left[ \langle\bar{q}q\rangle + 7 \langle\bar{s}s\rangle\right] }{768 \pi^4}  \int_{y_i}^{y_f}dy\int_{z_i}^{\zeta}dz  \,   y z \xi \omega (2 s + 3 \omega)   \\
\notag &&+\frac{m_c m_s g_s^2 \langle\bar{q}q\rangle^2}{41472 \pi^6}  \int_{y_i}^{y_f}dy\int_{z_i}^{\zeta}dz  \,  z \omega   \\
\notag &&-\frac{13 g_s^2 \left[2 \langle\bar{q}q\rangle^2 + \langle\bar{s}s\rangle^2\right]}{82944 \pi^6}  \int_{y_i}^{y_f}dy\int_{z_i}^{\zeta}dz  \,   y z \xi \omega (2 s + 3 \omega)     \, ,
\end{eqnarray}

\begin{eqnarray}
\notag \rho^1_b(5)&=&-\frac{m_c \left[12 \langle\bar{q}g_s\sigma Gq\rangle + \langle\bar{s}g_s\sigma Gs\rangle\right]}{4096 \pi^6}  \int_{y_i}^{y_f}dy\int_{z_i}^{\zeta}dz  \,  z \xi \omega^2   \\
\notag &&-\frac{m_c \left[2 \langle\bar{q}g_s\sigma Gq\rangle + \langle\bar{s}g_s\sigma Gs\rangle\right]}{32768 \pi^6} \int_{y_i}^{y_f}dy\int_{z_i}^{\zeta}dz  \,  \xi^2 \omega^2   \\
\notag &&-\frac{m_c \left[56 \langle\bar{q}g_s\sigma Gq\rangle-\langle\bar{s}g_s\sigma Gs\rangle\right]}{16384 \pi^6}  \int_{y_i}^{y_f}dy\int_{z_i}^{\zeta}dz  \,  \frac{z \xi^2 \omega^2}{y}   \\
\notag &&-\frac{m_s \left[18 \langle\bar{q}g_s\sigma Gq\rangle - 13 \langle\bar{s}g_s\sigma Gs\rangle\right]}{12288 \pi^6} \int_{y_i}^{y_f}dy\int_{z_i}^{\zeta}dz  \,   y z \xi \omega (2 s + 3 \omega)   \\
\notag &&-\frac{9 m_s \langle\bar{q}g_s\sigma Gq\rangle}{16384 \pi^6}  \int_{y_i}^{y_f}dy\int_{z_i}^{\zeta}dz  \,  z \xi^2 \omega (2 s + 3 \omega)    \, ,
\end{eqnarray}

\begin{eqnarray}
\notag \rho^1_b(4)&=&\frac{m_c^3 m_s  \langle g_s^2GG\rangle }{589824 \pi^8}   \int_{y_i}^{y_f}dy\int_{z_i}^{\zeta}dz  \,   \frac{\xi^3 \omega}{y^2}    - \frac{13 m_c^2   \langle g_s^2GG\rangle }{1179648  \pi^8}  \int_{y_i}^{y_f}dy\int_{z_i}^{\zeta}dz  \,   \frac{z \xi^4}{y^2} \omega (2 s+3 \omega)   \\
\notag &&+ \frac{m_c m_s  \langle g_s^2GG\rangle }{1179648  \pi^8 }   \int_{y_i}^{y_f}dy\int_{z_i}^{\zeta}dz  \,   \frac{z  \xi^3 }{y^2}\omega \left[2 s y+(-2+3 y) \omega\right]   \\
\notag &&+ \frac{m_c m_s   \langle g_s^2GG\rangle }{786432  \pi^8 }   \int_{y_i}^{y_f}dy\int_{z_i}^{\zeta}dz  \,   \frac{(6 y-\xi) \xi^2\omega^2}{y} - \frac{m_c m_s  \langle g_s^2GG\rangle }{65536  \pi^8 }  \int_{y_i}^{y_f}dy\int_{z_i}^{\zeta}dz  \,   \frac{z (y-\xi) \xi \omega^2}{y}   \\
\notag &&-  \frac{ \langle g_s^2GG\rangle }{3145728 \pi^8}   \int_{y_i}^{y_f}dy\int_{z_i}^{\zeta}dz  \,  (332 z - \xi) \xi^3 \omega^2 (s + \omega)   + \frac{ \langle g_s^2GG\rangle }{65536 \pi^8}   \int_{y_i}^{y_f}dy\int_{z_i}^{\zeta}dz  \,   y z \xi^2 \omega^2 (s + \omega)     \, ,
\end{eqnarray}

\begin{eqnarray}
\notag \rho^1_b(3)&=&- \frac{m_s \left[14 \langle\bar{q}q\rangle - 13 \langle\bar{s}s\rangle\right]}{4096 \pi^6}   \int_{y_i}^{y_f}dy\int_{z_i}^{\zeta}dz  \,   y z \xi^2 \omega^2 (s + \omega)    - \frac{m_c \left[28 \langle\bar{q}q\rangle + \langle\bar{s}s\rangle\right]}{12288 \pi^6}   \int_{y_i}^{y_f}dy\int_{z_i}^{\zeta}dz  \,  z \xi^2 \omega^3     \, ,
\end{eqnarray}

\begin{eqnarray}
\notag \rho^1_b(0)&=&- \frac{m_c m_s}{196608 \pi^8}   \int_{y_i}^{y_f}dy\int_{z_i}^{\zeta}dz  \,  z \xi^3 \omega^4    +\frac{13}{1966080 \pi^8}   \int_{y_i}^{y_f}dy\int_{z_i}^{\zeta}dz  \,   y z \xi^4 \omega^4 (5 s + 3 \omega)  \, .
\end{eqnarray}

The $c$ type integrals for $\rho_{QCD}^1(s)$,

\begin{eqnarray}
\notag \rho^1_c(13)&=&  \frac{13 m_c^3  \langle g_s^2GG\rangle  \langle\bar{q}q\rangle^2 \langle\bar{s}s\rangle}{41472  \pi^2 }  \int_{y_i}^{y_f}dy  \, \frac{1}{y^2 T^4}    - \frac{m_c^2 m_s  \langle g_s^2GG\rangle   \langle\bar{q}q\rangle^2 \langle\bar{s}s\rangle}{20736 \pi^2}   \int_{y_i}^{y_f}dy  \,   \frac{\zeta \widetilde{m}_c^2}{ T^6 y^2}   \\
\notag &&+ \frac{13 m_c  \langle g_s^2GG\rangle  \langle\bar{q}q\rangle^2 \langle\bar{s}s\rangle}{41472  \pi^2 }  \int_{y_i}^{y_f}dy  \, \frac{\zeta}{y^2} \frac{T^2 (-2+y)+\widetilde{m}_c^2 y}{T^4}   \\
\notag &&- \frac{m_c  \langle g_s^2GG\rangle  \langle\bar{q}q\rangle^2 \langle\bar{s}s\rangle}{27648  \pi^2 }   \int_{y_i}^{y_f}dy  \, \frac{1}{T^2 y}    \\
\notag &&- \frac{13 m_c  \langle g_s^2GG\rangle  \langle\bar{q}q\rangle^2 \langle\bar{s}s\rangle}{27648  \pi^2 }   \int_{y_i}^{y_f}dy  \, \frac{\zeta (\widetilde{m}_c^4+2 \widetilde{m}_c^2 T^2+2 T^4)}{T^6}   \\
\notag &&- \frac{13 m_c \langle\bar{q}g_s\sigma Gq\rangle \left[\langle\bar{q}g_s\sigma Gq\rangle \langle\bar{s}s\rangle+2 \langle\bar{q}q\rangle \langle\bar{s}g_s\sigma Gs\rangle\right]}{9216  \pi^2 }   \int_{y_i}^{y_f}dy  \, \frac{\zeta (\widetilde{m}_c^4+2 \widetilde{m}_c^2 T^2+2 T^4)}{T^6}   \\
\notag &&+  \frac{m_c \langle\bar{q}g_s\sigma Gq\rangle \left[ 7 \langle\bar{q}g_s\sigma Gq\rangle \langle\bar{s}s\rangle+20 \langle\bar{q}q\rangle \langle\bar{s}g_s\sigma Gs\rangle \right]}{9216   \pi^2 }   \int_{y_i}^{y_f}dy  \,   \frac{\widetilde{m}_c^2+T^2}{T^4}   \\
\notag &&+ \frac{7 m_c \langle\bar{q}g_s\sigma Gq\rangle \left[\langle\bar{q}g_s\sigma Gq\rangle \langle\bar{s}s\rangle+\langle\bar{q}q\rangle  \langle\bar{s}g_s\sigma Gs\rangle\right]}{2304  \pi^2 }   \int_{y_i}^{y_f}dy  \, \frac{\zeta}{y} \frac{\widetilde{m}_c^2+T^2}{T^4}   \\
\notag && - \frac{19 m_c \langle\bar{q}g_s\sigma Gq\rangle \left[ \langle\bar{q}g_s\sigma Gq\rangle \langle\bar{s}s\rangle+\langle\bar{q}q\rangle  \langle\bar{s}g_s\sigma Gs\rangle \right]}{12288  \pi^2 }  \int_{y_i}^{y_f}dy  \, \frac{1}{T^2 y}    \\
\notag &&+ \frac{m_s \langle\bar{q}g_s\sigma Gq\rangle \left[ 3 \langle\bar{q}g_s\sigma Gq\rangle \langle\bar{s}s\rangle+4 \langle\bar{q}q\rangle \langle\bar{s}g_s\sigma Gs\rangle \right]}{27648  \pi^2 }   \int_{y_i}^{y_f}dy  \, \frac{y \zeta (\widetilde{m}_c^6+3 \widetilde{m}_c^4 T^2+6 \widetilde{m}_c^2 T^4+6 T^6)}{T^8}   \\
\notag &&+ \frac{ m_s  \langle g_s^2GG\rangle  \langle\bar{q}q\rangle^2 \langle\bar{s}s\rangle}{55296  \pi^2 }   \int_{y_i}^{y_f}dy  \,   \frac{\widetilde{m}_c^2+T^2}{T^4}   + \frac{m_s \langle\bar{q}g_s\sigma Gq\rangle^2 \langle\bar{s}s\rangle}{18432  \pi^2 }   \int_{y_i}^{y_f}dy  \,   \frac{\widetilde{m}_c^2+T^2}{T^4 }  \\
\notag &&+ \frac{m_s  \langle g_s^2GG\rangle  \langle\bar{q}q\rangle^2 \langle\bar{s}s\rangle}{41472  \pi^2}   \int_{y_i}^{y_f}dy  \,  \frac{y \zeta (\widetilde{m}_c^6+3 \widetilde{m}_c^4 T^2+6 \widetilde{m}_c^2 T^4+6 T^6)}{T^8}   \\
\notag &&- \frac{5 m_s \langle\bar{q}g_s\sigma Gq\rangle \left[3 \langle\bar{q}g_s\sigma Gq\rangle \langle\bar{s}s\rangle+2 \langle\bar{q}q\rangle  \langle\bar{s}g_s\sigma Gs\rangle\right]}{55296  \pi^2 }   \int_{y_i}^{y_f}dy  \, \frac{\zeta (\widetilde{m}_c^4+2 \widetilde{m}_c^2 T^2+2 T^4)}{T^6}  \, ,
\end{eqnarray}

\begin{eqnarray}
\notag \rho^1_c(12)&=&  \frac{g_s^2 \langle\bar{q}q\rangle^2 \left[7 \langle\bar{q}q\rangle \langle\bar{s}s\rangle+ \langle\bar{s}s\rangle^2\right]}{23328  \pi^2 }   \int_{y_i}^{y_f}dy  \,   \frac{y  \zeta (\widetilde{m}_c^4+4 \widetilde{m}_c^2 T^2+6 T^4)}{T^4}   \\
\notag && -\frac{7  m_c m_s g_s^2 \langle\bar{q}q\rangle^3 \langle\bar{s}s\rangle}{23328 \pi^2}  \int_{y_i}^{y_f}dy  \, \frac{\zeta (\widetilde{m}_c^4+2 \widetilde{m}_c^2 T^2+2 T^4)}{T^6}       \, ,
\end{eqnarray}

\begin{eqnarray}
\notag \rho^1_c(11)&=& \frac{13 m_c \langle\bar{q}q\rangle \left[2 \langle\bar{q}g_s\sigma Gq\rangle \langle\bar{s}s\rangle+\langle\bar{q}q\rangle \langle\bar{s}g_s\sigma Gs\rangle\right]}{2304 \pi^2 }  \int_{y_i}^{y_f}dy  \, \frac{\zeta (\widetilde{m}_c^2+2 T^2)}{T^2}   \\
\notag && - \frac{m_c \langle\bar{q}q\rangle \left[14 \langle\bar{q}g_s\sigma Gq\rangle \langle\bar{s}s\rangle+13 \langle\bar{q}q\rangle \langle\bar{s}g_s\sigma Gs\rangle\right]}{4608 \pi^2}   \int_{y_i}^{y_f}dy  \,    \\
\notag && - \frac{7 m_c \langle\bar{q}q\rangle \langle\bar{q}g_s\sigma Gq\rangle \langle\bar{s}s\rangle}{576  \pi^2 }  \int_{y_i}^{y_f}dy  \, \frac{\zeta}{y}   \\
\notag &&+  \frac{7  m_c \langle\bar{q}g_s\sigma Gq\rangle  g_s^2  \langle\bar{s}s\rangle^2}{62208  \pi^4 }  \int_{y_i}^{y_f}dy  \,    \frac{\zeta (\widetilde{m}_c^2+2 T^2)}{T^2 }  \\
\notag &&+  \frac{m_c  g_s^2 \langle\bar{q}q\rangle^2 \left[14 \langle\bar{q}g_s\sigma Gq\rangle+\langle\bar{s}g_s\sigma Gs\rangle\right]}{124416  \pi^4 }   \int_{y_i}^{y_f}dy  \,    \frac{\zeta (\widetilde{m}_c^2+2 T^2)}{T^2}   \\
\notag &&+  \frac{5 m_s \langle\bar{q}q\rangle \langle\bar{q}g_s\sigma Gq\rangle \langle\bar{s}s\rangle}{4608  \pi^2 }   \int_{y_i}^{y_f}dy  \, \frac{\zeta(\widetilde{m}_c^2+2 T^2)}{T^2 }  \\
\notag &&- \frac{m_s \langle\bar{q}q\rangle \left[3 \langle\bar{q}g_s\sigma Gq\rangle \langle\bar{s}s\rangle+\langle\bar{q}q\rangle \langle\bar{s}g_s\sigma Gs\rangle\right]}{3456  \pi^2 }   \int_{y_i}^{y_f}dy  \,   \frac{y \zeta(\widetilde{m}_c^4+4 \widetilde{m}_c^2 T^2+6 T^4)}{T^4}  \\
\notag &&+ \frac{m_s  g_s^2 \langle\bar{q}q\rangle^2 \left[21 \langle\bar{q}g_s\sigma Gq\rangle-26 \langle\bar{s}g_s\sigma Gs\rangle\right]}{373248  \pi^4 }  \int_{y_i}^{y_f}dy  \,  \frac{y \zeta (\widetilde{m}_c^4+4 \widetilde{m}_c^2 T^2+6 T^4)}{T^4}  \, ,
\end{eqnarray}

\begin{eqnarray}
\notag \rho^1_c(10)&=&  \frac{m_c m_s \langle\bar{q}g_s\sigma Gq\rangle \left[39 \langle\bar{q}g_s\sigma Gq\rangle-28 \langle\bar{s}g_s\sigma Gs\rangle\right]}{36864  \pi^4 }   \int_{y_i}^{y_f}dy  \,  \frac{\zeta (\widetilde{m}_c^2+2 T^2)}{T^2}   \\
\notag &&+ \frac{m_c m_s \langle\bar{q}g_s\sigma Gq\rangle \left[-21 \langle\bar{q}g_s\sigma Gq\rangle+\langle\bar{s}g_s\sigma Gs\rangle\right]}{36864 \pi^4}  \int_{y_i}^{y_f}dy  \,    \\
\notag &&- \frac{7 m_c m_s \langle\bar{q}g_s\sigma Gq\rangle \left[3 \langle\bar{q}g_s\sigma Gq\rangle-2 \langle\bar{s}g_s\sigma Gs\rangle\right]}{9216  \pi^4  }  \int_{y_i}^{y_f}dy  \, \frac{\zeta}{y}   \\
\notag &&- \frac{m_c m_s  \langle g_s^2GG\rangle   \langle\bar{q}q\rangle \langle\bar{s}s\rangle}{36864  \pi^4 }   \int_{y_i}^{y_f}dy  \,   \frac{7 y+4 \zeta}{y}   \\
\notag &&+ \frac{m_c m_s   \langle g_s^2GG\rangle \langle\bar{q}q\rangle \left[13 \langle\bar{q}q\rangle-7 \langle\bar{s}s\rangle\right] }{55296  \pi^4 }  \int_{y_i}^{y_f}dy  \,    \frac{\zeta (\widetilde{m}_c^2+2 T^2)}{T^2}   \\
\notag &&+ \frac{m_c m_s \langle\bar{q}g_s\sigma Gq\rangle \langle\bar{s}g_s\sigma Gs\rangle}{9216 \pi^4 }  \int_{y_i}^{y_f}dy  \,   \frac{\zeta (\widetilde{m}_c^2+2 T^2)}{T^2}   \\
\notag &&+ \frac{\langle\bar{q}g_s\sigma Gq\rangle \left[2 \langle\bar{q}g_s\sigma Gq\rangle+13 \langle\bar{s}g_s\sigma Gs\rangle\right]}{12288 \pi^4}   \int_{y_i}^{y_f}dy  \,   y \zeta  \widetilde{m}_c^2    \\
\notag &&+  \frac{ \langle g_s^2GG\rangle  \langle\bar{q}q\rangle \left[\langle\bar{q}q\rangle+7 \langle\bar{s}s\rangle\right]}{27648 \pi^4 }  \int_{y_i}^{y_f}dy  \,   y \zeta \widetilde{m}_c^2  \, ,
\end{eqnarray}

\begin{eqnarray}
\notag \rho^1_c(9)&=&  \frac{m_s \langle\bar{q}q\rangle^2 \langle\bar{s}s\rangle}{576 \pi^2}   \int_{y_i}^{y_f}dy  \,  y \zeta \widetilde{m}_c^2    - \frac{m_s  g_s^2 \langle\bar{q}q\rangle^2 \left[7  \langle\bar{q}q\rangle-13  \langle\bar{s}s\rangle\right]}{31104 \pi^4}  \int_{y_i}^{y_f}dy  \,   y \zeta \widetilde{m}_c^2  \, .
\end{eqnarray}

The $d$ type integrals for $\rho_{QCD}^1(s)$,

\begin{eqnarray}
\notag \rho^1_d(10)&=&  \frac{m_c^3 m_s   \langle g_s^2GG\rangle \langle\bar{q}q\rangle \left[-13 \langle\bar{q}q\rangle+14 \langle\bar{s}s\rangle\right]}{55296  \pi^4 }   \int_{y_i}^{y_f}dy\int_{z_i}^{\zeta}dz  \, \frac{1}{T^2 y^2}    \\
\notag &&+ \frac{m_c^2   \langle g_s^2GG\rangle \langle\bar{q}q\rangle \left[\langle\bar{q}q\rangle+7 \langle\bar{s}s\rangle\right]}{13824 \pi^4}  \int_{y_i}^{y_f}dy\int_{z_i}^{\zeta}dz  \,   \frac{z \xi}{y^2} \frac{\overline{m}_c^2+2  T^2 }{T^2}    \\
\notag &&- \frac{m_c m_s   \langle g_s^2GG\rangle \langle\bar{q}q\rangle \left[13 \langle\bar{q}q\rangle-14 \langle\bar{s}s\rangle\right]}{55296  \pi^4 }   \int_{y_i}^{y_f}dy\int_{z_i}^{\zeta}dz  \,  \frac{z }{y^2} \frac{\overline{m}_c^2 y-2 T^2 \zeta}{T^2}   \\
\notag &&+ \frac{m_c m_s   \langle g_s^2GG\rangle \langle\bar{q}q\rangle \left[\langle\bar{q}q\rangle-\langle\bar{s}s\rangle\right]}{36864  \pi^4 }  \int_{y_i}^{y_f}dy\int_{z_i}^{\zeta}dz  \, \frac{1}{y}   \\
\notag &&+ \frac{19  m_c m_s \langle\bar{q}g_s\sigma Gq\rangle^2}{16384  \pi^4 }  \int_{y_i}^{y_f}dy\int_{z_i}^{\zeta}dz  \, \frac{1}{y}   \\
\notag &&- \frac{ \langle g_s^2GG\rangle  \langle\bar{q}q\rangle \left[\langle\bar{q}q\rangle+2 \langle\bar{s}s\rangle\right]}{36864 \pi^4} \int_{y_i}^{y_f}dy\int_{z_i}^{\zeta}dz  \,  \xi \overline{m}_c^2   \\
\notag &&+ \frac{ \langle g_s^2GG\rangle \langle\bar{q}q\rangle \left[\langle\bar{q}q\rangle+30 \langle\bar{s}s\rangle\right]}{36864 \pi^4}  \int_{y_i}^{y_f}dy\int_{z_i}^{\zeta}dz  \,  z  \overline{m}_c^2    \\
\notag &&- \frac{\langle\bar{q}g_s\sigma Gq\rangle \left[124 \langle\bar{q}g_s\sigma Gq\rangle+377 \langle\bar{s}g_s\sigma Gs\rangle\right]}{294912 \pi^4}  \int_{y_i}^{y_f}dy\int_{z_i}^{\zeta}dz  \,  z  \overline{m}_c^2    \\
\notag &&- \frac{\langle\bar{q}g_s\sigma Gq\rangle \left[6 \langle\bar{q}g_s\sigma Gq\rangle+5 \langle\bar{s}g_s\sigma Gs\rangle\right]}{73728 \pi^4}  \int_{y_i}^{y_f}dy\int_{z_i}^{\zeta}dz  \,  \xi  \overline{m}_c^2    \, ,
\end{eqnarray}

\begin{eqnarray}
\notag \rho^1_d(7)&=&  \frac{m_c^2 m_s   \langle g_s^2GG\rangle   \left[14 \langle\bar{q}q\rangle-13 \langle\bar{s}s\rangle\right]}{73728  \pi^6 } \int_{y_i}^{y_f}dy\int_{z_i}^{\zeta}dz  \,   \frac{z \xi^2 \overline{m}_c^2}{y^2 }  \, .
\end{eqnarray}

The $a$ type integrals for $\rho_{QCD}^0(s)$,

\begin{eqnarray}
\notag \rho^0_a(10)&=&  \frac{m_c \langle\bar{q}g_s\sigma Gq\rangle \left[\langle\bar{q}g_s\sigma Gq\rangle+12 \langle\bar{s}g_s\sigma Gs\rangle\right]}{3072 \pi^4}  \int_{y_i}^{y_f}dy \,  y   +  \frac{m_c   \langle g_s^2GG\rangle \langle\bar{q}q\rangle \left[\langle\bar{q}q\rangle+28 \langle\bar{s}s\rangle\right]}{27648 \pi^4}  \int_{y_i}^{y_f}dy \,  y\, ,
\end{eqnarray}

\begin{eqnarray}
\notag \rho^0_a(9)&=& - \frac{13 m_c^2 \langle\bar{q}q\rangle^2 \langle\bar{s}s\rangle}{288 \pi^2}  \int_{y_i}^{y_f}dy \,   - \frac{m_c^2 g_s^2 \langle\bar{q}q\rangle \left[7 \langle\bar{q}q\rangle^2+2  \langle\bar{q}q\rangle \langle\bar{s}s\rangle+7 \langle\bar{s}s\rangle^2\right]}{31104 \pi^4}  \int_{y_i}^{y_f}dy \, \\
\notag &&+ \frac{m_c m_s \langle\bar{q}q\rangle^2 \langle\bar{s}s\rangle}{576 \pi^2}  \int_{y_i}^{y_f}dy \,  y- \frac{m_c m_s g_s^2  \langle\bar{q}q\rangle^2 \left[28 \langle\bar{q}q\rangle-13 \langle\bar{s}s\rangle\right]}{31104 \pi^4}  \int_{y_i}^{y_f}dy \,  y\, ,
\end{eqnarray}

\begin{eqnarray}
\notag \rho^0_a(8)&=&  \frac{m_c^2 m_s \left[-78 \langle\bar{q}q\rangle \langle\bar{q}g_s\sigma Gq\rangle+9 \langle\bar{q}g_s\sigma Gq\rangle \langle\bar{s}s\rangle+7 \langle\bar{q}q\rangle \langle\bar{s}g_s\sigma Gs\rangle\right]}{4608 \pi^4}  \int_{y_i}^{y_f}dy \, \, .
\end{eqnarray}

The $b$ type integrals for $\rho_{QCD}^0(s)$,

\begin{eqnarray}
\notag \rho^0_b(10)&=&  \frac{ m_c  \langle g_s^2GG\rangle \langle\bar{q}q\rangle \left[\langle\bar{q}q\rangle+28 \langle\bar{s}s\rangle\right]}{27648  \pi^4 }    \int_{y_i}^{y_f}dy\int_{z_i}^{\zeta}dz  \,   \frac{z \xi (-2+3 y)}{y^2}\\
\notag &&+ \frac{m_c   \langle g_s^2GG\rangle \langle\bar{q}q\rangle \left[\langle\bar{q}q\rangle+2 \langle\bar{s}s\rangle\right]}{18432  \pi^4 }  \int_{y_i}^{y_f}dy\int_{z_i}^{\zeta}dz  \,    \frac{z^2-y \xi}{y z} \\
\notag &&- \frac{m_c   \langle g_s^2GG\rangle \langle\bar{q}q\rangle \left[\langle\bar{q}q\rangle-56 \langle\bar{s}s\rangle\right]}{18432 \pi^4}  \int_{y_i}^{y_f}dy\int_{z_i}^{\zeta}dz  \, \\
\notag &&- \frac{m_c \langle\bar{q}g_s\sigma Gq\rangle \left[3 \langle\bar{q}g_s\sigma Gq\rangle+4 \langle\bar{s}g_s\sigma Gs\rangle\right]}{6144 \pi^4}   \int_{y_i}^{y_f}dy\int_{z_i}^{\zeta}dz  \, \\
\notag &&- \frac{m_c \langle\bar{q}g_s\sigma Gq\rangle \left[3 \langle\bar{q}g_s\sigma Gq\rangle+52 \langle\bar{s}g_s\sigma Gs\rangle\right]}{12288  \pi^4 }  \int_{y_i}^{y_f}dy\int_{z_i}^{\zeta}dz  \, \frac{z}{y}\\
\notag &&- \frac{m_c \langle\bar{q}g_s\sigma Gq\rangle \left[\langle\bar{q}g_s\sigma Gq\rangle+2 \langle\bar{s}g_s\sigma Gs\rangle\right]}{6144  \pi^4 }  \int_{y_i}^{y_f}dy\int_{z_i}^{\zeta}dz  \,  \frac{\xi}{z}\, ,
\end{eqnarray}

\begin{eqnarray}
\notag \rho^0_b(8)&=&  \frac{m_c^2 m_s \langle\bar{q}g_s\sigma Gq\rangle \left[140 \langle\bar{q}q\rangle-29 \langle\bar{s}s\rangle\right]}{6144  \pi^4 }   \int_{y_i}^{y_f}dy\int_{z_i}^{\zeta}dz  \, \frac{1}{y}\\
\notag &&- \frac{m_c \left[3 \langle\bar{q}q\rangle \langle\bar{q}g_s\sigma Gq\rangle+24 \langle\bar{q}g_s\sigma Gq\rangle \langle\bar{s}s\rangle+28 \langle\bar{q}q\rangle \langle\bar{s}g_s\sigma Gs\rangle\right]}{3072 \pi^4}  \int_{y_i}^{y_f}dy\int_{z_i}^{\zeta}dz  \,   y (s+2 \omega)\\
\notag &&- \frac{m_c \langle\bar{q}g_s\sigma Gq\rangle \left[\langle\bar{q}q\rangle+2 \langle\bar{s}s\rangle\right]}{1536 \pi^4}  \int_{y_i}^{y_f}dy\int_{z_i}^{\zeta}dz  \,  \xi (s+2 \omega)\\
\notag &&- \frac{m_c \left[14 \langle\bar{q}g_s\sigma Gq\rangle \langle\bar{s}s\rangle+\langle\bar{q}q\rangle (\langle\bar{q}g_s\sigma Gq\rangle+14 \langle\bar{s}g_s\sigma Gs\rangle)\right]}{3072  \pi^4 }   \int_{y_i}^{y_f}dy\int_{z_i}^{\zeta}dz  \,   \frac{z \xi}{y}(s+2 \omega)\, ,
\end{eqnarray}

\begin{eqnarray}
\notag \rho^0_b(7)&=& \frac{m_c^3 m_s   \langle g_s^2GG\rangle  \left[56 \langle\bar{q}q\rangle-13 \langle\bar{s}s\rangle\right]}{147456 \pi^6} \int_{y_i}^{y_f}dy\int_{z_i}^{\zeta}dz  \,   \frac{(y+z) \xi^2}{y^3}  \\
\notag && + \frac{m_c^2   \langle g_s^2GG\rangle  \left[7 \langle\bar{q}q\rangle+\langle\bar{s}s\rangle\right]}{36864  \pi^6 }  \int_{y_i}^{y_f}dy\int_{z_i}^{\zeta}dz  \, \frac{\xi^2}{y^2}(s y-2 \zeta \omega) - \frac{m_c^2  \langle g_s^2GG\rangle  \langle\bar{q}q\rangle}{49152  \pi^6 }  \int_{y_i}^{y_f}dy\int_{z_i}^{\zeta}dz  \,   \frac{\xi^2 \omega}{y z} \\
\notag &&+  \frac{5 m_c^2   \langle g_s^2GG\rangle  \left[3 \langle\bar{q}q\rangle+\langle\bar{s}s\rangle\right]}{24576  \pi^6 }  \int_{y_i}^{y_f}dy\int_{z_i}^{\zeta}dz  \,   \frac{\xi \omega}{y}+  \frac{7  m_c^2  \langle g_s^2GG\rangle  \left[2 \langle\bar{q}q\rangle-\langle\bar{s}s\rangle\right]}{73728 \pi^6}   \int_{y_i}^{y_f}dy\int_{z_i}^{\zeta}dz  \,  \omega\\
\notag &&- \frac{m_c m_s   \langle g_s^2GG\rangle  \left[56 \langle\bar{q}q\rangle-13 \langle\bar{s}s\rangle\right]}{98304  \pi^6 }   \int_{y_i}^{y_f}dy\int_{z_i}^{\zeta}dz  \,   \frac{z \xi^2}{y^2}(s+2 \omega)\\
\notag &&- \frac{m_c m_s   \langle g_s^2GG\rangle  \left[4 \langle\bar{q}q\rangle-\langle\bar{s}s\rangle\right]}{196608  \pi^6 } \int_{y_i}^{y_f}dy\int_{z_i}^{\zeta}dz  \, \frac{\xi^2}{z}(s+2 \omega)\\
\notag &&+  \frac{7 m_c m_s   \langle g_s^2GG\rangle  \left[4 \langle\bar{q}q\rangle-\langle\bar{s}s\rangle\right]}{24576 \pi^6 }   \int_{y_i}^{y_f}dy\int_{z_i}^{\zeta}dz  \,  \xi (s+2 \omega)\\
\notag &&+ \frac{m_c m_s  \langle g_s^2GG\rangle  \left[2 \langle\bar{q}q\rangle-7 \langle\bar{s}s\rangle\right]}{49152  \pi^6 }   \int_{y_i}^{y_f}dy\int_{z_i}^{\zeta}dz  \,    \frac{z \xi}{y}(s+2 \omega)\\
\notag && - \frac{m_c m_s   \langle g_s^2GG\rangle  \left[28 \langle\bar{q}q\rangle-3 \langle\bar{s}s\rangle\right]}{147456 \pi^6}   \int_{y_i}^{y_f}dy\int_{z_i}^{\zeta}dz  \,   y (s+2 \omega)\, ,
\end{eqnarray}

\begin{eqnarray}
\notag \rho^0_b(6)&=&  \frac{m_c^2 m_s \langle\bar{q}q\rangle \left[26 \langle\bar{q}q\rangle-7 \langle\bar{s}s\rangle\right]}{768 \pi^4}   \int_{y_i}^{y_f}dy\int_{z_i}^{\zeta}dz  \,  \omega + \frac{m_c^2 m_s   g_s^2  \langle\bar{q}q\rangle^2}{20736 \pi^6}   \int_{y_i}^{y_f}dy\int_{z_i}^{\zeta}dz  \,  \omega\\
\notag &&- \frac{m_c \langle\bar{q}q\rangle \left[\langle\bar{q}q\rangle+28 \langle\bar{s}s\rangle\right]}{768 \pi^4} \int_{y_i}^{y_f}dy\int_{z_i}^{\zeta}dz  \,   y  \xi  \omega (s+\omega)\\
\notag &&- \frac{13  m_c g_s^2 \left[2 \langle\bar{q}q\rangle^2+\langle\bar{s}s\rangle^2\right]}{82944 \pi^6}  \int_{y_i}^{y_f}dy\int_{z_i}^{\zeta}dz  \,   y \xi \omega (s+\omega)\, ,
\end{eqnarray}

\begin{eqnarray}
\notag \rho^0_b(5)&=& - \frac{m_c^2 \left[13 \langle\bar{q}g_s\sigma Gq\rangle+2 \langle\bar{s}g_s\sigma Gs\rangle\right]}{8192 \pi^6}   \int_{y_i}^{y_f}dy\int_{z_i}^{\zeta}dz  \,  \xi \omega^2\\
\notag &&- \frac{m_c^2 \left[29 \langle\bar{q}g_s\sigma Gq\rangle+\langle\bar{s}g_s\sigma Gs\rangle\right]}{16384  \pi^6 } \int_{y_i}^{y_f}dy\int_{z_i}^{\zeta}dz  \,   \frac{\xi^2 \omega^2}{y}\\
\notag &&- \frac{m_c m_s \left[72 \langle\bar{q}g_s\sigma Gq\rangle-13 \langle\bar{s}g_s\sigma Gs\rangle\right]}{12288 \pi^6}  \int_{y_i}^{y_f}dy\int_{z_i}^{\zeta}dz  \,   y \xi \omega (s+\omega)\\
\notag &&- \frac{m_c m_s \langle\bar{q}g_s\sigma Gq\rangle}{4096  \pi^6 }  \int_{y_i}^{y_f}dy\int_{z_i}^{\zeta}dz  \,   \frac{(2 y+7 z) \xi^2}{y} \omega (s+\omega)\, ,
\end{eqnarray}

\begin{eqnarray}
\notag \rho^0_b(4)&=& - \frac{13 m_c^3   \langle g_s^2GG\rangle }{2359296  \pi^8 }  \int_{y_i}^{y_f}dy\int_{z_i}^{\zeta}dz  \, \frac{\xi^4}{y^2}\omega (s+\omega)+ \frac{m_c^2 m_s   \langle g_s^2GG\rangle }{147456  \pi^8 }   \int_{y_i}^{y_f}dy\int_{z_i}^{\zeta}dz  \, \frac{\xi^3}{y^2} \omega (s y-\zeta \omega)\\
\notag &&- \frac{m_c^2 m_s   \langle g_s^2GG\rangle  }{131072  \pi^8 }   \int_{y_i}^{y_f}dy\int_{z_i}^{\zeta}dz  \,   \frac{(4 y-5 \xi) \xi  \omega^2}{y}- \frac{13 m_c   \langle g_s^2GG\rangle }{7077888  \pi^8 }  \int_{y_i}^{y_f}dy\int_{z_i}^{\zeta}dz  \,    \frac{z  \xi^4 \omega}{y^2}\\
\notag &&+ \frac{m_c   \langle g_s^2GG\rangle }{9437184  \pi^8 }   \int_{y_i}^{y_f}dy\int_{z_i}^{\zeta}dz  \,   \frac{\xi^4}{z}  (3 s+2 \omega)  \omega^2 - \frac{3 m_c  \langle g_s^2GG\rangle }{262144 \pi^8}  \int_{y_i}^{y_f}dy\int_{z_i}^{\zeta}dz  \,  \xi^3 \omega^2 (3 s+2 \omega)\\
\notag &&- \frac{3 m_c  \langle g_s^2GG\rangle }{262144  \pi^8 }  \int_{y_i}^{y_f}dy\int_{z_i}^{\zeta}dz  \,    \frac{z (\xi^3)}{y} \omega^2 (3 s+2 \omega)- \frac{m_c  \langle g_s^2GG\rangle }{65536 \pi^8}  \int_{y_i}^{y_f}dy\int_{z_i}^{\zeta}dz  \,   y \xi^2 \omega^2 (3 s+2 \omega)\, ,
\end{eqnarray}

\begin{eqnarray}
\notag \rho^0_b(3)&=& - \frac{m_c m_s \left[56 \langle\bar{q}q\rangle-13 \langle\bar{s}s\rangle\right]}{24576 \pi^6}  \int_{y_i}^{y_f}dy\int_{z_i}^{\zeta}dz  \,   y \xi^2 \omega^2 (3 s+2 \omega)- \frac{m_c^2 \left[7 \langle\bar{q}q\rangle+\langle\bar{s}s\rangle\right]}{6144 \pi^6}  \int_{y_i}^{y_f}dy\int_{z_i}^{\zeta}dz  \,  \xi^2 \omega^3\, ,
\end{eqnarray}

\begin{eqnarray}
\notag \rho^0_b(0)&=&- \frac{m_c^2 m_s}{98304 \pi^8}  \int_{y_i}^{y_f}dy\int_{z_i}^{\zeta}dz  \,  \xi^3 \omega^4+ \frac{13 m_c}{3932160 \pi^8}  \int_{y_i}^{y_f}dy\int_{z_i}^{\zeta}dz  \,   y \xi^4 \omega^4 (5 s+2 \omega)\, .
\end{eqnarray}

The $c$ type integrals for $\rho_{QCD}^0(s)$,

\begin{eqnarray}
\notag \rho^0_c(13)&=&  \frac{m_c^3 m_s   \langle g_s^2GG\rangle  \langle\bar{q}q\rangle^2 \langle\bar{s}s\rangle }{82944  \pi^2 } \int_{y_i}^{y_f}dy  \,  \frac{y+\zeta}{y^3} \frac{-\widetilde{m}_c^2+T^2}{T^6}\\
\notag &&- \frac{13  m_c^2 \langle\bar{q}g_s\sigma Gq\rangle \left[\langle\bar{q}g_s\sigma Gq\rangle \langle\bar{s}s\rangle+2 \langle\bar{q}q\rangle  \langle\bar{s}g_s\sigma Gs\rangle\right]}{4608  \pi^2 }  \int_{y_i}^{y_f}dy  \, \frac{\widetilde{m}_c^4}{T^6}\\
\notag &&+ \frac{13  m_c^2  \langle g_s^2GG\rangle  \langle\bar{q}q\rangle^2 \langle\bar{s}s\rangle}{10368  \pi^2 }  \int_{y_i}^{y_f}dy  \, \frac{1}{y^2} \frac{2 T^2-\widetilde{m}_c^2 y}{T^4}\\
\notag &&- \frac{m_c^2   \langle g_s^2GG\rangle   \langle\bar{q}q\rangle^2 \langle\bar{s}s\rangle}{13824  \pi^2 }   \int_{y_i}^{y_f}dy  \, \left(\frac{1}{T^2 y \zeta}+\frac{13 \widetilde{m}_c^4}{T^6}\right) \\
\notag &&+ \frac{m_c^2 \langle\bar{q}g_s\sigma Gq\rangle \left[35 \langle\bar{q}g_s\sigma Gq\rangle \langle\bar{s}s\rangle+48 \langle\bar{q}q\rangle  \langle\bar{s}g_s\sigma Gs\rangle\right]}{4608 \pi^2}  \int_{y_i}^{y_f}dy  \, \frac{\widetilde{m}_c^2}{T^4 y} \\
\notag &&- \frac{19 m_c^2 \langle\bar{q}g_s\sigma Gq\rangle \left[\langle\bar{q}g_s\sigma Gq\rangle \langle\bar{s}s\rangle+\langle\bar{q}q\rangle \langle\bar{s}g_s\sigma Gs\rangle\right]}{6144  \pi^2 }   \int_{y_i}^{y_f}dy  \, \frac{1}{T^2 y \zeta} \\
\notag &&+ \frac{m_c m_s \langle\bar{q}g_s\sigma Gq\rangle \left[3 \langle\bar{q}g_s\sigma Gq\rangle \langle\bar{s}s\rangle+4 \langle\bar{q}q\rangle  \langle\bar{s}g_s\sigma Gs\rangle\right]}{55296  \pi^2 }  \int_{y_i}^{y_f}dy  \,   \frac{y \widetilde{m}_c^6}{T^8}\\
\notag &&+  \frac{  m_c m_s  \langle g_s^2GG\rangle \langle\bar{q}q\rangle^2 \langle\bar{s}s\rangle}{55296  \pi^2 }  \int_{y_i}^{y_f}dy  \,   \left(\frac{y^2+2 \zeta^2}{y^2 \zeta}\frac{\widetilde{m}_c^2}{T^4}+ \frac{2y \widetilde{m}_c^6}{3T^8}\right)\\
\notag &&- \frac{m_c m_s \left[3 \langle\bar{q}g_s\sigma Gq\rangle^2 \langle\bar{s}s\rangle+2 \langle\bar{q}q\rangle \langle\bar{q}g_s\sigma Gq\rangle \langle\bar{s}g_s\sigma Gs\rangle\right]}{27648  \pi^2 }  \int_{y_i}^{y_f}dy  \, \left(\frac{\widetilde{m}_c^4}{ T^6}+\frac{\zeta  \widetilde{m}_c^4 }{2 T^6 y}\right)\\
\notag &&+ \frac{m_c m_s \langle\bar{q}g_s\sigma Gq\rangle^2 \langle\bar{s}s\rangle}{18432  \pi^2 }  \int_{y_i}^{y_f}dy  \, \frac{\widetilde{m}_c^2}{T^4 \zeta} \, ,
\end{eqnarray}

\begin{eqnarray}
\notag \rho^0_c(12)&=& - \frac{7  m_c^2 m_s  g_s^2  \langle\bar{q}q\rangle^3 \langle\bar{s}s\rangle }{46656  \pi^2 }  \int_{y_i}^{y_f}dy  \, \frac{\widetilde{m}_c^4}{T^6}\\
\notag &&+ \frac{m_c g_s^2 \langle\bar{q}q\rangle^2\left[28 \langle\bar{q}q\rangle \langle\bar{s}s\rangle+\langle\bar{s}s\rangle^2\right]}{46656 \pi^2}   \int_{y_i}^{y_f}dy  \,  \frac{y (\widetilde{m}_c^4+2 \widetilde{m}_c^2 T^2+2T^4)}{T^4} \, ,
\end{eqnarray}

\begin{eqnarray}
\notag \rho^0_c(11)&=& - \frac{m_c^2 \langle\bar{q}q\rangle\left[70 \langle\bar{q}g_s\sigma Gq\rangle \langle\bar{s}s\rangle+13 \langle\bar{q}q\rangle \langle\bar{s}g_s\sigma Gs\rangle\right]}{2304 \pi^2} \int_{y_i}^{y_f}dy  \,  y \\
\notag &&+ \frac{13 m_c^2 \langle\bar{q}q\rangle\left[2 \langle\bar{q}g_s\sigma Gq\rangle \langle\bar{s}s\rangle+\langle\bar{q}q\rangle \langle\bar{s}g_s\sigma Gs\rangle\right]}{1152  \pi^2 }  \int_{y_i}^{y_f}dy  \,   \frac{\widetilde{m}_c^2+T^2}{T^2}\\
\notag &&+ \frac{m_c^2 \left[7 g_s^2  \langle\bar{q}q\rangle^2 \langle\bar{q}g_s\sigma Gq\rangle+7 \langle\bar{q}g_s\sigma Gq\rangle g_s^2 \langle\bar{s}s\rangle^2+2 g_s^2 \langle\bar{q}q\rangle^2 \langle\bar{s}g_s\sigma Gs\rangle\right]}{124416  \pi^4 }  \int_{y_i}^{y_f}dy  \,  \frac{\widetilde{m}_c^2+T^2}{T^2}\\
\notag &&- \frac{m_c m_s \langle\bar{q}q\rangle \left[3 \langle\bar{q}g_s\sigma Gq\rangle \langle\bar{s}s\rangle+\langle\bar{q}q\rangle \langle\bar{s}g_s\sigma Gs\rangle\right]}{6912 \pi^2} \int_{y_i}^{y_f}dy  \, \frac{y(\widetilde{m}_c^4+2 \widetilde{m}_c^2 T^2+2 T^4)}{T^4} \\
\notag &&+ \frac{m_c m_s  g_s^2 \langle\bar{q}q\rangle^2 \left[42 \langle\bar{q}g_s\sigma Gq\rangle-13 \langle\bar{s}g_s\sigma Gs\rangle\right]}{373248  \pi^4 }  \int_{y_i}^{y_f}dy  \,  \frac{y (\widetilde{m}_c^4+2 m_c^2 T^2+2 T^4)}{T^4}\\
\notag &&+  \frac{m_c m_s \langle\bar{q}q\rangle \langle\bar{q}g_s\sigma Gq\rangle \langle\bar{s}s\rangle}{4608 \pi^2 }  \int_{y_i}^{y_f}dy  \,   \frac{2 y+\zeta}{y} \frac{\widetilde{m}_c^2+T^2}{T^2} \, ,
\end{eqnarray}

\begin{eqnarray}
\notag \rho^0_c(10)&=&  \frac{m_c^2 m_s \langle\bar{q}g_s\sigma Gq\rangle \left[13 \langle\bar{q}g_s\sigma Gq\rangle-2 \langle\bar{s}g_s\sigma Gs\rangle \right]}{6144  \pi^4 }  \int_{y_i}^{y_f}dy  \,   \frac{\widetilde{m}_c^2+T^2}{T^2}\\
\notag &&+ \frac{m_c^2 m_s  \langle g_s^2GG\rangle \langle\bar{q}q\rangle \left[52 \langle\bar{q}q\rangle-7 \langle\bar{s}s\rangle \right]}{110592  \pi^4 }  \int_{y_i}^{y_f}dy  \,   \frac{ \widetilde{m}_c^2 + T^2 }{T^2}\\
\notag &&- \frac{m_c^2 m_s   \langle g_s^2GG\rangle  \langle\bar{q}q\rangle \langle\bar{s}s\rangle}{4096  \pi^4 }  \int_{y_i}^{y_f}dy  \, \frac{1}{y}\\
\notag &&+ \frac{m_c^2 m_s \langle\bar{q}g_s\sigma Gq\rangle \left[-210 \langle\bar{q}g_s\sigma Gq\rangle+29 \langle\bar{s}g_s\sigma Gs\rangle\right]}{36864  \pi^4 }  \int_{y_i}^{y_f}dy  \,  \frac{1}{y}\\
\notag &&+ \frac{m_c \langle\bar{q}g_s\sigma Gq\rangle \left[\langle\bar{q}g_s\sigma Gq\rangle+12 \langle\bar{s}g_s\sigma Gs\rangle\right]}{6144 \pi^4}   \int_{y_i}^{y_f}dy  \,  y  \widetilde{m}_c^2 \\
\notag &&+ \frac{m_c   \langle g_s^2GG\rangle \langle\bar{q}q\rangle  \left[\langle\bar{q}q\rangle+28 \langle\bar{s}s\rangle\right]}{55296 \pi^4}  \int_{y_i}^{y_f}dy  \,  y \widetilde{m}_c^2\, ,
\end{eqnarray}

\begin{eqnarray}
\notag \rho^0_c(9)&=& - \frac{m_c m_s g_s^2  \langle\bar{q}q\rangle^2 \left[28 \langle\bar{q}q\rangle-13 \langle\bar{s}s\rangle\right]}{62208 \pi^4}   \int_{y_i}^{y_f}dy  \,  y   \widetilde{m}_c^2 + \frac{m_c m_s \langle\bar{q}q\rangle^2 \langle\bar{s}s\rangle}{1152 \pi^2}  \int_{y_i}^{y_f}dy  \,  y  \widetilde{m}_c^2\, .
\end{eqnarray}

The $d$ type integrals for $\rho_{QCD}^0(s)$,

\begin{eqnarray}
\notag \rho^0_d(10)&=&  \frac{m_c^3   \langle g_s^2GG\rangle \langle\bar{q}q\rangle \left[\langle\bar{q}q\rangle+28  \langle\bar{s}s\rangle\right]}{55296  \pi^4 }   \int_{y_i}^{y_f}dy\int_{z_i}^{\zeta}dz  \,   \frac{\xi}{y^2} \frac{\overline{m}_c^2+T^2}{T^2}\\
\notag &&- \frac{m_c^2 m_s   \langle g_s^2GG\rangle \langle\bar{q}q\rangle \left[26 \langle\bar{q}q\rangle-7 \langle\bar{s}s\rangle\right]}{27648  \pi^4 }  \int_{y_i}^{y_f}dy\int_{z_i}^{\zeta}dz  \, \frac{1}{y^2} \frac{T^2 (-2+y)+\overline{m}_c^2 y}{T^2} \\
\notag &&+ \frac{m_c^2 m_s   \langle g_s^2GG\rangle \langle\bar{q}q\rangle \left[2 \langle\bar{q}q\rangle-\langle\bar{s}s\rangle\right]}{36864  \pi^4 }  \int_{y_i}^{y_f}dy\int_{z_i}^{\zeta}dz  \, \frac{1}{y z} \\
\notag &&+ \frac{19 m_c^2 m_s \langle\bar{q}g_s\sigma Gq\rangle^2}{8192  \pi^4 }   \int_{y_i}^{y_f}dy\int_{z_i}^{\zeta}dz  \, \frac{1}{y z} \\
\notag &&+ \frac{m_c   \langle g_s^2GG\rangle \langle\bar{q}q\rangle \left[\langle\bar{q}q\rangle+28 \langle\bar{s}s\rangle\right]}{55296  \pi^4 }  \int_{y_i}^{y_f}dy\int_{z_i}^{\zeta}dz  \,    \frac{z \xi}{y^2} \frac{\overline{m}_c^4 y+\overline{m}_c^2 T^2 (-2+4 y)}{T^2}\\
\notag &&+ \frac{m_c   \langle g_s^2GG\rangle \langle\bar{q}q\rangle \left[\langle\bar{q}q\rangle+2 \langle\bar{s}s\rangle\right]}{36864  \pi^4 }  \int_{y_i}^{y_f}dy\int_{z_i}^{\zeta}dz  \,  \frac{(z^2-y \xi) \overline{m}_c^2 }{y z} \\
\notag &&- \frac{ \langle g_s^2GG\rangle  m_c \langle\bar{q}q\rangle \left[\langle\bar{q}q\rangle-56 \langle\bar{s}s\rangle\right]}{36864 \pi^4}  \int_{y_i}^{y_f}dy\int_{z_i}^{\zeta}dz  \,   \overline{m}_c^2 \\
\notag &&- \frac{m_c \langle\bar{q}g_s\sigma Gq\rangle \left[ 3 \langle\bar{q}g_s\sigma Gq\rangle+4 \langle\bar{s}g_s\sigma Gs\rangle \right]}{12288 \pi^4}   \int_{y_i}^{y_f}dy\int_{z_i}^{\zeta}dz  \,  \overline{m}_c^2\\
\notag &&- \frac{m_c \langle\bar{q}g_s\sigma Gq\rangle \left[3 \langle\bar{q}g_s\sigma Gq\rangle+52 \langle\bar{s}g_s\sigma Gs\rangle \right]}{24576  \pi^4 }  \int_{y_i}^{y_f}dy\int_{z_i}^{\zeta}dz  \,   \frac{z  \overline{m}_c^2 }{y}\\
\notag &&- \frac{m_c \langle\bar{q}g_s\sigma Gq\rangle \left[\langle\bar{q}g_s\sigma Gq\rangle+2 \langle\bar{s}g_s\sigma Gs\rangle\right]}{12288  \pi^4 }  \int_{y_i}^{y_f}dy\int_{z_i}^{\zeta}dz  \,   \frac{\xi \overline{m}_c^2}{z}\, ,
\end{eqnarray}

\begin{eqnarray}
\notag \rho^0_d(7)&=&  \frac{m_c^3 m_s  \langle g_s^2GG\rangle   \left[56 \langle\bar{q}q\rangle-13 \langle\bar{s}s\rangle \right]}{294912  \pi^6 }   \int_{y_i}^{y_f}dy\int_{z_i}^{\zeta}dz  \,   \frac{(y+z) \xi^2 \overline{m}_c^2}{y^3}\, .
\end{eqnarray}

\section*{Acknowledgements}
This work is supported by National Natural Science Foundation, Grant Number 12175068 and Youth Foundation of NCEPU, Grant Number 93209703.

\end{document}